\documentclass[leqno, 12pt]{article}
\usepackage{oxford3}
\usepackage{kokoszka}
\usepackage{graphicx}
\usepackage{amsfonts}
\usepackage{amsmath}
\usepackage{amssymb}
\usepackage{enumitem}
\usepackage{color}
\usepackage{epsfig,subfigure}
\usepackage{epstopdf}
\usepackage{booktabs}
\usepackage{float, lscape,multirow}

\renewcommand{\baselinestretch}{1.02}
\begin{document}

\title{Functional diffusion driven stochastic volatility model}

\author{
Piotr Kokoszka$^a$\footnote{Corresponding author.
E-mail address: Piotr.Kokoszka@colostate.edu}
\and
Neda Mohammadi$^a$
\and
Haonan Wang$^a$
\and
Shixuan Wang$^b$
}
\date{}
\maketitle

${}^a$ Department of Statistics, Colorado State University, USA

${}^b$ Department of Economics, University of Reading, UK

\begin{abstract}
We propose a stochastic volatility model for  time series 
of curves. It is motivated by dynamics of intraday price curves 
that exhibit both between days dependence  and  intraday 
price evolution. The curves are suitably normalized to 
stationary in a function space and are functional analogs of 
point-to-point daily returns. The between curves dependence 
is modeled by a latent autoregression. 
The within curves behavior is modeled by a diffusion 
process.   We establish the properties of the model 
and propose several approaches to its estimation. 
These approaches are justified by asymptotic arguments 
that involve an interplay between between the latent 
autoregression and the intraday diffusions. The asymptotic 
framework combines the increasing number of daily curves and 
the refinement of the discrete grid on which each daily curve is 
observed. Consistency rates for the 
estimators of the intraday volatility curves are derived
as well as the asymptotic normality of the 
estimators of the latent autoregression. 
The estimation approaches  are 
further explored and compared by an application to intraday price curves of 
over seven thousand U.S. stocks and an  informative simulation study.

\medskip

\noindent{\it JEL classification:} C51,  
C58. 

\medskip

\noindent{\it Keywords:} Functional data analysis, Intraday price curves,
It{\^o} diffusion process, stochastic volatility.

\end{abstract}

\section{Introduction}\label{s:intro}
Time dependent volatility is one of the main features of  financial time series.
Diffusion models for price evolution have been been employed for 
over one hundred years, starting perhaps with \citetext{bachelier:1900}, 
with a robust development since the 1970s. This paper proposes 
a model that combines a diffusion model for the intraday price 
evolution with a stochastic volatility paradigm for day-to-day 
dependence. We combine the tools of time series analysis, 
functional data analysis and stochastic calculus. 

The concept of univariate conditional  heteroscedasticity traces back at least 
to  \citetext{engle:1982} who proposed autoregressive conditional heteroscedastic (ARCH) models and the influential contribution of \citetext{bollerslev_generalized_1986} who proposed generalized autoregressive conditional heteroscedastic (GARCH) models. Later on,
\citetext{bollerslev_capital_1988},
\citetext{bollerslev_modelling_1990},
\citetext{engle_multivariate_1995},
\citetext{engle_dynamic_2002}, among many others,
utilized GARCH models in analysing multivariate heteroscedastic time series. An overview of  multivariate GARCH models is provided in \citetext{bauwens_multivariate_2006}  and \citetext{silvennoinen_multivariate_2009}.
Bayesian inference for multivariate GARCH models is addressed in
\citetext{vrontos_full-factor_2003}.
\citetext{hormann:horvath:reeder:2013}
study conditional heteroscedasticity in the framework of functional
data analysis (FDA). In particular, they propose a functional
version of the ARCH model, which is extended to functional GARCH models by
\citetext{aue_functional_2017}. The common feature of all above
models is that the random volatility is measurable with respect to
\textit{past events}, i.e.  conditionally on  \textit{past observations}
the volatility process is deterministic. To illustrate this point,  in the
basic univariate ARCH(1), model
\[
\left\{
\begin{array}{ll}
    r_i&= g_i w_i, \ \ \ w_i \sim \ {\rm iid} \ N(0, 1);\\
g_i^2&=  \eta_1 r^2_{i-1} + \eta_0,
\end{array}
\right.
\]
the current volatility $g_i^2$ is a function of the previous observation
$r_{i-1}$. More complex functions lead to various models in the
ARCH family.
An alternative approach to model the randomness in   volatility  is
the so called stochastic volatility (SV), where in contrast to ARCH-type models,
conditionally on past events the volatility process is not fully observable.
The simplest univariate formulation of stochastic volatility is
\begin{align} \label{e:SVr:SVg}
\left\{
\begin{array}{ll}
    r_i &= g_i w_i, \ \ \ w_i \sim \ {\rm iid} \ N(0, \sg_w^2);\\
g_i &= \exp\lbr  \fg \log g_{i-1} + \eg_i \rbr, \ \ \ \eg_i \sim \ {\rm iid}
\ N(0, \sg_\eg^2),
\end{array}
\right .
\end{align}
where $\fg$  is a parameter  satisfying $|\fg| < 1$.
To draw an analogy to linear time series models,
GARCH models are analogous to ARMA models,
while the stochastic volatility models are analogous to state space models.
The stochastic volatility model was introduced by
\citetext{taylor_financial_1982} and further developed by
\citetext{taylor_modeling_1994},
\citetext{ghysels_stochastic_1996},
\citetext{shephard_statistical_1996} and
\citetext{taylor_modelling_2008}.
\citetext{harvey_multivariate_1994},
\citetext{danielsson_multivariate_1998}, and
\citetext{asai_multivariate_2006}
extend the stochastic volatility models to multivariate settings.
A Bayesian approach to multivariate stochastic volatility models
is investigated in \citetext{yu_multivariate_2006}.
There have been thousands of contributions to ARCH and SV modeling,
and a number of excellent  monographs have been published.
We listed only selected papers emphasizing those dealing
with multivariate models. We propose a {\em functional} SV model.

In its simplest form, our model is given by equations
analogous to equations \refeq{SVr:SVg}:
\begin{align}
\label{e:R:g}
\left\{
\begin{array}{ll}
    R_i(t) =& g_i\int_0^t \sg(u) d W_i(u), \quad t \in [0,1],   \\
     \log g_i =& \varphi \log g_{i-1} + \varepsilon_i, \quad \eg_i \sim \ {\rm iid}
\ \mathcal{W N}(0, \sg_\eg^2), \quad  i \in \mathbb{Z},
\end{array}
\right .
\end{align}
where $\mathcal{W N}(0, \sg_\eg^2)$ denotes white noise with mean zero
and finite second moment $\sg_\eg^2$. The $W_i(\cdot)$s are independent
standard Brownian motions  independent of the error sequence
$\{\varepsilon_i\}$. The random coefficients $g_i$ are positive with
probability one, $\sigma (\cdot)$ is a nonparametric function with
$\Vert\sigma \Vert_{\infty} < \infty$ and $\varphi$ is a scalar
with $\vert \varphi \vert  < 1$, see Section \ref{s:est} for a full discussion.
Setting $t=1$ in \eqref{e:R:g}, we obtain
\begin{align*}
R_i(1) = g_i \int_0^1 \sg(u) dW_i(u),
\end{align*}
retrieving the univariate model \eqref{e:SVr:SVg} with
$r_i = R_i(1) $, $w_i = \int_0^1 \sg(u) dW_i(u)$
and $\sg_w^2 = \int_0^1 \sg^2(u) du$. This  property shows that
the proposed model \eqref{e:R:g} extends the well established univariate
model \eqref{e:SVr:SVg}.

The data that motivate model \refeq{R:g} are intraday price curves
suitably transformed to form a stationary sequence of curves. Detailed
definitions are given in Section \ref{s:emp}. Basically,
$R_i(t)$ is the cumulative return on trading day $i$ up to intraday
time $t$. If the exchange opening time is rescaled to the unit
interval, $R_i(1)$ is the return on day $i$, except that we
compare the closing price to the opening price, rather than to the
closing price on the previous day.
The AR(1) formulation appearing in the second equation in \eqref{e:R:g}
models the dependence structure between the daily curves.
The strength of this dependence  is  quantified through $\varphi$,
or more parameters, as explained in Section \ref{s:ext}.
\citetext{muller_functional_2011} proposed a framework
that uses diffusions to model volatility, but assumed
i.i.d curves. It is broadly believed that there is day-to-day dependence
in price data. Uncorrelated stochastic volatility curves are also studied in
\citetext{jang_functional_2021}  who apply  dimension reduction
through basis expansions. This technique, however,
ignores the roughness of the sample paths which is a
crucial property for price processes; continuous time price models are
diffusions with nowhere differentiable paths.
\citetext{chong_statistical_2022} consider the roughness to
be  a crucial  feature of financial continuous time models and
propose stochastic volatility models driven by fractional Brownian
motion with the Hurst index  $H < 1/2$.
Their inference  targets the Hurst index $H$ and they establish a
minimax theory for this parameter. Their work completes the results
of \citetext{rosenbaum_estimation_2008} that focus on a similar
inference problem with $H > 1/2$.
The between curve dynamic dependence is not
addressed in \citetext{rosenbaum_estimation_2008} and
\citetext{chong_statistical_2022} because they consider a single
time interval.

The proposed model is comprehensive in that it
models the dependence between the curves through a latent autoregression,
roughness of sample paths through the stochastic integral with respect to
Brownian motion and  intraday volatility  through the product
$g_i \sigma(\cdot)$, where the $g_i$ exhibit day-today dependence.
Our main objective is to establish conditions for the existence
of solutions to model equations and develop inference for model
parameters, which include the autoregressive parameters and the
function $\sigma(\cdot)$. It is thus a parametric-nonparametric estimation
problem that involves challenges not encountered in previous
research, as explained in Section \ref{s:est}. Estimation must take into account the fact that the daily trajectories
are observed at discrete time points within a day.
The availability of replications indexed by $i$ (day)  suggests  to
approach the problem from a functional data analysis  perspective.
However, the roughness of the trajectories makes
established FDA approaches that assume smoothness less attractive.
There is comprehensive research on rough trajectories, with
a typical model assuming that smooth trajectories are
observed with i.i.d. randomly scattered measurement errors, see
\citetext{yao_functional_2005},
\citetext{hall:muller:wang:2006} and
\citetext{li_uniform_2010}, among many others.
These approaches are effective for biomedical data.
For continuous time price data, roughness is however a fundamental  property
modeled by diffusions rather than caused by completely random
errors. Our strategy is built on approximating the (latent) quadratic variation
process that usually satisfies some desired regularity property.
Utilizing the approximating quadratic variation processes in lieu
of the hidden curves paves the way to  employing FDA techniques.
Consistency of the realized quadratic variation process and hence
estimated objects are addressed under decay of step size, $\Delta$ say,
and growth of sample size $N$. \citetext{galbraith_garch_2015}
study GARCH models and propose an argument based on the realized
quadratic variation process. Due to the nature of the GARCH model,
\citetext{galbraith_garch_2015}  do not require the  consistency of
the realized quadratic variation process and treat the induced
discrepancy as a noise term.
This substantially differentiates their approach from the current study.

The remainder of the paper is organized as follows. In Section \ref{s:prelim},
for ease of reference and to fix notation and
terminology, we present relevant information related to stochastic integrals.
Section \ref{s:est} is dedicated to the explanation of the proposed model
and estimation approaches in the simplest case of the latent autoregression
of order 1, which already contains  key model and estimation features,
and allows us to focus
on them. Extension to AR($p)$ latent autoregression is developed in
Section~\ref{s:ext}. An Application to U.S. intraday stock prices and
a simulation study are presented in Section \ref{s:emp}. We conclude
in Section~\ref{s:sum} with
a brief summary and discussion of directions for future research.
Online Supplementary Material contains the proofs of
the results of Sections \ref{s:est} and \ref{s:ext}, as well
as additional information related to the empirical analysis in
Section~\ref{s:emp}.

\section{Preliminaries}\label{s:prelim}
 For ease of reference, we collect in this section useful  facts  related to  It{\^o} diffusion processes. More details are provided in the monographs of  \citetext{Oksendal03}, \citetext{ait-sahalia_high-frequency_2014}, and \citetext{karatzas_brownian_1991}, among others.  Let $(\Omega, \mathcal{F}, \mathbb{P})$  be the underlying probability space. We define the $\mathbb{R}$--valued It{\^o}  diffusion (diffusion in short) process by
\begin{align}\label{e:ito:sde}
    dX(t) =& \mu(t,X(t))dt + \sigma(t,X(t))dW(t),\qquad 0 < t,\\ \nonumber
    X(t) = &  X(0),\qquad  t = 0,
\end{align}
where $W$ denotes a standard Brownian motion. Integrals  with respect to $W$ should be understood in the sense of the stochastic It{\^o} integral. We assume that the initial distribution is independent of the $\sigma$-algebra $\mathcal{F}_{\infty}$ generated by $\left\{W(t)\right\}_{t \geq 0}$.
 The function $\mu(\cdot,\cdot)$,  the so called drift (viscosity), and the function   $\sigma(\cdot,\cdot)$, the so called diffusion (volatility),  are  Borel  measurable. The following theorem,  a consequence of Theorem 5.2.1 in \citetext{Oksendal03}, provides sufficient conditions for  existence and uniqueness of the process $X$ satisfying the stochastic differential equation \eqref{e:ito:sde}.

 \begin{theorem}[Existence and Uniqueness]\label{thm:exi:uni}
Let $T$ be a positive number and functions $\mu(\cdot, \cdot) : [0,T]\times \mathbb{R} \longmapsto \mathbb{R}$ and $\sigma(\cdot, \cdot) : [0,T]\times \mathbb{R} \longmapsto \mathbb{R}$ be measurable functions satisfying  the linear growth condition
\begin{equation}\label{e:lin:gro}
  |\mu(t,x)| + | \sigma(t,x) | \leq L(1+|x|), \quad x\in \mathbb{R}, \: t\in[0,T],
\end{equation}
and Lipschitz continuity in the space variable i.e.
\begin{equation}\label{e:lipsch}
  |\mu(t,x)| - \mu(t,y)| + | \sigma(t,x) - \sigma(t,y) | \leq L |x - y|,  \quad x,y\in \mathbb{R}, \: t\in[0,T],
\end{equation}
for some constant $L>0$. Let moreover $X(0)$ be a random variable independent of the $\sigma$-algebra $\mathcal{F}_{\infty}$ generated by $\left\{W(t)\right\}_{t\geq 0 }$ and such that $\mathbb{E} \left[X(0) ^2\right] < \infty$.
Then the stochastic differential equation \eqref{e:ito:sde} admits a unique  solution with {time-continuous} trajectories and adapted to the filtration $\left\{\mathcal{F}^{X_0}_t\right\}$ generated by $X(0)$ and the standard Brownian motion $\left\{W(s)\right\}_{s \leq t}$. Moreover,
\begin{align*}
    \mathbb{E}\left[\int_0^T |X(t)|^2 dt \right] < \infty.
\end{align*}
\end{theorem}

We now present the It{\^o} isometry that is  one of the most useful results in the context of stochastic calculus.
Define $\mathcal{V} = \mathcal{V}(S,T)$ to be the class of functions
\begin{align*}
   f: [0,\infty) \times \Omega \longrightarrow \mathbb{R},
\end{align*}
satisfying
\begin{itemize}
    \item [(i)] $(t,\omega) \mapsto f(t,\omega)$  is $\mathcal{B} \times \mathcal{F}$ measurable, where $\mathcal{B}$ denotes the Borel $\sigma$--algebra on $[0,\infty)$,

    \item [(ii)] $f(t,\cdot)$ is $\mathcal{F}_t$ adapted, where $\mathcal{F}_t$  is the $\sigma$--algebra generated by $\left\{W(s)\right\}_{s \leq t}$,

    \item [(iii)] $\mathbb{E}\left[\int_S^T |f(t,\omega)|^2 dt \right] < \infty.$
\end{itemize}
If $f \in \mathcal{V}$,  then, according to Corollary 3.1.7 in \citetext{Oksendal03},
\begin{align*}
 \mathbb{E}\left[\int_S^T f(t,\omega) dW(t)\right]^2    = \mathbb{E}\left[\int_S^T |f(t,\omega)|^2 dt \right].
\end{align*}
The quadratic variation process and its empirical counterpart, also known as realized quadratic variation process, play a fundamental role in the study of It{\^o} semimartingales. Our inferential procedure and consequently our consistency results heavily rely on calculation of realized quadratic variation and its convergence to the true process.
Let $X(\cdot)$ satisfy model \eqref{e:ito:sde}  and $ \{t_{k}\} $ be an
equispaced  partition of the unit interval with  step size $\Delta$. The realized quadratic variation process at point $t$ is defined through the following sum of squared  increments:
  \begin{align}\label{e:real:<,>}
  \sum_{k}\vert X(t_{k})-X(t_{k-1}) \vert^2 \mathbb{I}\{t_{k}\leq t\} ,  \quad   t \in [0,1].
\end{align}
The above sum is tightly related to the  quadratic variation  process  which will be denoted by $\lip X,X\rip_t$, $t \in [0,1]$. According to  Proposition 3.2.17 in \citetext{karatzas_brownian_1991}, under the conditions of Theorem \ref{thm:exi:uni},  the process $\lip X,X\rip_t$ can be defined by
\begin{align}\label{e:<,>}
    \lip X,X\rip_{t} = \int_0^{t} \sigma^2 (u,X(u)) du, \quad t \in [0,1].
\end{align}
We are now ready to present Theorem    \ref{thm:empQV} which examines convergence of the realized quadratic variation to the theoretical counterpart $\lip X,X\rip_t$, see Theorem  1.14 and relation (3.23) in \citetext{ait-sahalia_high-frequency_2014} for a more general statement.

\begin{theorem}\label{thm:empQV}
  Assume the conditions of Theorem \ref{thm:exi:uni}, and let $\Pi_N = \{t_{N,k}\}$ be a sequence of partitions of the unit interval with step size $\Delta(N)$ that tends to zero, as $N$ increases. Then the realized quadratic variation process tends to the quadratic variation process uniformly in probability, i.e. as $ \Delta(N) \longrightarrow 0$
  \begin{align}\label{e:emp->}
 \underset{0 \leq t \leq 1}{\sup} \left \vert \sum_{k}\vert X(t_{N,k})-X(t_{N,k-1}) \vert^2 \mathbb{I}\{t_{N,k}\leq t\} - \int_0^t \sigma^2 (u,X(u)) du \right \vert \overset{\mathbb{P}}{\longrightarrow } 0,
\end{align}
where $\overset{\mathbb{P}}{\longrightarrow }$ denotes convergence in probability.
\end{theorem}

We close this section by stating a  time change result as a
corollary of Dambis--Dubins--Schwarz theorem which expresses
\textit{any} continuous local martingale as a time change of a Brownian motion,
see e.g. Section 5.3.2 in \citetext{le2016brownian}.
\begin{corollary}\label{cor:tim:chng}
Assume the setting of Theorem \ref{thm:exi:uni} and set the drift function
equal to zero, $\mu(\cdot,\cdot) = 0$. Then $X(t)$ has the same distribution
as  $W\left( \lip X,X\rip_t\right)$, i.e.
\[
\lbr \int_0^t \sigma(u,X(u)) dW(u), \ t\in [0,1] \rbr
 \eqd
 \lbr W\left( \int_0^t \sigma^2(u,X(u)) du\right), \ t\in [0,1] \rbr,
\]
where the equality in distribution is in the space $C([0,1])$ of
continuous functions.
\end{corollary}

\section{Development and estimation of order 1 model}\label{s:est}
In this section, we focus on the model defined in the Introduction by
equations \refeq{R:g}. It already contains the most essential elements of the
proposed framework and the important issues related to its properties
and estimation are easier to explain. An extension to higher order
latent autoregressions is presented in Section \ref{s:ext}. For ease of
reference,  we display equations \refeq{R:g} as
\begin{align}
\label{e:R1}
R_i(t) =& g_i\int_0^t \sg(u) d W_i(u), \quad t \in [0,1], \quad  i \in \mathbb{Z},\\
 \label{e:g:AR1}   \log g_i =& \varphi \log g_{i-1} + \varepsilon_i, \quad \eg_i \sim \ {\rm iid}
\ \mathcal{W N}(0, \sg_\eg^2),
\end{align}
where $\mathcal{W N}(0, \sg_\eg^2)$ denotes nondeterministic
white noise with mean zero and finite second moment $\sg_\eg^2$.

We will use the following assumptions. Not all of them are needed for every
result, as specified in the following, but all results are valid if all
conditions listed below hold.
\begin{enumerate}
 \item \label{itm:sig} The function $\sigma(\cdot)$ is nonnegative and deterministic with $\underset{0 \leq t \leq 1}{\sup}  \sigma(t) = \Vert\sigma \Vert_{\infty} < \infty$,
\item \label{itm:g>0} the scalar coefficients $g_i $ are nonnegative with probability one. This assumption is equivalent to setting  $g_i = \exp(x_i)$, for a real--valued random sequence $\{x_i\}$, and formulating the AR(1) model \eqref{e:g:AR1}  in terms of $\{x_i\}$,
\item \label{itm:BM} the random processes $W_i(\cdot)$ are independent standard Brownian motions (Wiener processes),
\item \label{itm:ind} the sequences $\{W_i\}$ and $\{g_i\}$  are   independent,
\item \label{itm:phi} the autoregressive coefficient
$\fg$ in \refeq{g:AR1} satisfies  $\vert \fg \vert  < 1$,
\item \label{itm:int:sig} The function $\sigma(t)$, $t \in [0,1]$, is non--zero almost everywhere with respect to Lebesgue measure, i.e.   $\mathcal{L}eb \{ t: \sigma(t) = 0 \} =0$, where $\mathcal{L}eb$ denotes  the Lebesgue measure restricted to the unit interval $[0,1]$.
\item \label{itm:err4} $\mathbb{E}\left( \varepsilon_0^4\right) = \eta \sigma^2_{\varepsilon} < \infty$.
\end{enumerate}

\begin{remark}\label{rmk:alpha}
Assumptions \ref{itm:sig} and \ref{itm:int:sig} together  imply that there is no subinterval of $[0,1]$ on which the volatility $\sigma(\cdot)$ is infinity or zero. Assumptions \ref{itm:int:sig} also implies that  for any fixed $\alpha \in (0,1)$,
function $G(t) = \int_0^t\sigma^2(u) du$  is bounded away from zero on the restricted domain $t \in [\alpha,1]$:
\begin{align}\label{e:alpha}
    \underset{\alpha \leq t \leq 1}{\inf}  \int_0^t\sigma^2(u) du \geq \int_0^{\alpha}\sigma^2(u) du > 0.
\end{align}
We will use relation \eqref{e:alpha} in our proofs.
We comment on this point further in Remark \ref{r:ag}.
\end{remark}

The proposed model \eqref{e:R1}--\eqref{e:g:AR1} decomposes
the full random behavior of the  curves $R_i$ into the
between curves dynamics quantified by the
$g_i$ and the within curve dynamics
described by the stochastic integrals
$\int_0^{t} \sigma(u) dW_i(u)$, $t \in [0,1]$.
The  between curves dynamics is  regulated  by the dependence
between the $g_i$s that replaces the independence of curves assumption
used in previous research discussed in the Introduction.
The within curve dynamics is expressed  in terms of the  diffusion
$\int_0^{t} \sigma(u) dW_i(u)$,
which  models the roughness of the trajectories.

Our first  theorem establishes the  existence and uniqueness of
a strictly stationary  functional sequence $R_i$ satisfying
\eqref{e:R1}--\eqref{e:g:AR1}, as well
as the identifiability of model components. We present the proof
here because it is short and we use in the following
the relations it contains. For the sake of compactness,
we denote $ \lip R_i, R_i\rip_{t}$ by $Q_i(t)$
from now on. According to \eqref{e:<,>},
\begin{align}\label{e:QV}
 Q_i(t) :=   \lip R_i,R_i\rip_t = g_i^2 \int_0^t \sigma^2(u) du
 = g_i^2 \exp\{H(t)\}, \quad t \in (0,1], \quad
 i=1,2,\ldots, N,
\end{align}
where
\begin{align}\label{e:Ht}
    H(t) = \log \int_0^t \sigma^2(u) du  =  \log G(t),\qquad t \in (0,1].
\end{align}

\begin{theorem}\label{t:ss1}
Suppose  conditions \ref{itm:sig}--\ref{itm:phi} hold.
Then, there exists a unique strictly stationary
functional sequence $R_i$ satisfying  \refeq{R1}--\refeq{g:AR1}.
Moreover, the scalar $g_i$ is distinguishable from the
function $\sg$ in \refeq{g:AR1}.
\end{theorem}
\noindent{\sc Proof:}
First observe that condition \ref{itm:phi}, the finiteness of the second
moment of the $\varepsilon_i$ and their iid property imply the existence,
the  uniqueness and the strict stationarity of of the sequence
$\{ \log g_i \}$ satisfying \eqref{e:g:AR1} and hence the same
properties of the sequence $\{ g_i\}$, cf. condition \ref{itm:g>0}.

We now argue that condition \ref{itm:sig} implies existence and
uniqueness of the It{\^o} integral $\int_0^t \sg(u) d W_i(u)$, for each $i$.
This  follows from Theorem \ref{thm:exi:uni} because conditions
\refeq{lin:gro} and \refeq{lipsch} hold. Equation \refeq{R1} thus
defines the functional sequence $\{R_i\}$ directly.

Observe  that the integrated volatility $\int_0^t \sigma^2(u) du$
appearing in \eqref{e:QV} is identifiable from $g_i^2$ if and only
if the term $\log  \int_0^t \sigma^2(u) du$ is identifiable in the sum
\begin{equation} \label{e:logQ}
\log Q_i(t) = 2\log g_i  + \log  \int_0^t \sigma^2(u) du
    = 2\log g_i  + H(t) .
\end{equation}
The summands appearing in \refeq{logQ}  are identifiable due
to the assumption $\mathbb{E}(\varepsilon_i) = 0$
which implies   $\mathbb{E}(\log g_i)  = 0$.
Thus, $H(t)$ is identifiable via
\begin{align}\label{e:LQH}
    \mathbb{E} \log Q_i(t) = H(t).
\end{align}

\rightline{\QED}

We now turn to estimation.
We assume one has access to discrete observations
$  R_i(t_{i,k})$, $i=1,2,\ldots,N$, $k =0,1,\ldots, m $. We assume that the design points $t_{i,k}$ are regular and the same for all  curves, that is $t_{i,k} = t_k = \Delta k$, for some positive $\Delta $ that   decays to zero.
We also assume  $t_0 =0$ and $t_m =1$.
We aim to develop  inference for the  vector
\begin{align} \label{e:parameter}
    \btheta &=  [G(\cdot), \fg, \sg_\eg^2],
\end{align}
where
\begin{align*}
  G(t) = \int_0^{t}\sg^2(u) du, \quad t \in [0,1],
\end{align*}
is the cumulative volatility. We focus on the function $G$ because it
is sufficiently smooth, whereas under our general assumptions,  $\sg$
can be basically any measurable function. To recover $\sg$
from $G$ we need to assume that $\sg$ is continuous. It can then be
computed by differentiating $G$ numerically.

The  dense (high frequency)
sampling regime of observations  suggest working with the quadratic variation
processes $Q_i(t)$, $t \in [0,1]$, which can be estimated in such a framework.
Representation \eqref{e:logQ}  paves the way to explaining
the idea behind our inferential procedure. Assume one has access
to the latent quadratic variation processes $\{Q_i(t), t \in (0,1]\}$,
for $i=1, 2, \ldots, N$. Equation \eqref{e:logQ}
together with \eqref{e:g:AR1} leads to the family of
AR(1) models, indexed by time $t$,
\begin{align}\label{e:AR-QV}
   \log Q_i (t) - H(t)
   = \fg \lb \log Q_{i-1} (t)- H(t) \rb + 2 \eg_i.
\end{align}
Each AR(1) model is defined by the time
 index $i$, and we have a family of such models  indexed by continuous time $t$. However, by \eqref{e:g:AR1}, these models share the common innovation terms $\varepsilon_i$ which establish a connection between them that is explored in our estimation procedure.  Now, one can apply any of the well--known estimation techniques, see for example Chapter 8 of  \citetext{brockwell:davis:1991}, to obtain  the  oracle estimates (oracle only because the $Q_i(t)$ are not observable).
To focus on a specific simple approach, we set, cf. \eqref{e:LQH},
\begin{align}\label{e:tild:H}
    \widetilde{H}(t) = \frac{1}{N}\sum_{i=1}^{N} \log  Q_i(t),\qquad t \in (0,1],
\end{align}
or equivalently
\begin{align*}
    \widetilde{G}(t)  = \exp \left( \frac{1}{N}\sum_{i=1}^{N} \log  Q_i(t) \right),\qquad t \in (0,1].
\end{align*}
Next, we define, respectively,  the lag zero and lag one oracle empirical autocovariances
\begin{align}\label{e:gam_0}
    \gamma_{0,N} = & \frac{1}{4N}\sum_{i=1}^{N} \left(\log Q_i(t) - \frac{1}{N}\sum_{i=1}^N \log Q_i(t)\right)^{2}\\ \nonumber
    =& \frac{1}{N}\sum_{i=1}^{N} \left(\log g_i - \frac{1}{N}\sum_{i=1}^N \log g_i\right)^{2}.
    \end{align}
    and
    \begin{align}\label{e:gam_1}
        \gamma_{1,N} = & \frac{1}{4N}\sum_{i=1}^{N-1} \left(\log Q_i(t) - \frac{1}{N}\sum_{i=1}^N \log Q_i(t) \right)\left(\log Q_{i+1}(t) - \frac{1}{N}\sum_{i=1}^N \log Q_i(t) \right)\\ \nonumber
    =& \frac{1}{N}\sum_{i=1}^{N-1} \left(\log g_i - \frac{1}{N}\sum_{i=1}^N \log g_i \right)\left(\log g_{i+1} - \frac{1}{N}\sum_{i=1}^N \log g_i \right).
\end{align}
 We now define  the oracle Yule-Walker estimators
\begin{align*}
   \Tilde{\varphi} =       \gamma_{0,N}^{-1}     \gamma_{1,N},\qquad
    \Tilde{\sigma}^2_{\varepsilon} =      \gamma_{0,N} -  \Tilde{\varphi}     \gamma_{1,N} .
\end{align*}
Altogether, we propose the following vector of oracle estimators
\begin{align}\label{e:tild}
  \tilde{\btheta} =  \left[\widetilde{G}(t), t \in [0,1],\quad
   \tilde{\varphi}, \quad \tilde{\sigma}^2_\eg \right].
\end{align}

Notice, however, that the $\log g_i$ and the quadratic variation processes $\{Q_i(t), t \in [0,1]\}$,  for $i=1, 2, \ldots, N$,   are unobservable. Motivated by \eqref{e:emp->}, we replace $\{Q_i(t), t \in [0,1]\}$   by their realized  counterparts
\begin{align}
\nonumber
    \widehat{Q}_i(t)= & \sum_{k=1}^{m}\vert R_i(t_{k})-R_i(t_{k-1}) \vert^2 \mathbb{I}\{t_{k}\leq t\}  \\  \label{e:Qhat}
    =& g_i^2 \sum_{k}  \left\vert  \int_{t_{k-1}}^{t_k}\sigma(u) dW_i(u)\right\vert^2 \mathbb{I}\{t_{k}\leq t\}, \qquad t \in [0,1], \quad i = 1,2, \ldots N.
\end{align}
We can compute
\begin{align}\label{e:hatH}
    \widehat{H}(t) = \frac{1}{N}\sum_{i=1}^{N} \log  \widehat{Q}_i(t),\qquad t \in (0,1],
\end{align}
or equivalently
\begin{align}\label{e:hatG}
    \widehat{G}(t)  = \exp \left( \frac{1}{N}\sum_{i=1}^{N} \log  \widehat{Q}_i(t) \right),\qquad t \in (0,1].
\end{align}
Analogously to \eqref{e:gam_0} and \eqref{e:gam_1}, we define

\begin{align}\label{e:c0hQ}
    \Gamma_{0,N}(t)  =& \frac{1}{4N}\sum_{i=1}^N \left(\log \widehat{Q}_i(t)-\frac{1}{N}\sum_{i=1}^{N} \log \widehat{Q}_i(t)\right)^2,
\end{align}
and
\begin{align}\label{e:c1hQ}
    \Gamma_{1,N}(t)  =& \frac{1}{4N}\sum_{i=1}^{N-1} \left(\log \widehat{Q}_i(t)-\frac{1}{N}\sum_{i=1}^{N} \log \widehat{Q}_i(t)\right)\left(\log \widehat{Q}_{i+1}(t)-\frac{1}{N}\sum_{i=1}^{N} \log \widehat{Q}_i(t)\right).
\end{align}
Notice that in contrast to $    \gamma_{0,N}$ and $    \gamma_{1,N}$, $\Gamma_{0,N}(t) $ and $\Gamma_{1,N}(t) $ are not necessarily independent of $t$. Plugging them in the Yule-Walker equations defines a family of estimators,  indexed by time $t$, for the scalars $\varphi$ and $\sigma^2_{\varepsilon}$. In order to address this issue and remove the dependence on $t$,
we propose the following three estimation procedures. Procedures B and C
involve averaging over the interval $[\ag, 1] \subset [0,1]$, for
some $0 < \ag < 1$. Our theory explains that averaging over
the whole interval $[0,1]$ is not possible. This is due to the structure of the
model and is further elaborated on in Remark \ref{r:ag}. It turns out, cf.
Section \ref{s:emp},  that averaging over $[0,1]$ is not possible in practice
either. Theorems stated at the end of this section show that
all three procedures lead to consistent estimators with the $N^{-1/2}$
convergence rate (asymptotic normality). In Section \ref{s:emp}, we
investigate which approach works best in finite samples, as well as
the effect of the truncation parameter $\ag$.

\medskip
\noindent{PROCEDURE A:}
We define
\begin{align}\label{e:check}
 \check{\btheta} =   \left[\widehat{G}(t), t \in (0,1],\quad
   \check{\varphi}, \quad \check{\sigma}^2_\eg \right],
\end{align}
where
\begin{align}\label{e:chk:pi:sg}
    \check{\varphi} =  \Gamma_{0,N}^{-1}(1)  \Gamma_{1,N}(1) ,
    \qquad  \check{\sigma}^2_\eg
    =  \Gamma_{0,N}(1)  - \check{\varphi}\Gamma_{1,N}(1) .
\end{align}
The estimators   \eqref{e:chk:pi:sg} are motivated by the fact that  the \textit{total} variability in the stochastic
volatility model is accumulated at point $t=1$. The curve $\widehat{G}$ is defined by \eqref{e:hatG}.

\medskip
\noindent{PROCEDURE B:} In the second procedure, we first  average,  over $t \in [\alpha,1]$,   the autocovariance functions  $\Gamma_{0,N}(t)$ and $\Gamma_{1,N}(t)$, that is for fixed $\alpha \in (0,1)$, we define
\begin{align}
    \nonumber
    \bar{\Gamma}_{0,N} := \frac{1}{1-\alpha} \int_{\alpha}^1 \Gamma_{0,N}(t) dt, \qquad   \bar{\Gamma}_{1,N} := \frac{1}{1-\alpha} \int_{\alpha}^1 \Gamma_{1,N}(t) dt.
\end{align}
We then  plug the integrated autocovariances in the  Yule-Walker equations and propose
\begin{align}\label{e:bar:pi:sg}
    \bar{\varphi} =   \bar{\Gamma}_{0,N}^{-1}  \bar{\Gamma}_{1,N} , \qquad
    \bar{\sigma}^2_{\varepsilon} = \bar{\Gamma}_{0,N} - \bar{\varphi} \bar{\Gamma}_{1,N} .
\end{align}
The above, together with \eqref{e:hatG} gives the vector of estimates
\begin{align}\label{e:bar}
 \bar{\btheta} =   \left[\widehat{G}(t), t \in (0,1],\quad
   \bar{\varphi}, \quad \bar{\sigma}^2_\eg \right].
\end{align}

\medskip
\noindent{PROCEDURE C:} This procedure is motivated by \citetext{yao_functional_2005} who, in a different context, obtain a family of estimates,  $\{\hat{\sigma}^2(t)\}_{t \in \mathcal{T}}$ say, for a scalar parameter $\sigma^2$ and propose
$\hat{\sigma}^2 = \frac{1}{\vert \mathcal{T}\vert}\int_{\mathcal{T}} \hat{\sigma}^2(t) dt$. Here, we first plug the autocovariance functions  $\Gamma_{0,N}(t)$ and $\Gamma_{1,N}(t)$     in the Yule--Walker estimates and then take average over $t$ i.e. we define the integrated estimates
\begin{align}\label{e:hat:pi:sg}
    \hat{\varphi} = \frac{1}{1- \alpha}\int_{\alpha}^1\Gamma_{0,N}^{-1}(t)  \Gamma_{1,N}(t)dt ,
    \qquad  \hat{\sigma}^2_\eg
    = \frac{1}{1- \alpha} \int_{\alpha}^1 \left(\Gamma_{0,N}(t)  - \hat{\varphi} \Gamma_{1,N}(t) \right) dt,
\end{align}
where $\alpha$ is any fixed positive  number. The above, together with \eqref{e:hatG}, produce the final estimator vector
\begin{align}\label{e:hat}
   \hat{\btheta} = \left[\widehat{G}(t), t \in (0,1],\quad
   \hat{\varphi}, \quad \hat{\sigma}^2_\eg \right].
\end{align}

\begin{remark}\label{r:ag}
Model \eqref{e:R1}--\eqref{e:g:AR1} links  the parameters $\varphi$ and $\sigma^2_{\varepsilon}$  to the process $\{g_i\}$ only, while  the  observational scheme  provides the product $g_i\int_0^t \sg(u) d W_i(u)$, at discrete times, and does not distinguish $g_i$ and $\int_0^t \sg(u) d W_i(u)$.  This fact is more apparent in the oracle identity \eqref{e:QV}.
To perform  inference, we apply the logarithmic function $\log (\cdot)$ to the processes  $Q_i(\cdot)$ (in practice to the realized processes $\widehat{Q}_i(\cdot)$) and impose the  condition $\mathbb{E} \log g_i =0$. To establish consistency of this method, we require the function  $G(t) =  \int_0^t \sigma^2(u)du$ to be bounded away from zero. This forces us to remove the subinterval $[0,\alpha)$ from our analysis when we want to draw inference about $G(\cdot)$ or $H(\cdot)$ or when we want to apply  Procedures B or C, see the proofs of Theorems \ref{thm:hat:H}, \ref{thm:bar:phi:sig} and \ref{thm:hat:phi:sig}. Procedure A uses $t=1$ only, where boundedness away from zero holds true due to $G(1) = \int_0^1 \sigma^2(u)du > 0$.
\end{remark}

The asymptotic properties of the estimates   \eqref{e:check}, \eqref{e:bar} and \eqref{e:hat} are addressed under assumptions on the
 growth of  sample size $N$ (the number of curves) and the decay of the step size $\Delta$,  see Theorems \ref{thm:hat:H}--\ref{thm:hat:phi:sig}. It is worth mentioning that replacing $\{Q_i(t), t \in [0,1]\}$ by its empirical counterpart $\widehat{Q}_i(t)$
introduces an additional error term, $u_i(t)=   \log \widehat{Q}_i(t)- \log Q_i(t)$ say,  in the AR(1)  model \eqref{e:AR-QV}. That is
\begin{align}\label{e:AR:QVhat}
   \nonumber
   \log \widehat{Q}_i (t) - H(t) =& \fg \log \widehat{Q}_{i-1} (t) - \fg H(t) + 2 \eg_i +u_i(t) - \fg u_i(t)
   \\
   =:& \fg \log \widehat{Q}_{i-1} (t)- \fg H(t)+ \delta_i(t).
\end{align}
By  careful inspection, we deduce that the error terms $\delta_i(t)$, $i=1,2,\ldots,N$, are not necessarily independent. Moreover, the error terms $\delta_i(t)$ encompass three different terms which makes $\sigma^2_{\eg}$ nonidentifiable.
To overcome this issue   we establish the \textit{uniform consistency} of $\log \widehat{Q}_i(t)$. 
Roughly speaking, applying the results of    Lemma \ref{lem:2nd:moment}, 
 we  conclude that the error terms $u_i(t) $  tend to zero \textit{sufficiently fast}, so \eqref{e:AR:QVhat} is a sufficiently good approximation to model \eqref{e:AR-QV}. \citetext{galbraith_garch_2015} confront a similar issue in analysing GARCH models.  Differently from the current study, the essence of their problem allows them to  interpret the inaccuracy caused by this effect as an error term. Their procedure thus does not require 
 decay of the $u_i(t) $s to zero.

We now turn to the large sample justification of the procedures proposed above.
Recall the stochastic volatility model \eqref{e:R1}--\eqref{e:g:AR1} with the unknown parameter vector   $\btheta$ given in
\eqref{e:parameter}.  We study the  limiting behavior of the  estimates  $\check{\btheta}$, $\bar{\btheta}$ and $\hat{\btheta}$  proposed in  \eqref{e:check}, \eqref{e:bar} and  \eqref{e:hat} , respectively. Theorem \ref{thm:hat:H}, in particular, addresses convergence of $\widehat{H}(\cdot) - H(\cdot)$ uniformly in $L^1$ sense. Convergence of $\check \varphi$ and $\check \sigma^2_{\varepsilon}$ as well as their rate of
convergence are established in Theorem \ref{thm:chk:phi:sig}.
Theorems    \ref{thm:bar:phi:sig} and \ref{thm:hat:phi:sig} investigate, respectively,     $\bar \varphi$ and $\bar \sigma^2_{\varepsilon}$ and $\hat \varphi$ and $\hat \sigma^2_{\varepsilon}$.
The proofs  are deferred to Section \ref{s:proofs}.   The proofs fundamentally rely on the decay of the error induced by replacing the quadratic variation processes by their realized variants. This is quantified in Propositions \ref{prop:volatility}, \ref{prop:Gam:gam} and \ref{prop:unif:Gam:gam}. These together with consistency of oracle estimates $\widetilde{H}$ and $\left(    \gamma_{0,N},     \gamma_{1,N}\right)$  addressed in Propositions \ref{prop:tild:G} and \ref{prop:gam:kapp} entail the main results of the current section. Recall that $H(t) = \log \int_0^{t}\sg^2(u) du$ and $ \widehat{H}(t)$ and $ \widehat{G}(t)$ are given by \eqref{e:hatH} and \eqref{e:hatG}, respectively.

\begin{theorem}\label{thm:hat:H}
    Assume the stochastic volatility model defined by \eqref{e:R1}--\eqref{e:g:AR1} and  conditions \ref{itm:sig}--\ref{itm:int:sig} and recall \eqref{e:hatH} and \eqref{e:hatG}. Then,  for any fixed   $0 < \alpha < 1$,
       \begin{align}
       \label{e:cons:Hhat-H}
    \mathbb{E} \underset{t \in [\alpha,1]}{\sup} \left\vert  \widehat{H}(t)-  H(t)\right\vert  = & O\left( N^{-\frac{1}{2}} + \Delta^{\frac{1}{2}} \right),\\ \label{e:cons:Ghat-G}
       \underset{t \in [\alpha,1]}{\sup} \left\vert  \widehat{G}(t)-  G(t)\right\vert  = & O_P\left( N^{-\frac{1}{2}} + \Delta^{\frac{1}{2}} \right). 
   \end{align}
\end{theorem}

Before stating the next results, we need to define the lag--$h$ autocovariances:
\begin{equation}\label{e:kappas}
    \kappa_h :=  \mathbb{E}[ \log g_0 \log g_h]  =
    \frac{1}{4}\mathbb{E}\left[\left(\log Q_{0}(t) -  H(t) \right)\left(\log Q_h(t) -  H(t) \right)\right],  \quad h \in \mathbb{Z}.
\end{equation}

In the following theorems, no assumptions are needed
for the interplay between $N$ and $\Delta$.

\begin{theorem}\label{thm:chk:phi:sig}
 Assume the stochastic volatility model defined by \eqref{e:R1}--\eqref{e:g:AR1} satisfying conditions \ref{itm:sig}--\ref{itm:err4} and recall  \eqref{e:chk:pi:sg}. Then,
    \begin{align*}
        N^{1/2} \left( \check{\varphi} - \varphi\right)  \overset{\mathcal{L}aw}{\longrightarrow} \mathcal{N} \left(0, \nu \right),\quad \text{as }
 N \rightarrow \infty \text{ and } \Delta \rightarrow 0,
    \end{align*}
    and
\begin{align*}
     N^{1/2} \left(  \check{\sigma}^2_{\varepsilon} -  \sigma^2_{\varepsilon} \right)  \overset{\mathcal{L}aw}{\longrightarrow} \mathcal{N} \left(0, \tau \right),\quad \text{as }
 N \rightarrow \infty \text{ and } \Delta \rightarrow 0,
\end{align*}
 where
 \begin{align}\label{e:nu:tau}
     \nu = \left( -\kappa_0^{-2}\kappa_1 , \kappa_0^{-1}\right) V \left( -\kappa_0^{-2}\kappa_1 , \kappa_0^{-1}\right)^{\top}
\quad \mathrm{and} \quad
\tau = \left( 1+\kappa_0^{-2}\kappa_1^2 , -\kappa_0^{-1}\right) V \left( 1+\kappa_0^{-2}\kappa_1^2 , -\kappa_0^{-1}\right)^{\top},
  \end{align}
 and $V$ is a $2$ by $2$ matrix with entries
\begin{align}\label{e:matrixV}
   V_{k,l} =  (\eta - 3) \kappa_k \kappa_l + \sum_{h= - \infty}^{\infty} \left( \kappa_h \kappa_{h-k+l}+\kappa_{h-k}\kappa_{h+l}\right), \quad k , l = 0,\;1.
    \end{align}
\end{theorem}

\begin{theorem}\label{thm:bar:phi:sig}
Assume the stochastic volatility model defined by \eqref{e:R1}--\eqref{e:g:AR1} satisfying conditions \ref{itm:sig}--\ref{itm:err4} and recall  \eqref{e:bar:pi:sg}. Then the limiting results of Theorem \ref{thm:chk:phi:sig} holds true for $
\bar{\varphi}$ and $\bar{\sigma}^2$ as well, i.e.  for any $0< \ag < 1$,
    \begin{align*}
        N^{1/2} \left( \bar{\varphi} - \varphi\right)  \overset{\mathcal{L}aw}{\longrightarrow} \mathcal{N} \left(0, \nu \right),\quad \text{as }
 N \rightarrow \infty \text{ and } \Delta \rightarrow 0,
    \end{align*}
    and
\begin{align*}
     N^{1/2} \left(  \bar{\sigma}^2_{\varepsilon} -  \sigma^2_{\varepsilon} \right)  \overset{\mathcal{L}aw}{\longrightarrow} \mathcal{N} \left(0, \tau \right),\quad \text{as }
 N \rightarrow \infty \text{ and } \Delta \rightarrow 0,
\end{align*}
where $ \nu$ and $\tau$ are defined through \eqref{e:nu:tau}-\eqref{e:matrixV}.
\end{theorem}

\begin{theorem}\label{thm:hat:phi:sig}
 Assume the stochastic volatility model defined by \eqref{e:R1}--\eqref{e:g:AR1} and  conditions \ref{itm:sig}--\ref{itm:err4} and recall \eqref{e:hat:pi:sg}.
 Then,  for any $0< \ag < 1$,
 \begin{align} \label{e:lim:hatphi}
 & N^{\frac{1}{2}}\left( \hat{\varphi}- \varphi \right)\overset{\mathcal{L}aw}{\longrightarrow} \mathcal{N} \left(0, \pi \right)\quad \text{as }
 N \rightarrow \infty \text{ and } \Delta \rightarrow 0,
 \end{align}
 and
\begin{align*}
  &  N^{\frac{1}{2}}\left( \hat{\sigma}^2_{\varepsilon}- \sigma^2_{\varepsilon} \right)\overset{\mathcal{L}aw}{\longrightarrow} \mathcal{N} \left(0, \rho \right),\quad \text{as }
 N \rightarrow \infty \text{ and } \Delta \rightarrow 0,
\end{align*}
where
\begin{align*}
    \pi = \left( 0 , \kappa_0^{-1}\right) V \left( 0 , \kappa_0^{-1}\right)^{\top}
\quad \mathrm{and} \quad
\rho = \left( 1, -\varphi\right) V \left( 1 , -\varphi\right)^{\top},
\end{align*}
and matrix $V$ is given in \eqref{e:matrixV}.
\end{theorem}

The proofs of all results stated in this section are given in Section
\ref{s:proofs} of the Supplementary Material. We explain here briefly
where main challenges  requiring novel approaches occur.
At a heuristic level, the proposed method
involves the unobservable quadratic variation processes $Q_i(\cdot)$ which makes it  infeasible. Substituting $Q_i(\cdot)$ with discretely
observed curves $\widehat{Q}_i(\cdot)$ and studying  the decay of the error induced by this approximation locate the problem at the interface between  FDA and SDE.  Proposition \ref{prop:Gam:gam} and Corollaries \ref{cor:Gam:kapp} and \ref{cor:int:Gam:kapp}  establish consistency of the empirical autocovariances of the proxy processes $\widehat{Q}_i(\cdot)$. These key results pave the way for  applying  the delta method to obtain consistency results claimed in Theorems \ref{thm:chk:phi:sig} and \ref{thm:bar:phi:sig}.  Theorem \ref{thm:hat:phi:sig} is however more demanding and relies on uniform consistency of the empirical autocovariances of the proxy processes $\widehat{Q}_i(\cdot)$ and the reciprocal empirical variance of   $\widehat{Q}_i(\cdot)$. These are proved in  Proposition \ref{prop:unif:Gam:gam} and  Lemma \ref{lem:Gamm^-1}.
It is worth mentioning that discarding an arbitrarily narrow interval $[0,\alpha)$  is required  to
transfer the problem to the level of $\log Q_i(\cdot)$ at which the
between curves dependence is formulated. It is an insight  that
is not obvious from model formulation
and is utilized in the proofs of Lemma \ref{lem:2nd:moment}, Corollary \ref{cor:unif:log:2nd} and Theorem \ref{thm:hat:H}.

\section{Extension to order $p$ latent autoregression}\label{s:ext}
This section extends the model formulated in Section \ref{s:est}
by replacing the order 1 autoregression in \refeq{g:AR1} by
an AR($p$) model. This increases the flexibility of the model.
The fundamental  properties and estimation approaches remain the
same, but the limiting covariance structure has to be worked out
carefully.
An extension to a more general ARMA structure is more challenging
and is not pursued in this paper. It is well known that adding moving average
terms, while conceptually simple, often requires theoretically and practically
nontrivial modifications. This point is well explained in a high-dimensional
context in \citetext{wilms:2023} who give numerous relevant references.

For completeness, we begin with model equations, noting that
equation \refeq{AR(p)} is the same as \refeq{R1}:
\begin{equation}\label{e:AR(p)}
R_i(t) =g_i\int_0^t \sg(u) d W_i(u), \quad t \in [0,1], \quad  i \in \mathbb{Z},
\end{equation}
\begin{equation}  \label{e:g:ARp}
\log g_i = \varphi_1 \log g_{i-1}
 + \varphi_2 \log g_{i-2} +\ldots + \varphi_p \log g_{i-p}
 +\varepsilon_i,   \quad \eg_i \sim \ {\rm iid}
\ \mathcal{WN}(0, \sg_\eg^2),
\end{equation}
the  autoregressive  polynomial has  no zeros in the closed complex unit disk,
i.e.
\begin{align}\label{e:causal}
1-  \varphi_1 z - \varphi_2 z^2 -\ldots - \varphi_p z^p  \neq 0, \
\varphi_p \neq 0,
\quad \mathrm{for} \; \;\; \vert z \vert \leq 1.
\end{align}
It is well-known that \refeq{causal} is equivalent to
the existence of a stationary causal solution to \refeq{g:ARp},
see e.g Theorem 3.1.1. in \citetext{brockwell:davis:1991}.
Notice again  that we could equivalently formulate this model by replacing
$\log g_i$ with $x_i$ and $g_i$ with $\exp (x_i)$.

We begin by stating an extension of Theorem \ref{t:ss1}. The proof
is analogous, so it is omitted.

\begin{theorem}\label{t:ssp} Under \refeq{AR(p)}-\refeq{g:ARp},
if  conditions \ref{itm:sig}--\ref{itm:ind} of Section \ref{s:est}
and \refeq{causal} hold,
then  there exists a unique strictly stationary
functional sequence $R_i$ satisfying  \refeq{R1}--\refeq{g:AR1}.
Moreover, the scalar $g_i$ is distinguishable from the
function $\sg$ in \refeq{AR(p)}.
\end{theorem}

Turning to estimation, define  the vector of autoregressive coefficients:
\begin{align*}
    \boldsymbol{\varphi}= \left(\varphi_1,\varphi_2,\ldots, \varphi_p \right)^{\top}.
\end{align*}
Recall \eqref{e:kappas} and define the vector $\boldsymbol{\kappa}_p$
and the matrix  $ \mathbf{\Xi}_p$:
\begin{align*}
    \boldsymbol{\kappa}_p= \left(\kappa_1,\kappa_2,\ldots, \kappa_p \right)^{\top}, \qquad
    \mathbf{\Xi}_p = \left( \kappa_{i-j} \right)_{i,j = 1}^p.
\end{align*}
Define also
\begin{align}\label{e:ARp:gamm}
    \boldsymbol{\gamma}_{p,N}= \left(    \gamma_{1,N},\gamma_{2,N},\ldots, \gamma_{p,N} \right)^{\top}, \qquad
    \mathbf{\Psi}_{p,N} = \left( \gamma_{i-j,N} \right)_{i,j = 1}^p,
\end{align}
where $\gamma_{h,N}  = \gamma_{-h,N} $ and, for $h =0,1,\ldots,p$ ,
\begin{align}\label{e:ARp:gam_h}
\gamma_{h,N} = &  \frac{1}{4N}\sum_{i=1}^{N-h} \left(\log Q_i(t) - \frac{1}{N}\sum_{i=1}^N \log Q_i(t) \right)\left(\log Q_{i+h}(t) - \frac{1}{N}\sum_{i=1}^N \log Q_i(t) \right)\\ \nonumber
 =& \frac{1}{N}\sum_{i=1}^{N-h} \left(\log g_i - \frac{1}{N}\sum_{i=1}^N \log g_i \right)\left(\log g_{i+h} - \frac{1}{N}\sum_{i=1}^N \log g_i \right).
\end{align}
The above definitions allow us to defined the  oracle Yule-Walker estimates:
\begin{align*}
   \Tilde{\boldsymbol{\varphi}} =   \mathbf{\Psi}_{p,N}^{-1}  \boldsymbol{\gamma}_{p,N}, \qquad
    \Tilde{\sigma}^2_{\varepsilon} =     \gamma_{0,N} -  \Tilde{\boldsymbol{\varphi}}^{\top} \boldsymbol{\gamma}_{p,N}.
\end{align*}

For each $t \in [0,1]$, the realized counterpart of \eqref{e:ARp:gamm} can be written in the form
\begin{align*}
    \boldsymbol{\Gamma}_{p,N}(t)= \left(\Gamma_{1,N}(t) ,\Gamma_{2,N}(t),\ldots, \Gamma_{p,N}(t) \right)^{\top},\qquad
   \mathbf{\Sigma}_{p,N} (t)= \left( \Gamma_{i-j,N}(t) \right)_{i,j = 1}^p, \quad t \in [0,1],
\end{align*}
where $\Gamma_{h,N}(t) = \Gamma_{-h,N}(t)$ and, for $h =0,1,\ldots,p$,
\begin{align}\label{e:ARp:Gamm}
    \Gamma_{h,N}(t) =&  \frac{1}{4N}\sum_{i=1}^{N-h} \left(\log \widehat{Q}_i(t)-\frac{1}{N}\sum_{i=1}^{N} \log \widehat{Q}_i(t)\right)\left(\log \widehat{Q}_{i+h}(t)-\frac{1}{N}\sum_{i=1}^{N} \log \widehat{Q}_i(t)\right).
\end{align}

In contrast to the sequence $\gamma_{h,N}$, the above empirical
autocovariances may depend on $t$. If we directly plug them in the
Yule-Walker equations, we obtain   a \textit{family} of  estimators,
indexed by $t$, for the constant parameters
$ \boldsymbol{\varphi}$ and $\sigma^2_{\varepsilon}$.
To overcome this issue, similarly to Section \ref{s:est},
we propose the following three procedures whose asymptotic
properties are established in
Theorems \ref{t:p}--\ref{t:pC} below.

\medskip
\noindent{PROCEDURE A:}
We define
\begin{align*}
 \check{\btheta} =   \left[\widehat{G}(t), t \in [0,1],\quad
   \check{\boldsymbol{\varphi}}, \quad \check{\sigma}^2_\eg \right],
\end{align*}
where $\widehat{G}(t)$ is the same as \eqref{e:hatG} and
$\check{\varphi}$ and $\check{\sigma}^2_\eg$ are obtained by
using  on the terminal time $t=1$:
\begin{align}\label{e:AR:chk:pi:sg}
    \check{\boldsymbol{\varphi}} = \mathbf{\Sigma}_{p,N}^{-1}(1)  \boldsymbol{\Gamma}_{p,N}(1) ,
    \qquad  \check{\sigma}^2_\eg
    =  \Gamma_{0,N}(1)  -  \check{\boldsymbol{\varphi}}^{\top} \boldsymbol{\Gamma}_{p,N}(1).
\end{align}

\medskip
\noindent{PROCEDURE B:} We define
\begin{align*}
 \bar{\btheta} =   \left[\widehat{G}(t), t \in [0,1],\quad
   \bar{\boldsymbol{\varphi}}, \quad \bar{\sigma}^2_\eg \right],
\end{align*}
where $\widehat{G}(t)$ is the same as \eqref{e:hatG}
and $\bar{\boldsymbol{\varphi}}$ and $\bar{\sigma}^2_{\varepsilon}$
are defined by
\begin{align}\label{e:AR:bar:pi:sg}
   \bar{\boldsymbol{\varphi}} =    \bar{\mathbf{\Sigma}}_{p,N}^{-1} \bar{\boldsymbol{\Gamma}}_{p,N}, \qquad
    \bar{\sigma}^2_{\varepsilon}
    =  \bar\Gamma_{0,N}
    -  \bar{\boldsymbol{\varphi}}^{\top} \bar{\boldsymbol{\Gamma}}_{p,N},
\end{align}
where
\begin{align*}
    \bar{\boldsymbol{\Gamma}}_{p,N} =  \frac{1}{1-\alpha} \int_{\alpha}^1 \boldsymbol{\Gamma}_{p,N}(t) dt, \quad
    \bar\Gamma_{0,N}= \frac{1}{1-\alpha} \int_{\alpha}^1 \Gamma_{0,N}(t) dt, \quad
     \bar{\mathbf{\Sigma}}_{p,N} =  \frac{1}{1-\alpha} \int_{\alpha}^1 \mathbf{\Sigma}_{p,N}(t) dt,
\end{align*}
and where   $\alpha$ is any fixed number in the interval  $(0,1)$.

\medskip
\noindent{PROCEDURE C:} We define
\begin{align*}
   \hat{\btheta} = \left[\widehat{G}(t), t \in [0,1],\quad
   \hat{\boldsymbol{\varphi}}, \quad \hat{\sigma}^2_\eg \right].
\end{align*}
where $\widehat{G}(t)$ is the same as \eqref{e:hatG}
and $ \hat{\boldsymbol{\varphi}}$ and $\hat{\sigma}^2_\eg$ are
\begin{align}\label{e:AR:hat:pi:sg}
    \hat{\boldsymbol{\varphi}} = \frac{1}{1- \alpha}\int_{\alpha}^1\mathbf{\Sigma}_{p,N}^{-1}(t)  \boldsymbol{\Gamma}_{p,N}(t)dt ,
    \qquad  \hat{\sigma}^2_\eg
    = \frac{1}{1- \alpha} \int_{\alpha}^1 \left(\Gamma_{0,N}(t)
    -  \hat{\boldsymbol{\varphi}}^{\top}
    \boldsymbol{\Gamma}_{p,N}(t) \right)dt,
\end{align}
where   $\alpha$ is any fixed number in  the interval $(0,1)$.

\begin{theorem}\label{t:p}
Assume the stochastic volatility model defined by
\eqref{e:AR(p)}--\eqref{e:g:ARp} and  conditions
\ref{itm:sig}--\ref{itm:int:sig} except that
we replace condition \ref{itm:phi} by \eqref{e:causal}.
Define   $\widehat{H}(t)$ and $\widehat{G}(t)$  by
\eqref{e:hatH} and \eqref{e:hatG}, respectively. Then, for any fixed  $0 < \alpha < 1$,
       \begin{align}
       \label{e:ARpcons:Hhat-H}
    \mathbb{E} \underset{t \in [\alpha,1]}{\sup} \left\vert  \widehat{H}(t)-  H(t)\right\vert  = & O\left( N^{-\frac{1}{2}} + \Delta^{\frac{1}{2}} \right),\\ \label{e:ARp:cons:Ghat-G}
       \underset{t \in [\alpha,1]}{\sup} \left\vert  \widehat{G}(t)-  G(t)\right\vert  = & O_P\left( N^{-\frac{1}{2}} + \Delta^{\frac{1}{2}} \right).
   \end{align}
\end{theorem}

Before investigating asymptotic
properties of procedures A and B.
we need to introduce  the following notation.
Consider the domain $\mathcal{D} \subset \mathbb{R}^p$ defined by
\begin{align*}
 \mathcal{D} = \left \{ \left(x_0,\ldots,x_{p-1} \right) \;|\;
 X = \lb x_{i-j}\rb_{i,j = 1}^p \text{ is positive definite}\right\},
\end{align*}
and  the functions
\begin{align*}
&f  :    \mathcal{D} \times \mathbb{R} \longrightarrow \mathbb{R}^p, \;
    f (x_0,\ldots,x_p) = X^{-1}a,\\
&g : \mathcal{D} \times \mathbb{R} \longrightarrow \mathbb{R}, \;
    g (x_0,\ldots,x_p) = x_0 - \left( X^{-1}a\right)^{T}a , \qquad
\end{align*}
where $
X = \lb x_{i-j}\rb_{i,j = 1}^p, \ a = \left(x_1,\ldots,x_p \right)^{\top}$.
The above functions $f(\cdot)$ and $g(\cdot)$ are continuously
differentiable.

In theorems \ref{t:pA}, \ref{t:pB} and \ref{t:pC}, we
assume the stochastic volatility model
\eqref{e:AR(p)}--\eqref{e:g:ARp} and  conditions
\ref{itm:sig}--\ref{itm:err4} of Section \ref{s:est}, except that we replace
condition \ref{itm:phi} by \eqref{e:causal},  if $p>1$.
We do not assume  any interplay between  $N \rightarrow \infty$
and  $\Delta \rightarrow 0$. In Theorems \ref{t:pB} and \ref{t:pC},
the same asymptotic distribution holds for any $\ag\in (0,1)$.

\begin{theorem}\label{t:pA}
For the estimators defined by  \refeq{AR:chk:pi:sg} (Procedure A),
    \begin{align*}
        N^{1/2} \left( \check{\boldsymbol{\varphi}}
        - \boldsymbol{\varphi}\right)  \overset{\mathcal{L}aw}{\longrightarrow}
        \mathcal{N} \left(0, \boldsymbol{\nu}_p\right),\quad \text{as }
 N \rightarrow \infty \text{ and } \Delta \rightarrow 0,
    \end{align*}
    and
\begin{align*}
     N^{1/2} \left(  \check{\sigma}^2_{\varepsilon} -  \sigma^2_{\varepsilon} \right)  \overset{\mathcal{L}aw}{\longrightarrow} \mathcal{N} \left(0, \tau_p \right),\quad \text{as }
 N \rightarrow \infty \text{ and } \Delta \rightarrow 0,
\end{align*}
 where
 \begin{align}\label{e:ARp:nu:tau}
   \boldsymbol{\nu}_p = \left(\nabla f\left( \kappa_0, \boldsymbol{\kappa}_p\right)\right)^{\top} V \nabla f\left( \kappa_0, \boldsymbol{\kappa}_p\right), \quad \tau_p = \left(\nabla g\left( \kappa_0 ,\boldsymbol{\kappa}_p\right)\right)^{\top} V \nabla g\left( \kappa_0 ,\boldsymbol{\kappa}_p\right)
  \end{align}
 and $V$ is a $(p+1) \times (p+1)$ matrix with entries
\begin{align}\label{e:ARp:matrixV}
V_{k,l} =  (\eta - 3) \kappa_k \kappa_l + \sum_{h= - \infty}^{\infty} \left( \kappa_h \kappa_{h-k+l}+\kappa_{h-k}\kappa_{h+l}\right), \quad k , l = 0,\ldots, p.
    \end{align}
\end{theorem}

\begin{theorem}\label{t:pB}
The estimators defined by  \refeq{AR:bar:pi:sg} (Procedure B)
have the same asymptotic distribution as the
estimators  \refeq{AR:chk:pi:sg} (Procedure A), i.e.
\begin{align*}
        N^{1/2} \left( \bar{\boldsymbol{\varphi}} - \boldsymbol{\varphi}\right)  \overset{\mathcal{L}aw}{\longrightarrow} \mathcal{N} \left(0, \boldsymbol{\nu}_p \right),\quad \text{as }
 N \rightarrow \infty \text{ and } \Delta \rightarrow 0,
    \end{align*}
    and
\begin{align*}
     N^{1/2} \left(  \bar{\sigma}^2_{\varepsilon} -  \sigma^2_{\varepsilon} \right)  \overset{\mathcal{L}aw}{\longrightarrow} \mathcal{N} \left(0, \tau_p \right),\quad \text{as }
 N \rightarrow \infty \text{ and } \Delta \rightarrow 0,
\end{align*}
 where $ \boldsymbol{\nu}_p$ and $\tau_p$ are defined through \eqref{e:ARp:nu:tau}-\eqref{e:ARp:matrixV}.
\end{theorem}

\begin{theorem}\label{t:pC}
For the estimators \eqref{e:AR:hat:pi:sg} (procedure C),
 \begin{align} \label{e:ARp:lim:hatphi}
 & N^{\frac{1}{2}}\left( \hat{\boldsymbol{\varphi}}- \boldsymbol{\varphi} \right)\overset{\mathcal{L}aw}{\longrightarrow} \mathcal{N} \left(0,  \boldsymbol{\pi}_p\right)\quad \text{as }
 N \rightarrow \infty \text{ and } \Delta \rightarrow 0,
 \end{align}
 and
\begin{align*}
  &  N^{\frac{1}{2}}\left( \hat{\sigma}^2_{\varepsilon}- \sigma^2_{\varepsilon} \right)\overset{\mathcal{L}aw}{\longrightarrow} \mathcal{N} \left(0, \rho_p \right),\quad \text{as }
 N \rightarrow \infty \text{ and } \Delta \rightarrow 0,
\end{align*}
where
\begin{align*}
\boldsymbol{\pi}_p =  \mathbf{\Xi}_p^{-1}  W V   W^{\top}\mathbf{\Xi}_p^{-1} ,\quad \rho_p =    \left( 1, -\boldsymbol{\varphi}^{\top}\right) V \left(\begin{array}{c}
         1  \\
          -\boldsymbol{\varphi}
    \end{array} \right),
\end{align*}
with $W$  a $p\times (p+1)$ matrix in the form:
\begin{align}\label{e:mtrx:W}
    W = \left(
    \begin{array}{cccccc}
         0 & 1 & 0 & 0 &\ldots & 0 \\
         0 & 0 & 1 & 0 &\ldots & 0 \\
         0 & 0 & 0 & 1 &\ldots & 0 \\
         \vdots & \vdots & \vdots & \vdots &\ddots & \vdots \\
         0 & 0 & 0 & 0 &\ldots & 1 \\
    \end{array}
    \right),
\end{align}
and matrix $V$ is given in \eqref{e:ARp:matrixV}.
\end{theorem}

\section{Empirical Analysis} \label{s:emp}
We first present in Section \ref{ss:pro} the estimation of our model on intraday
price curves, suitably transformed to stationarity, of a large number of U.S.
stocks. Such an analysis will explain the meaning of
the model elements introduced in previous sections and will suggest relevant
parameter settings for a simulation study presented in Section \ref{ss:sim}.

\subsection{Application to U.S. stocks}\label{ss:pro}
By way of introduction, we begin with the
analysis of curves derived from price data of  Apple Inc. (Permno: 14593).
The sample period is from Jan 3, 2016 to Dec 31, 2021, corresponding
to $N=4021$ trading days. In each trading day $i$,
we have the opening price $P_i(t_0)$
and the following 78 of 5-min intraday price observations
$P_i(t_k), k = 1,...,78$, the last trading prices in every 5-min  time interval.
A different time resolution could be used, but the five minute resolution
allows us to analyze most stocks traded in U.S. because
not all of them are traded as frequently as Apple.
Additionally, the intraday price data at 5-min frequency provides
a good balance between informative signals and effects of
market microstructure errors, see e.g. \citetext{barndorff2002econometric}.
Thus, the step size is $\Delta = 1/78$ and
the design points are $t_k= k\times\Delta, k = 0, 1,...,m$,
where $m=78$.

\begin{figure}
	\centering
	\includegraphics[width=0.9\linewidth]{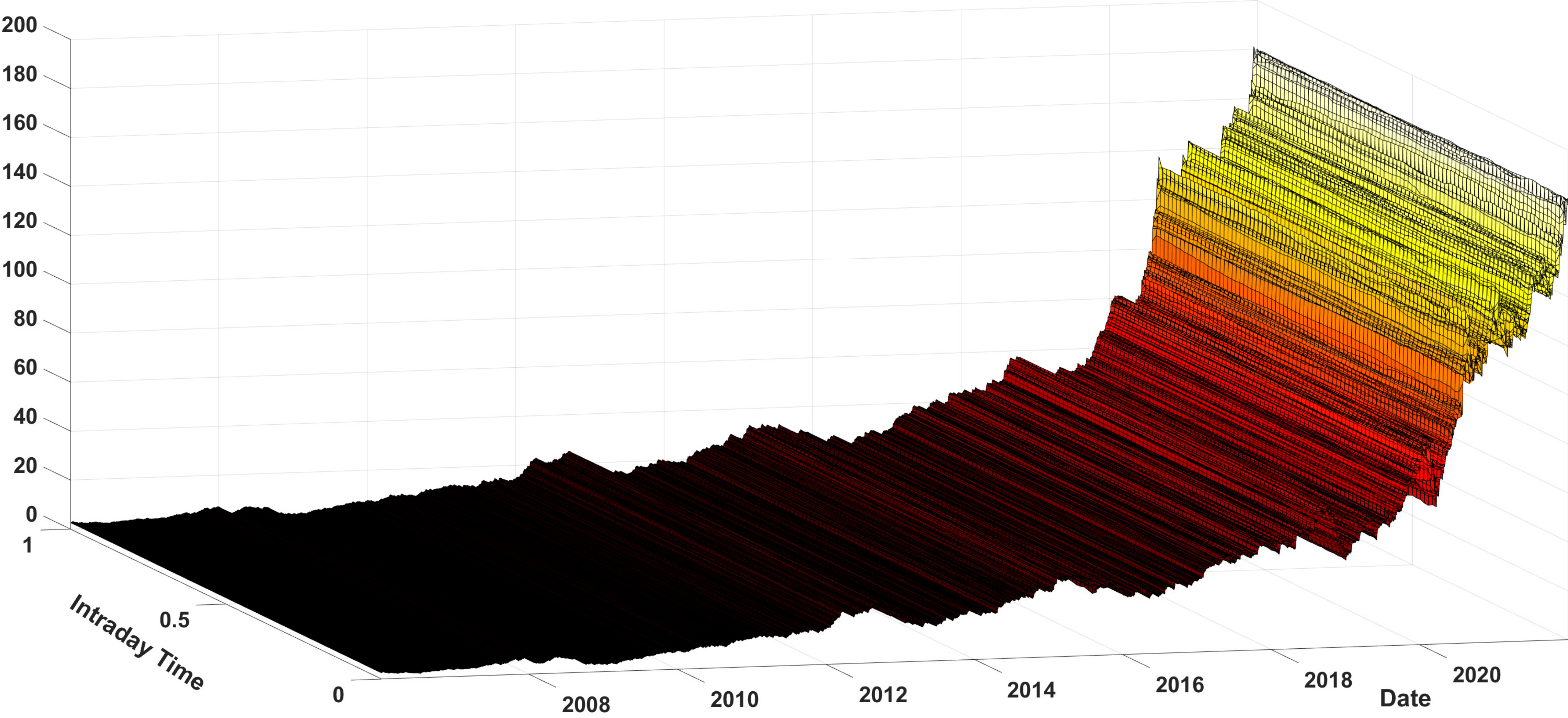}
	\includegraphics[width=0.9\linewidth]{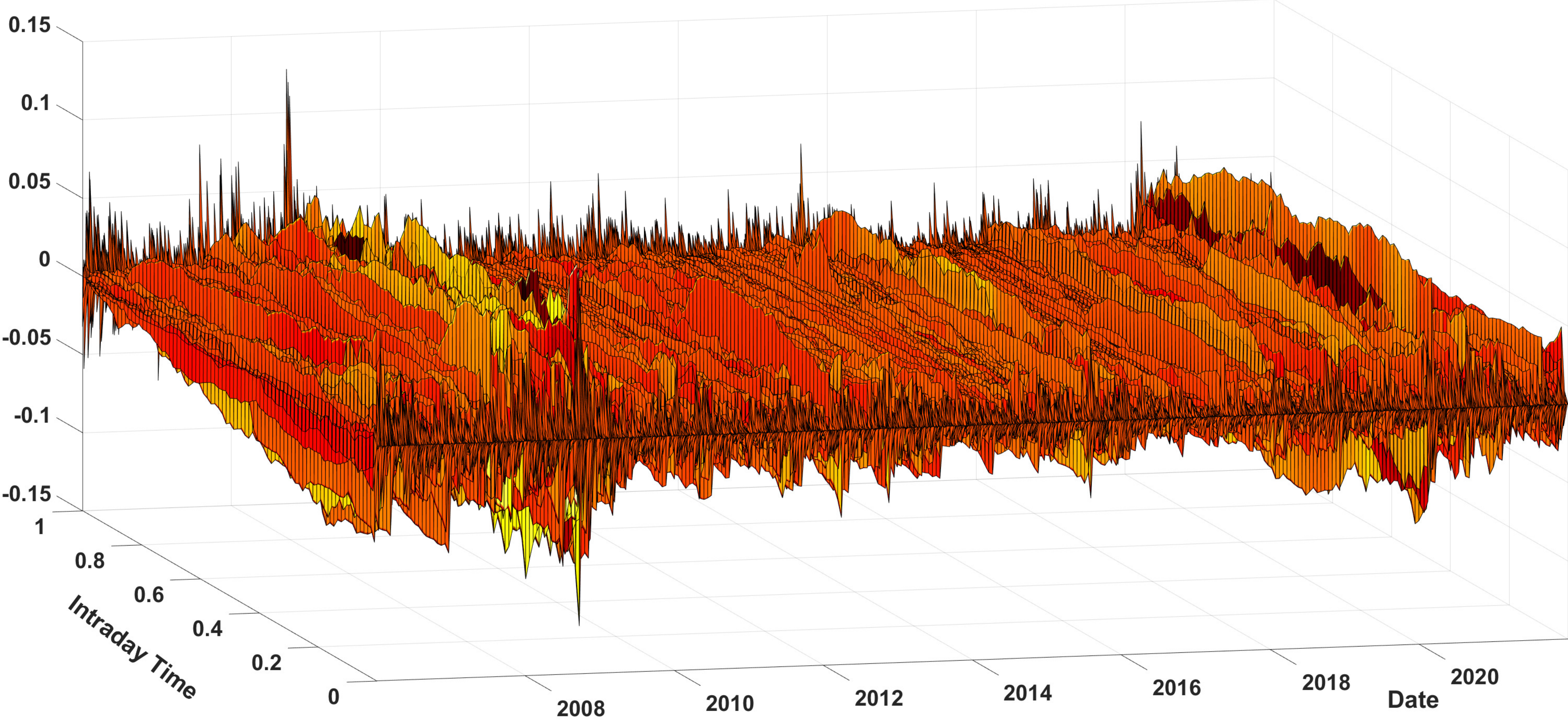}
	\includegraphics[width=0.9\linewidth]{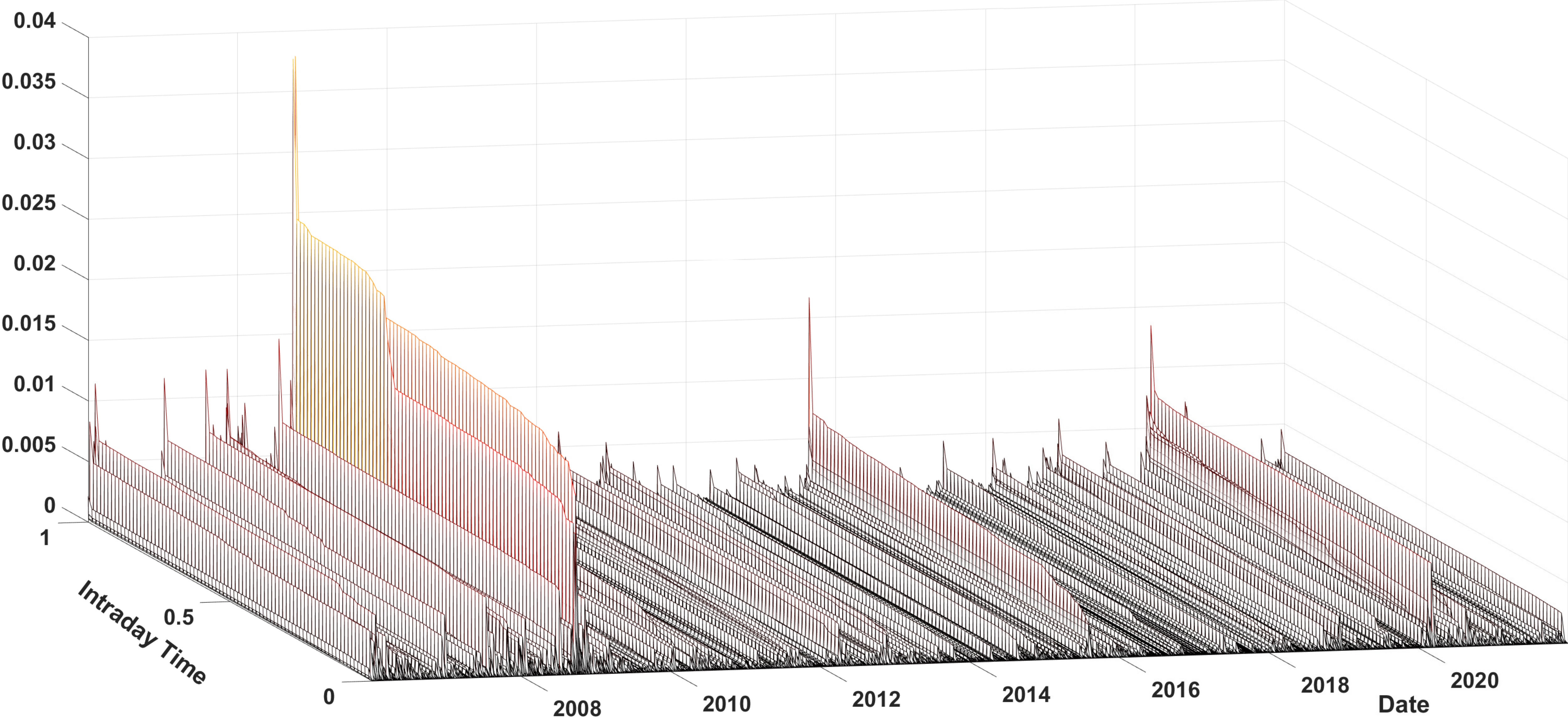}
	\caption{Time series of functional objects derived
from intraday Apple prices. Upper Panel: Intraday Prices $P_i(t_k)$;
Middle Panel: the CIDRs $R_i(t_k)$;
Lower Panel: Realized Quadratic Variation $\widehat{Q}_i(t)$.
	\label{fig:apple}}
\end{figure}

\begin{figure}
	\centering
	\includegraphics[width=0.4\textwidth]{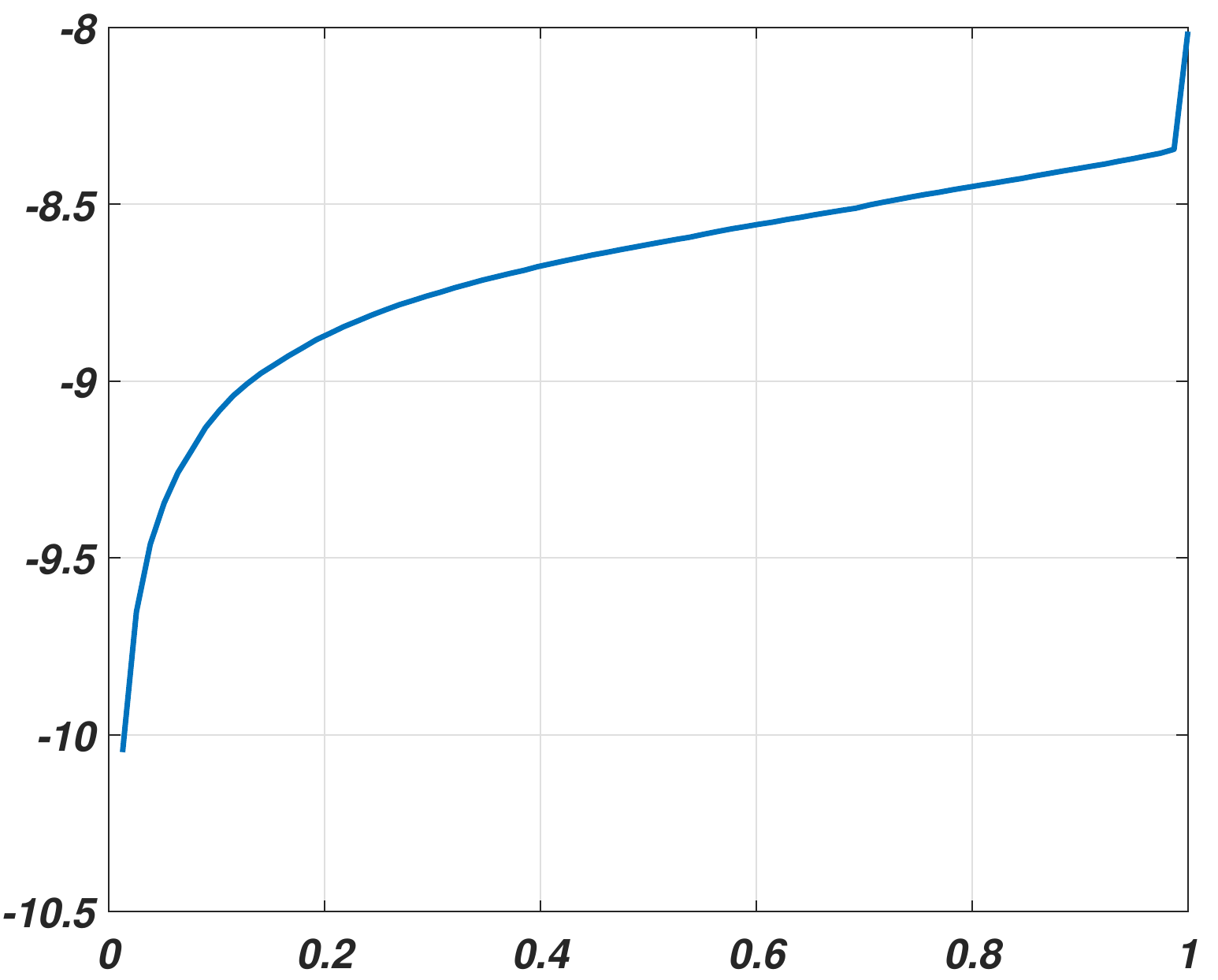}
	\includegraphics[width=0.4\textwidth]{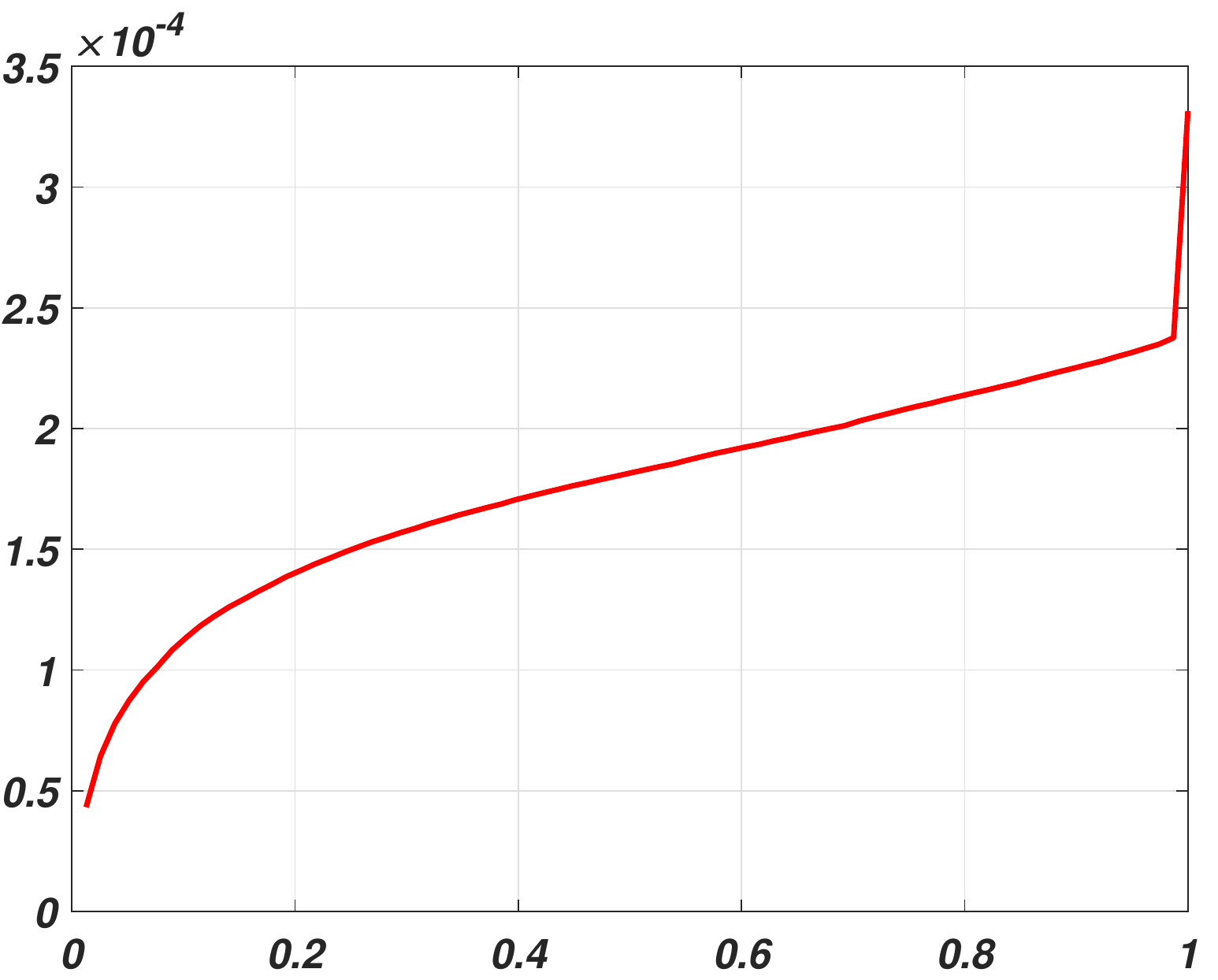}
	\caption{Cumulative intraday volatility curves
for Apple; Left: $\widehat{H}$; Right: $\widehat{G}$.
	\label{f:appleHG}}
\end{figure}

We calculate the cumulative intraday return (CIDR) curves  as
\[
R_i(t_k) = \log(P_i(t_k)) - \log(P_i(t_0)),
\qquad k = 0,1,...,m, \ i = 1, 2, ...,N.
\]
By definition, the CIDR curves always start from zero, i.e. $R_i(t_0) = 0$,
and are scale invariant. This generally leads to a stationary
sequence of curves, as investigated in \citetext{horvath:kokoszka:rice:2014}.
We calculate the realized version of the quadratic variation as
\[
\widehat{Q}_i (t) = \sum_{k=1}^{K}
\left| R_i (t_k) - R_i (t_{k-1}) \right|^2 \mathbb{I}\left\lbrace t_k \leq t\right\rbrace, \qquad t \in[0,1], \qquad i = 1,2,...,N.
\]
Figure \ref{fig:apple} shows the Apple intraday price $P_i(t_0)$ (upper panel),
the CIDRs $R_i(t_k)$ (middle panel) and the
realized quadratic variation $\widehat{Q}_i (t)$ (lower panel).
Note that we can have $\widehat{Q}_i(t) = 0$
for a few 5-min long intervals at the beginning of the trading day.
This is because the price can remain  the same as the opening
price in the absence of any trades for some time after the opening.
In such cases, $\log \widehat{Q}_i(t)$ is not computable.
This means that the truncation at $\ag>0$
required by our theory is often practically needed. In the definitions below
we assume that $\ag$ is such that $\log \widehat{Q}_i(t)$ can be computed
for $t>\ag$. For the stocks we consider, $\ag = \Dg = 1/78$ is generally
sufficient. With this caveat, we can calculate
$\widehat{H}_i (t)$ and $\widehat{G}_i (t)$ given, respectively,
by \refeq{hatH} and \refeq{hatG}. Figure \ref{f:appleHG}
shows both curves for Apple.

\begin{table}[htbp]
	\centering
	\caption{Estimation results for Apple with $\alpha=\Delta$}
    \begin{tabular}{lp{1cm}p{2cm}p{2cm}p{2cm}}
	\toprule
	\toprule
	&       & Proc. A & Proc. B & Proc. C \\
	\midrule
	$\varphi$ &       & 0.537 & 0.512 & 0.535 \\
	$\sigma^2$ &       & 0.262 & 0.281 & 0.276 \\
	\bottomrule
	\bottomrule
	\end{tabular}%
	\label{tab:apple_est}
\end{table}

After computing the
covariances $\Gamma_{0,N} (t)$ and $\Gamma_{1,N} (t)$,
given, respectively, by \refeq{c0hQ} and \refeq{c1hQ}, we can
compute all estimators in Procedures A, B and C  introduced
in Section \ref{s:est}.
Table \ref{tab:apple_est} shows the estimation results of the
three procedures with $\alpha=\Delta$.
We see that they yield  similar estimates.
This is encouraging because it indicates that they all could
close to the true values of these parameters. This will be
investigated in Section \ref{ss:sim}, but before we do it,
we need to get a more comprehensive picture of possible parameter
ranges. For this purpose, we repeat the same analysis  for 7293 stocks
in the U.S. stock markets.
The original dataset includes all U.S. stocks from 2006 to 2021.
To ensure data quality, the intraday price data is cleaned based on
the rules explained in Section \ref{s:dataclean} in the Supplementary Material.
The summary statistics of the estimators are  presented in
Table \ref{tab:summ_stat}. We see that the three procedure
produce estimates in similar ranges.

\begin{table}[htbp]
	\centering
	\caption{Summary statistics of the three estimators based on
7293 stocks}
	\begin{tabular}{lrrrrrrrr}
		\toprule
		\toprule
		&       & \multicolumn{3}{c}{$\varphi$} &       & \multicolumn{3}{c}{$\sigma^2$} \\
		\cmidrule{3-9}          &       & \multicolumn{1}{c}{Proc. A} & \multicolumn{1}{c}{Proc. B} & \multicolumn{1}{c}{Proc. C} &       & \multicolumn{1}{c}{Proc. A} & \multicolumn{1}{c}{Proc. B} & \multicolumn{1}{c}{Proc. C} \\
		\midrule
		Mean  &       & 0.557 & 0.504 & 0.529 &       & 0.234 & 0.292 & 0.287 \\
		SD    &       & 0.123 & 0.121 & 0.117 &       & 0.060 & 0.072 & 0.072 \\
		Skewness &       & 0.119 & 0.257 & 0.158 &       & 4.453 & 3.340 & 3.434 \\
		Kurtosis &       & 2.857 & 2.967 & 2.967 &       & 39.421 & 22.813 & 23.662 \\
		\midrule
		Min   &       & 0.119 & 0.093 & 0.112 &       & 0.059 & 0.133 & 0.131 \\
		Q. 25\% &       & 0.470 & 0.419 & 0.447 &       & 0.205 & 0.250 & 0.246 \\
		Median &       & 0.551 & 0.495 & 0.522 &       & 0.227 & 0.278 & 0.272 \\
		Q. 75\% &       & 0.642 & 0.584 & 0.608 &       & 0.252 & 0.314 & 0.308 \\
		Max   &       & 0.961 & 0.941 & 0.943 &       & 0.955 & 1.020 & 1.019 \\
		\bottomrule
		\bottomrule
	\end{tabular}
	\label{tab:summ_stat}
\end{table}

We conclude this section with an investigation of the impact of
the truncation parameter $\ag\in (0,1)$. This parameter is absent
in Procedure A, but we included it for comparison.
We repeated the comprehensive analysis
with $\alpha \in \left\lbrace 5\Delta, 20 \Delta, 40\Delta\right\rbrace$.
Since $\Delta = 1/78$, we have
$\alpha= \left\lbrace 0.0641, 0.2564, 0.5128\right\rbrace$.
Table \ref{tab:sensitivity_alpha} presents selected, most informative,  summary
statistics   based on different values of $\alpha$.
It shows that larger $\alpha$ leads to larger estimates of
$\varphi$ and smaller estimates of $\sigma$,
but all those estimates are similar.

\begin{table}[htbp]
	\centering
	\caption{Estimation results for different $\alpha$.}\medskip
		\begin{tabular}{lrrrrrrrr}
			\toprule
			&       & \multicolumn{3}{c}{$\varphi$} &       & \multicolumn{3}{c}{$\sigma^2$} \\
			\midrule
			&       & \multicolumn{1}{c}{Proc. A} & \multicolumn{1}{c}{Proc. B} & \multicolumn{1}{c}{Proc. C} &       & \multicolumn{1}{c}{Proc. A} & \multicolumn{1}{c}{Proc. B} & \multicolumn{1}{c}{Proc. C} \\
			\midrule
			\multicolumn{9}{l}{{$\alpha=5\Delta$}} \\
			\midrule
			Mean  &       & 0.557 & 0.532 & 0.542 &       & 0.234 & 0.263 & 0.261 \\
			SD    &       & 0.123 & 0.120 & 0.119 &       & 0.060 & 0.071 & 0.071 \\
			Q. 25\% &       & 0.470 & 0.447 & 0.459 &       & 0.205 & 0.224 & 0.222 \\
			Median &       & 0.551 & 0.526 & 0.537 &       & 0.227 & 0.249 & 0.247 \\
			Q. 75\% &       & 0.642 & 0.613 & 0.622 &       & 0.252 & 0.283 & 0.281 \\
			\midrule
			\midrule
			\multicolumn{9}{l}{{$\alpha=20\Delta$}} \\
			\midrule
			Mean  &       & 0.557 & 0.565 & 0.567 &       & 0.234 & 0.232 & 0.231 \\
			SD    &       & 0.123 & 0.119 & 0.119 &       & 0.060 & 0.067 & 0.067 \\
			Q. 25\% &       & 0.470 & 0.482 & 0.485 &       & 0.205 & 0.196 & 0.196 \\
			Median &       & 0.551 & 0.561 & 0.564 &       & 0.227 & 0.219 & 0.218 \\
			Q. 75\% &       & 0.642 & 0.647 & 0.649 &       & 0.252 & 0.248 & 0.248 \\
			\midrule
			\midrule
			\multicolumn{9}{l}{{$\alpha=40\Delta$}} \\
			Mean  &       & 0.557 & 0.586 & 0.587 &       & 0.234 & 0.213 & 0.213 \\
			SD    &       & 0.123 & 0.118 & 0.118 &       & 0.060 & 0.065 & 0.065 \\
			Q. 25\% &       & 0.470 & 0.505 & 0.505 &       & 0.205 & 0.180 & 0.180 \\
			Median &       & 0.551 & 0.584 & 0.585 &       & 0.227 & 0.201 & 0.201 \\
			Q. 75\% &       & 0.642 & 0.668 & 0.669 &       & 0.252 & 0.228 & 0.228 \\
			\bottomrule
			\bottomrule
		\end{tabular}
	\label{tab:sensitivity_alpha}
\end{table}

\subsection{A simulation study}\label{ss:sim}
The purpose of  this section is to obtain more detailed insights into
the finite sample performance of the proposed estimators and to compare them.
Functional time series are generated according to \refeq{R:g},  but with
$i = 1,\ldots, N$. We use $\epsilon_i \sim i.i.d. \
\mathcal{N}(0, \sigma_\varepsilon^2)$.
Based on  the results of Section~\ref{ss:pro},
we use  $\varphi = 0.55$ and $\sigma_\varepsilon^2 = 0.25$.

We use four intraday volatility functions:
\begin{itemize}\setlength\itemsep{0em}
		\item Flat: $\sigma (u) = 0.2$, the same intraday volatility throughout
                 the day.
		\item Slope: $\sigma (u) = 0.1 + 0.2u$, intraday volatility increases
in a linear manner.
		\item Sine: $\sigma (u) = 0.1\sin(2\pi u) + 0.2$,
higher volatility in the morning, but lower volatility in the afternoon.
		\item U-shape: $\sigma (u) = (u-0.5)^2 + 0.1145299$.
This choice is most relevant as it reflects the stylized fact that the volatility
is typically highest at the beginning and the end of a trading day.
\end{itemize}
The coefficients in the four
$\sigma$ functions are set to ensure the same level of average daily volatility.
Figure \ref{fig:simsigmahg} in Section \ref{s:HG} in the Supplementary Material
displays the four choices of $\sigma$ function and
 their corresponding (theoretical)  $H$ and $G$ functions.
 The $H$ and $G$ functions under
 the U-shape $\sigma$ function exhibit a generally similar pattern to
 those shown in Figure \ref{fig:apple}, with a minor difference at the final
 5-min interval of the trading day.

To simulate the integral $\int_{0}^{t} \sigma (u) d W (u)$,
we use the time change formula explained in Section~\ref{s:prelim},
cf. Corollary \ref{cor:tim:chng}, i.e.
we set
\begin{equation} \label{e:tc}
\int_{0}^{t} \sigma(u) d W(u)
= W\left( \int_{0}^{t} \sigma^2(u)du\right).
\end{equation}
Corresponding to the analysis in Section \ref{ss:pro}, 	
the continuous time $t$ in $[0,1]$ is discretized as
$\left[t_0, t_1, ..., t_K \right] $, where $t_k=  k\Delta$, $k=1,...,K$.
The stepsize is chosen to be
$\Delta=1/78$ which corresponds to 5-min frequency.  Using \refeq{tc},
we generate
\[
\int_{0}^{t_k} \sigma(u) d W(u) = \sum_{s=1}^{k} d(t_k), \quad k=1,...,K,
\]
where $d(t_k) \sim \mathcal{N}(0, G(t_k) - G(t_{k-1}))$ are independent
random variables.

We consider sample sizes
$N = 100, 500, 1000, 2000$, with the larger sizes being most
relevant (we used over 4,000 trading days  in Section \ref{ss:pro}).

To compare the different procedures, we use the following evaluation
metrics.
For the estimators  of $\varphi$ and ${\sigma}_\varepsilon^2$,
we calculate the empirical bias (B) and the empirical
mean root squared error (RMSE). For Procedure A, they are defined as
	$$
	\mbox{B}(\check{\varphi})
= \frac{1}{R} \sum_{r=1}^{R} \check{\varphi}_r - \varphi,
\qquad \mbox{RMSE} (\check{\varphi})
= \left\lbrace \frac{1}{R} \sum_{r=1}^{R}
(\check{\varphi}_r - \varphi)^2 \right\rbrace^{1/2},
	$$
	and
	$$
	\mbox{B}(\check{\sigma}_\varepsilon^2)
= \frac{1}{R} \sum_{r=1}^{R} \check{\sigma}_{\varepsilon,r}^2
- \sigma_\varepsilon^2, \qquad \mbox{RMSE} (\check{\sigma}_\varepsilon^2)
= \left\lbrace \frac{1}{R} \sum_{r=1}^{R} (\check{\sigma}_{\varepsilon,r}^2 - \sigma_\varepsilon^2)^2 \right\rbrace^{1/2},
	$$
where the subscript $r$ denotes the $r$-th simulation repetition.
For Procedures B  and  C, the above metrics are defined analogously.

As for evaluating the estimation of $G(t)$, we know $\sigma (u)$
since we simulate the data, and thus we can compute the analytical
value of
	$$
	G(t) = \int_{0}^{t} \sigma^2 (u) du.
	$$
	Then we can compute the functional empirical bias (fB)
and the functional empirical root mean squared error (fRMSE),
	$$
	\mbox{fB}(\widehat{G}) = \left( \frac{1}{R} \sum_{r=1}^{R} \int_{0}^{1} (\widehat{G}_r(t) - G(t)) dt\right),
	$$
	and
	$$
		\mbox{fRMSE}(\widehat{G})
= \left( \frac{1}{R} \sum_{r=1}^{R} \int_{0}^{1}
(\widehat{G}_r(t) - G(t))^2 dt\right) ^{1/2}.
	$$
The above two measures are very close to  zero even for 	$N=100$,
and decrease with $N$ even further. We therefore report only
the functional relative error
	$$
		\mbox{fRE}(\widehat{G}) = \left( \frac{1}{R}
\sum_{r=1}^{R} \dfrac{\int_{0}^{1} (\widehat{G}_r(t) - G(t))^2 dt}
{\int_{0}^{1} G^2(t) dt}\right) ^{1/2}.
	$$

Table \ref{tab:est_phi_sigma} shows the estimation error of
$\varphi$ and ${\sigma}_\varepsilon^2$ under the four different $\sigma$
functions with $\alpha = \Delta$. Generally, the consistency of the
three procedures for estimating $\varphi$ and ${\sigma}_\varepsilon^2$
is well supported by the simulation results because the RMSE decreases
with the increase in sample size. Comparing between the three procedures,
Procedure A outperforms the other two procedures in term of lowest bias
and RMSE. When considering the four shapes of the $\sigma$ functions,
there is no substantial distinction in the results of the estimation of
$\varphi$ and ${\sigma}_\varepsilon^2$, although the U-shape has
marginally higher bias and RMSE than the others. Lastly,
our estimation procedures have some minor bias
(underestimate $\varphi$ and overestimate ${\sigma}_\varepsilon^2$),
which is consistent with the Yule-Walker estimator of the scalar AR(1) process,
as noted by \citetext{shaman1988bias}. We provide a bias-corrected version
of our three procedures in Section \ref{s:bs_est} in the Supplementary material.
They improve the accuracy in  small sample sizes, but do not make much
difference for large $N$.

\begin{table}[htbp]
	\centering
	\caption{Estimation error for the estimation of $\varphi$ and $\sigma_\varepsilon^2$ with $\alpha = \Delta$}
	 \resizebox{0.8\columnwidth}{!}{ \begin{tabular}{llrccccccc}
	 		\toprule
	 		\toprule
	 		&       &       & \multicolumn{3}{c}{$\varphi$} &       & \multicolumn{3}{c}{$\sigma_\varepsilon^2$} \\
	 		\midrule
	 		&       &       & Proc. A & Proc. B & Proc. C &       & Proc. A & Proc. B & Proc. C \\
	 		\midrule
	 		\textbf{Flat} &       &       &       &       &       &       &       &       &  \\
	 		\midrule
	 		\multirow{4}[2]{*}{B} & $N=100$ &       & -0.040 & -0.083 & -0.069 &       & 0.006 & 0.045 & 0.043 \\
	 		& $N=500$ &       & -0.016 & -0.060 & -0.046 &       & 0.008 & 0.048 & 0.045 \\
	 		& $N=1000$ &       & -0.014 & -0.057 & -0.043 &       & 0.008 & 0.049 & 0.046 \\
	 		& $N=2000$ &       & -0.011 & -0.055 & -0.041 &       & 0.008 & 0.049 & 0.046 \\
	 		\midrule
	 		\multirow{4}[2]{*}{RMSE} & $N=100$ &       & 0.095 & 0.120 & 0.110 &       & 0.038 & 0.060 & 0.058 \\
	 		& $N=500$ &       & 0.042 & 0.071 & 0.060 &       & 0.018 & 0.051 & 0.049 \\
	 		& $N=1000$ &       & 0.030 & 0.063 & 0.051 &       & 0.014 & 0.050 & 0.047 \\
	 		& $N=2000$ &       & 0.022 & 0.058 & 0.045 &       & 0.012 & 0.050 & 0.047 \\
	 		\midrule
	 		\midrule
	 		\textbf{Slope} &       &       &       &       &       &       &       &       &  \\
	 		\midrule
	 		\multirow{4}[2]{*}{B} & $N=100$ &       & -0.045 & -0.087 & -0.073 &       & 0.008 & 0.047 & 0.045 \\
	 		& $N=500$ &       & -0.019 & -0.062 & -0.048 &       & 0.010 & 0.050 & 0.047 \\
	 		& $N=1000$ &       & -0.015 & -0.058 & -0.044 &       & 0.011 & 0.051 & 0.048 \\
	 		& $N=2000$ &       & -0.014 & -0.057 & -0.043 &       & 0.011 & 0.051 & 0.048 \\
	 		\midrule
	 		\multirow{4}[2]{*}{RMSE} & $N=100$ &       & 0.097 & 0.121 & 0.112 &       & 0.038 & 0.060 & 0.058 \\
	 		& $N=500$ &       & 0.042 & 0.072 & 0.060 &       & 0.019 & 0.053 & 0.050 \\
	 		& $N=1000$ &       & 0.031 & 0.064 & 0.051 &       & 0.015 & 0.052 & 0.049 \\
	 		& $N=2000$ &       & 0.024 & 0.060 & 0.047 &       & 0.014 & 0.051 & 0.049 \\
	 		\midrule
	 		\midrule
	 		\textbf{Sine} &       &       &       &       &       &       &       &       &  \\
	 		\midrule
	 		\multirow{4}[2]{*}{B} & $N=100$ &       & -0.046 & -0.087 & -0.073 &       & 0.008 & 0.046 & 0.044 \\
	 		& $N=500$ &       & -0.020 & -0.061 & -0.047 &       & 0.011 & 0.049 & 0.047 \\
	 		& $N=1000$ &       & -0.017 & -0.058 & -0.044 &       & 0.012 & 0.051 & 0.048 \\
	 		& $N=2000$ &       & -0.016 & -0.057 & -0.043 &       & 0.012 & 0.051 & 0.048 \\
	 		\midrule
	 		\multirow{4}[2]{*}{RMSE} & $N=100$ &       & 0.098 & 0.122 & 0.112 &       & 0.038 & 0.060 & 0.058 \\
	 		& $N=500$ &       & 0.044 & 0.072 & 0.061 &       & 0.020 & 0.052 & 0.050 \\
	 		& $N=1000$ &       & 0.032 & 0.064 & 0.052 &       & 0.017 & 0.052 & 0.050 \\
	 		& $N=2000$ &       & 0.025 & 0.060 & 0.047 &       & 0.014 & 0.051 & 0.049 \\
	 		\midrule
	 		\midrule
	 		\textbf{U-Shape} &       &       &       &       &       &       &       &       &  \\
	 		\midrule
	 		\multirow{4}[2]{*}{B} & $N=100$ &       & -0.047 & -0.091 & -0.077 &       & 0.010 & 0.051 & 0.048 \\
	 		& $N=500$ &       & -0.021 & -0.065 & -0.052 &       & 0.012 & 0.054 & 0.051 \\
	 		& $N=1000$ &       & -0.018 & -0.062 & -0.048 &       & 0.013 & 0.054 & 0.052 \\
	 		& $N=2000$ &       & -0.017 & -0.061 & -0.047 &       & 0.013 & 0.054 & 0.052 \\
	 		\midrule
	 		\multirow{4}[2]{*}{RMSE} & $N=100$ &       & 0.098 & 0.125 & 0.116 &       & 0.038 & 0.064 & 0.062 \\
	 		& $N=500$ &       & 0.044 & 0.076 & 0.065 &       & 0.021 & 0.057 & 0.054 \\
	 		& $N=1000$ &       & 0.032 & 0.068 & 0.055 &       & 0.017 & 0.056 & 0.053 \\
	 		& $N=2000$ &       & 0.025 & 0.064 & 0.051 &       & 0.015 & 0.055 & 0.053 \\
	 		\bottomrule
	 		\bottomrule
	 	\end{tabular}}%
	\label{tab:est_phi_sigma}%
\end{table}

Table \ref{tab:est_Gt} presents the estimation error of the $G$ function.
As we use the same estimation for $G$ in the three procedures,
there is no need to compare them.  It can be observed that the fRE
decreases with the larger sample size. Specifically, the fRE decreases from approximately
22\% for a sample of 100 to almost 5\% for a sample of 2000.  This provides evidence supporting the consistency of our estimation of the $G$ function. Again, there is no obvious different pattern between the four shapes of $\sigma$ functions for the estimation results of the  $G$ function.

\begin{table}[htbp]
	\centering
	\caption{Functional relative error (fRE) for the estimation of $G(t)$}
	\begin{tabular}{lrcccc}
		\toprule
		\toprule
		&       & $N=100$ & $N=500$ & $N=1000$ & $N=2000$ \\
		\midrule
		Flat  &       & 22.5\% & 10.1\% & 7.3\% & 5.3\% \\
		Slope &       & 22.1\% & 10.0\% & 7.1\% & 5.3\% \\
		Sine  &       & 22.5\% & 9.9\% & 7.4\% & 5.5\% \\
		U-shape &       & 22.1\% & 10.5\% & 7.9\% & 6.2\% \\
		\bottomrule
		\bottomrule
	\end{tabular}%
	\label{tab:est_Gt}%
\end{table}

The choice of $\alpha$ affects the estimation of $\varepsilon$
and $\sigma_{\varepsilon}$ in Procedures B and C, but not much.
Table \ref{tab:sim_sense_alpha} shows the bias and the RMSE of
estimation error of $\varphi$ and $\sigma_\varepsilon^2$ under the  U-shaped
$\sigma(u)$ with$\alpha \in \left\lbrace 5\Delta, 20\Delta, 40\Delta\right\rbrace $.
We can observe that the bias and RMSE exhibit a marginal reduction when
a larger value of $\alpha$ is employed. Lastly, even setting
$\alpha = 40 \Delta$,  Procedures B and C still still underperform Procedure A.

\begin{table}[htbp]
	\centering
	\caption{Sensitivity to $\alpha$}
	 \resizebox{0.7\columnwidth}{!}{\begin{tabular}{llrccccc}
	 		\toprule
	 		\toprule
	 		&       &       & \multicolumn{2}{c}{$\varphi$} &       & \multicolumn{2}{c}{$\sigma^2$} \\
	 		\midrule
	 		&       &       & Proc. B & Proc. C &       & Proc. B & Proc. C \\
	 		\midrule
	 		$\alpha=5\Delta$ &       &       &       &       &       &       &  \\
	 		\midrule
	 		\multirow{4}[2]{*}{B} & $N=100$ &       & -0.068 & -0.066 &       & 0.028 & 0.028 \\
	 		& $N=500$ &       & -0.041 & -0.040 &       & 0.031 & 0.031 \\
	 		& $N=1000$ &       & -0.039 & -0.038 &       & 0.032 & 0.032 \\
	 		& $N=2000$ &       & -0.038 & -0.036 &       & 0.032 & 0.032 \\
	 		\midrule
	 		\multirow{4}[2]{*}{RMSE} & $N=100$ &       & 0.112 & 0.111 &       & 0.048 & 0.048 \\
	 		& $N=500$ &       & 0.057 & 0.056 &       & 0.036 & 0.036 \\
	 		& $N=1000$ &       & 0.048 & 0.047 &       & 0.034 & 0.034 \\
	 		& $N=2000$ &       & 0.042 & 0.041 &       & 0.033 & 0.033 \\
	 		\midrule
	 		\midrule
	 		$\alpha=20\Delta$ &       &       &       &       &       &       &  \\
	 		\midrule
	 		\multirow{4}[2]{*}{B} & $N=100$ &       & -0.058 & -0.058 &       & 0.019 & 0.019 \\
	 		& $N=500$ &       & -0.032 & -0.032 &       & 0.022 & 0.022 \\
	 		& $N=1000$ &       & -0.029 & -0.029 &       & 0.022 & 0.022 \\
	 		& $N=2000$ &       & -0.027 & -0.027 &       & 0.022 & 0.022 \\
	 		\midrule
	 		\multirow{4}[2]{*}{RMSE} & $N=100$ &       & 0.106 & 0.106 &       & 0.044 & 0.044 \\
	 		& $N=500$ &       & 0.051 & 0.051 &       & 0.028 & 0.028 \\
	 		& $N=1000$ &       & 0.040 & 0.040 &       & 0.025 & 0.025 \\
	 		& $N=2000$ &       & 0.033 & 0.033 &       & 0.024 & 0.024 \\
	 		\midrule
	 		\midrule
	 		$\alpha=40\Delta$ &       &       &       &       &       &       &  \\
	 		\midrule
	 		\multirow{4}[2]{*}{B} & $N=100$ &       & -0.051 & -0.051 &       & 0.014 & 0.014 \\
	 		& $N=500$ &       & -0.027 & -0.027 &       & 0.018 & 0.018 \\
	 		& $N=1000$ &       & -0.024 & -0.024 &       & 0.018 & 0.018 \\
	 		& $N=2000$ &       & -0.023 & -0.023 &       & 0.018 & 0.018 \\
	 		\midrule
	 		\multirow{4}[2]{*}{RMSE} & $N=100$ &       & 0.102 & 0.102 &       & 0.040 & 0.040 \\
	 		& $N=500$ &       & 0.047 & 0.047 &       & 0.024 & 0.024 \\
	 		& $N=1000$ &       & 0.037 & 0.037 &       & 0.022 & 0.022 \\
	 		& $N=2000$ &       & 0.030 & 0.030 &       & 0.020 & 0.020 \\
	 		\bottomrule
	 		\bottomrule
	 	\end{tabular}}%
	\label{tab:sim_sense_alpha}%
\end{table}%

\clearpage

\section{Summary and further work} \label{s:sum}
We have proposed a model for  a sequence of curves
of normalized intraday asset prices. The curves are functional
analogs of point-to-point daily returns. The model extends
the extensively used stochastic volatility paradigm to the
setting of functional time series. It includes day-to-day dynamics
that quantify the dependence between the daily curves as
well as a diffusion model for the evolution of the curves within
each day. We have developed estimation methodology for this
model and justified it both theoretically and via  a data application 
and an informative  simulation study.  

The formulation of the model framework and effective
estimation methodology opens up paths toward further
developments. In particular, order selection, the selection
of $p$ in Section \ref{s:ext},  may be practically relevant.
Related to this, goodness-of-fit tests that consider
the suitability of any model in the model family we introduced
is another important problem. The ability of the model
to predict future curves should be explored. All these tasks
will require new theoretical derivations and extensive
numerical studies. It is hoped that this paper will motivate
research in these and other similar directions, including more extensive 
applications. 

\bigskip

\noindent{\bf Acknowledgements} \
We are grateful for Dr. Jian Chen's help on collecting and cleaning the U.S. stock data.
This research was partially
supported by the United States NSF grant DMS--2123761.

\bigskip
\bibliographystyle{oxford3}
\renewcommand{\baselinestretch}{0.9}
\small
\bibliography{neda}

\appendix

\newpage

\renewcommand{\baselinestretch}{1.0}
\normalsize

\setcounter{page}{1}

\begin{center}
{\it \Large Supplementary Material}
\end{center}

\section{Proofs of the results of Section \ref{s:est}}\label{s:proofs}
\subsection{Fundamental Results}
In the current  subsection, we state Lemma \ref{lem:2nd:moment} and Proposition \ref{prop:unif:2nd} and the resulting Corollary 
\ref{cor:unif:log:2nd}  which are the cornerstones of the proofs of our main results.
The  results of this  subsection concern the dynamics within each random
curve indexed by $i$. They establish bounds in terms of $\Delta$
that do not depend on $i$ nor on the sample size $N$.

\begin{lemma}\label{lem:2nd:moment} 
 Consider the volatility process $\int_0^t\sigma(u)dW_i(u)$   
 in  \eqref{e:R1}.  Assume conditions \ref{itm:sig}--\ref{itm:int:sig} 
 and define  the associated oracle empirical quadratic variation process
   \begin{align} \label{e:V_i}
        V_i (t) := \sum_{k=1}^m \left \vert \int_{t_{k-1}}^{t_k}\sigma(u)dW_i(u)\right\vert^2 \mathbb{I}\{t_k  \leq t \} ,\quad t \in [0,1], \quad i =1,2,\ldots N.
    \end{align}
Define also 
    \begin{align}\label{e:X_i}
    X_i(t) &:= \log V_i (t) = \log \sum_{k=1}^m \left \vert \int_{t_{k-1}}^{t_k}\sigma(u)dW_i(u)\right\vert^2 \mathbb{I}\{t_k \leq t \}, \quad t \in (0,1], \quad i=1,2,\ldots,N.
\end{align}
Then, for any fixed  $0 < \alpha < 1$,
     \begin{align}\label{e:Var(xi)}
        \underset{t \in [\alpha,1]}{\sup} \mathbb{E}\left[ X_i(t) - H(t)\right]^2 = O(\Delta),  \quad i=1,2,\ldots,N,
     \end{align}
     and
     \begin{align} \label{e:1th:moment}
  \underset{t \in [\alpha,1]}{\sup}  \mathbb{E} \left\vert  X_i(t) - H(t)  \right \vert & = O\left( \Delta^{\frac{1}{2}} \right),  \quad i=1,2,\ldots,N.
\end{align}
 \end{lemma}

\noindent{\sc Proof of Lemma \ref{lem:2nd:moment}.} Fix $ 0 < \alpha < 1 $. Observe that for  any $t \geq \alpha $
\begin{align*}
    \mathbb{E}\left[ X_i(t) -  H(t)\right]^2 = &  \mathbb{E}\left[ \log \left(\frac{V_i(t)}{G(t)}\right) \right]^2\\
    \leq & \mathbb{E}\left[ \frac{V_i(t)}{G(t)}-1 \right]^2\\
    = & \mathbb{E}\left[ \frac{V_i(t)- G(t)}{G(t)} \right]^2\\
    = & \frac{1}{G^2(t)} \mathbb{E}\left[ V_i(t)- G(t) \right]^2.
\end{align*}
Assumption \ref{itm:int:sig} guarantees the boundedness of function $G(t)$ away from zero on the domain $[\alpha , 1]$, see Remark \ref{rmk:alpha}. So it is enough to obtain the desired result for $\mathbb{E}\left[ V_i(t)- G(t) \right]^2$. To do so,  observe that
\begin{align}
    \label{e:(V-G)^2}
\mathbb{E}\left[ V_i(t)- G(t) \right]^2 = &\mathbb{E}\left[ \sum_{k=1}^m \left \vert \int_{t_{k-1}}^{t_k}\sigma(u)dW_i(u)\right\vert^2 \mathbb{I}\{t_k  \leq t \} - G(t) \right]^2\\ \label{e:t_k:t}
\leq &
2\mathbb{E}\left[ \sum_{k=1}^m \left \vert \int_{t_{k-1}}^{t_k}\sigma(u)dW_i(u)\right\vert^2 \mathbb{I}\{t_k  \leq t \} - G(t_k) \right]^2
+
2 \left[ G(t_k) - G(t) \right]^2
\end{align}
By assumption \ref{itm:sig}, the second term $ \left[ G(t_k) - G(t) \right]^2$ is dominated by $\Vert \sigma^2\Vert_{\infty}^2 \Delta^2$. We now turn to the first summand in \eqref{e:t_k:t}:
\begin{align}
\nonumber
\mathbb{E}\left[ \sum_{k=1}^m \left \vert \int_{t_{k-1}}^{t_k}\sigma(u)dW_i(u)\right\vert^2 \mathbb{I}\{t_k  \leq t \} - G(t_k) \right]^2= & \mathrm{Var} \left[ \sum_{k=1}^m \left \vert \int_{t_{k-1}}^{t_k}\sigma(u)dW_i(u)\right\vert^2 \mathbb{I}\{t_k  \leq t \}  \right]\\ \label{e:Var:sum}
= &   \sum_{k=1}^m   \mathrm{Var} \left [ \left \vert \int_{t_{k-1}}^{t_k}\sigma(u)dW_i(u)\right\vert^2 \mathbb{I}\{t_k  \leq t \}  \right ]\\ \nonumber
\leq & \sum_{k=1}^m   \mathrm{Var} \left \vert \int_{t_{k-1}}^{t_k}\sigma(u)dW_i(u)\right\vert^2 \\ \label{e:time:change}
= & \sum_{k=1}^m   \mathrm{Var} \left [ W_1\left( \int_{t{_k-1}}^{t_k}
\sigma^2(u)du\right)\right]^2\\ \nonumber
\leq & \sum_{k=1}^m   \mathbb{E} \left [ W_1\left( \int_{t{_k-1}}^{t_k}
\sigma^2(u)du\right)\right]^4\\ \nonumber
\leq & 3 \sum_{k=1}^m  \left( \int_{t_{k-1}}^{t_k}
\sigma^2(u)du\right)^2\\ \nonumber
\leq & 3 \Vert \sigma^2\Vert_{\infty}^2 m \Delta^2 \\ \label{e:bound:Delt}
= &  3 \Vert \sigma^2\Vert_{\infty}^2 \Delta.
\end{align}
Equation \eqref{e:Var:sum} is a result of the independence of 
non--overlapping increments of the Wiener process. 
Equation \eqref{e:time:change} is a consequence of
 Corollary \ref{cor:tim:chng}. This completes the proof of \eqref{e:Var(xi)}.
Combining  \eqref{e:Var(xi)} and Lyapunov's inequality,
we get \eqref{e:1th:moment}.

\rightline{\QED}

\begin{proposition}\label{prop:unif:2nd}
Consider the volatility process $\int_0^t\sigma(u)dW_i(u)$ appearing  
in  \eqref{e:R1}. Assume   conditions \ref{itm:sig}--\ref{itm:phi} hold  
and recall \eqref{e:V_i}. Then, 
     \begin{align}\label{e:unif:Var(xi)}
          \mathbb{E}\left[ \underset{t \in [0,1]}{\sup} 
          \left \vert V_i(t) -  G(t)\right \vert\right]^2
    = &  O(\Delta).
     \end{align}
\end{proposition}

\noindent{\sc Proof of Proposition \ref{prop:unif:2nd}} We drop the index $i$ 
and prove the result  in a general form.
We  first obtain the desired uniform result \eqref{e:unif:Var(xi)} 
by restricting the time domain  to the discrete points  
$t_0 = 0 < t_1 < \ldots < t_m = 1$. This is obtained by an application of  
Doob's maximal inequality to  the discrete time sequence 
$V(t_j) - G(t_j)$, $j =0,1,\ldots,m$,  see \citetext{dasgupta:2011} 
Theorem 14.7, for example. To do so, observe that the random variables
\begin{align*}
    \left \vert \int_{t_{k-1}}^{t_k}\sigma(u)dW(u)\right\vert^2 - \int_{t_{k-1}}^{t_k}\sg^2(u) du , \quad k = 1,2 , \ldots m,
\end{align*}
form a sequence of mean zero independent random variables. Hence the sequence
\begin{align*}
    V(t_j) - G(t_j) =& \sum_{k=1}^j \left \vert \int_{t_{k-1}}^{t_k}\sigma(u)dW(u)\right\vert^2 - \int_0^{t_j}\sg^2(u) du\\
    =& \sum_{k=1}^j \left (\left \vert \int_{t_{k-1}}^{t_k}\sigma(u)dW(u)\right\vert^2 - \int_{t_{k-1}}^{t_k}\sg^2(u) du \right), \quad j = 1,2 , \ldots m,
\end{align*}
forms a  martingale.
Applying Doob's maximal inequality, we obtain 
\begin{align*}
    \mathbb{E}\left[ \underset{j \in \{1,2,\ldots,m\}}{\sup} \left \vert V(t_j) -  G(t_j)\right \vert\right]^2 \leq & 4 \mathbb{E} \left \vert V(t_m) -  G(t_m)\right \vert^2\\
    = & 4 \mathbb{E} \left \vert V(1) -  G(1)\right \vert^2.
\end{align*}
Following the lines \eqref{e:(V-G)^2}--\eqref{e:bound:Delt}, 
with $t=1$, we conclude
\begin{align}\label{e:Vj-Gj}
    \mathbb{E}\left[ \underset{j \in \{1,2,\ldots,m\}}{\sup} \left \vert V(t_j) -  G(t_j)\right \vert\right]^2
    = &  O(\Delta).
\end{align}
For an arbitrary  $t \in [0,1]$, we have
\begin{align}\label{e:V-G}
    \left \vert V(t) -  G(t)\right \vert^2 \leq 2  \left \vert  V(t) -  G(t) - V(t_j) + G(t_j)\right \vert^2 + 2  \left \vert V(t_j) - G(t_j) \right \vert^2,
\end{align}
where $t_j \leq t < t_{j+1}$ that implies $\vert t_j - t \vert < \Delta$. Hence, definition \eqref{e:V_i} gives $V(t_j) = V(t)$. This leads   to
\begin{align*}
 \left \vert V(t) -  G(t)\right \vert^2 \leq & 2\left \vert     G(t) -G(t_j) \right \vert^2 + 2  \left \vert V(t_j) - G(t_j) \right \vert^2.
\end{align*}
Assumption \ref{itm:sig} implies
\begin{align*}
 \underset{t \in [0,1]}{\sup} \left \vert V(t) -  G(t)\right \vert^2 \leq & 2 \Delta^2 \Vert\sigma \Vert_{\infty}^2 + 2 \underset{j \in \{1,2,\ldots,m\}}{\sup} \left \vert V(t_j) - G(t_j) \right \vert^2.
\end{align*}
 Now, by applying \eqref{e:Vj-Gj}, we obtain
 \begin{align}
     \mathbb{E}\left[ \underset{t \in [0,1]}{\sup} \left \vert V(t) -  G(t)\right \vert\right]^2
    = &  O(\Delta),
 \end{align}
 as desired.

\rightline{\QED}

\begin{corollary}\label{cor:unif:log:2nd}
    Assume the setting of Proposition \ref{prop:unif:2nd} together with Assumption \ref{itm:int:sig}. Then,  for any fixed  $0 < \alpha < 1$,
    \begin{align} \label{e:XH2}
    \mathbb{E} \underset{t \in [\alpha,1]}{\sup} \left\vert X_i(t) - H(t) \right \vert^2 & = O\left( \Delta \right),  \quad i=1,2,\ldots,N.
\end{align}
and
    \begin{align} \label{e:XH1}
    \mathbb{E} \underset{t \in [\alpha,1]}{\sup} \left\vert X_i(t) - H(t) \right \vert & = O\left( \Delta^{\frac{1}{2}} \right),  \quad i=1,2,\ldots,N.
\end{align}
\end{corollary}
\noindent{\sc Proof of Corollary \ref{cor:unif:log:2nd}}
Fix $\alpha \in (0,1)$. Following the lines of the proof of Lemma \ref{lem:2nd:moment} we have
\begin{align*}
 \mathbb{E}  \underset{t \in [\alpha,1]}{\sup} \left \vert X_i(t) - H(t) \right \vert^2 = &
    \mathbb{E}  \underset{t \in [\alpha,1]}{\sup} \left \vert \log V_i(t) - \log G(t) \right \vert^2 \\= &  \mathbb{E} \underset{t \in [\alpha,1]}{\sup} \left\vert \log \left(\frac{V_i(t)}{G(t)}\right) \right\vert^2\\
    \leq & \mathbb{E}  \underset{t \in [\alpha,1]}{\sup}\left\vert \frac{V_i(t)}{G(t)}-1 \right\vert^2\\
    = & \mathbb{E}  \underset{t \in [\alpha,1]}{\sup} \left\vert \frac{V_i(t)- G(t)}{G(t)} \right \vert^2\\
    \leq & \frac{1}{ \underset{t \in [\alpha,1]}{\inf} G^2(t)} \mathbb{E}  \underset{t \in [\alpha,1]}{\sup} \left\vert V_i(t)- G(t) \right\vert^2.
\end{align*}
Assumption \ref{itm:int:sig} guarantees the boundedness away from zero
of  $ G^2(t)$ for $t \in [\alpha,1]$. 
Proposition \ref{prop:unif:2nd} entails \eqref{e:XH2}, 
and an application of  Lyapunov's inequality implies \eqref{e:XH1}.

\rightline{\QED}

 \subsection{Proof of Theorem \ref{thm:hat:H}}
We split the proof of   Theorem \ref{thm:hat:H} into two  steps. In step 1, we assume the latent processes $Q_i(t)$, $i=1,2,\ldots, N$, are observable, so we  apply functional (or univariate) time series techniques to obtain the  asymptotic result and the rate of convergence in terms of $N$, see Proposition \ref{prop:tild:G} below.  Indeed, Proposition \ref{prop:tild:G} entails uniform   convergence of  $\frac{1}{N}\sum_{i=1}^{N} \log Q_i(t)$ in the mean square sense with the parametric rate $\sqrt{N}$.
In Step 2, we study the difference between the \textit{average} of the serially dependent empirical quadratic variation processes,
$\frac{1}{N}\sum_{i=1}^{N} \log \widehat{Q}_i(t)$, and its oracle counterpart $\frac{1}{N}\sum_{i=1}^{N}\log Q_i(t)$. This entails a rate in terms of $\Delta$, see Proposition \ref{prop:volatility} below. Combining these results, we obtain   Theorem \ref{thm:hat:H}. Recall \eqref{e:Ht} and \eqref{e:tild:H}.

\begin{proposition}\label{prop:tild:G}
Assume the stochastic volatility model defined by \eqref{e:R1}--\eqref{e:g:AR1} and  conditions \ref{itm:sig}--\ref{itm:phi}. Then
    \begin{align*}
         \mathbb{E} \underset{t \in (0,1]}{\sup }\left \vert \widetilde{H}(t) -  H(t)\right \vert^2  = O(N^{-1}).
    \end{align*}
\end{proposition}

\noindent{\sc Proof of Proposition \ref{prop:tild:G}}
Observe that
\begin{align}
\nonumber N \mathbb{E} \underset{t \in (0,1]}{\sup } \left \vert \widetilde{H}(t) -  H(t) \right \vert^2 = & N  \mathbb{E} \underset{t \in (0,1]}{\sup } \left \vert \frac{1}{N}\sum_{i=1}^{N} \log Q_i(t) - H(t)\right \vert^2 \\
\label{e:logQ/G}
   =& N  \mathbb{E} \underset{t \in (0,1]}{\sup } \left \vert  \frac{1}{N}\sum_{i=1}^{N} \log \left( g^2_i \int_0^t\sigma^2(u)du \right) - \log  \int_0^t\sigma^2(u)du\right \vert^2
   \\ \nonumber 
   =& N  \mathbb{E} \left \vert  \frac{2}{N}\sum_{i=1}^{N} \log g_i \right \vert^2
   \\  \label{eq:covg}
   =& \frac{4}{N} \sum_{\vert h \vert \leq N-1} \mathrm{Cov}\left(\log g_{\vert h \vert+1}, \log g_1 \right) (N -\vert h \vert )
   \\ \nonumber
   =& 4 \sum_{\vert h \vert \leq N-1} \mathrm{Cov}\left(\log g_{\vert h \vert +1}, \log g_1 \right) \left( 1- \frac{\vert h \vert}{N}\right)
   \\ \nonumber 
   \leq & \frac{4}{1-\vert \varphi \vert^2} \mathrm{Var} \left ( \varepsilon_1\right) \sum_{\vert h \vert} \vert \varphi \vert^{\vert h \vert}
   \\ \nonumber
   < & \infty.
   \end{align}
   Where the  equality \eqref{e:logQ/G} follows  from  \eqref{e:QV} and \eqref{eq:covg} is a consequence of stationarity of the process $\{\log g_i\}_i$. This completes the proof.

\rightline{\QED}
\begin{proposition}\label{prop:volatility}
Assume the stochastic volatility model defined by \eqref{e:R1}--\eqref{e:g:AR1} satisfying conditions \ref{itm:sig}--\ref{itm:int:sig} and consider $\widehat{Q}_i(t)$   in \eqref{e:Qhat}.  Then,  for any fixed   $0 < \alpha < 1$,
   \begin{align*}
   \mathbb{E}  \underset{t \in [\alpha,1]}{\sup} \left\vert  \widehat{H}(t) -   \widetilde{H}(t) \right\vert  = O\left( \Delta^{\frac{1}{2}} \right).
   \end{align*}
\end{proposition}

 \noindent{\sc Proof of Proposition \ref{prop:volatility}.}
 Using definition \eqref{e:Qhat}  and equation \eqref{e:logQ}, one may write
\begin{align}
 \label{e:lgQ:lgQ}
 \widehat{H}(t) -   \widetilde{H}(t) = &\frac{1}{N}\sum_{i=1}^{N} \log \widehat{Q}_i(t)-  \frac{1}{N}\sum_{i=1}^{N}\log Q_i(t)
 \\ \nonumber = &  \frac{1}{N} \sum_{i=1}^{N} 2\log g_i + \frac{1}{N} \sum_{i=1}^{N} \log \sum_{k=1}^m \left \vert \int_{t_{k-1}}^{t_k}\sigma(u)dW_i(u)\right\vert^2 \mathbb{I}\{t_k
 \leq t \}
 \\ \nonumber
 & - \frac{1}{N} \sum_{i=1}^{N} 2\log g_i - \frac{1}{N} \sum_{i=1}^{N}  H(t)
 \\ \nonumber
 =&  \frac{1}{N} \sum_{i=1}^{N} \log \sum_{k=1}^m \left \vert \int_{t_{k-1}}^{t_k}\sigma(u)dW_i(u)\right\vert^2 \mathbb{I}\{t_k
 \leq t \}  -   H(t)
 \end{align}
 Recalling \eqref{e:X_i}, the difference in \eqref{e:lgQ:lgQ} can be rewritten as:
 \begin{align}
\nonumber  
 \frac{1}{N} \sum_{i=1}^{N} \left[X_i(t) -   H(t) \right].
 \end{align}
 Applying  Corollary \ref{cor:unif:log:2nd} together with the fact that the processes $\{X_i(t)\}_t$, $i=1,2,\ldots,N$,  form an independent sequence over index $i$,  we conclude
     \begin{align}\label{e:step2}
         \mathbb{E}\underset{t \in [\alpha,1]}{\sup} \left\vert\frac{1}{N}\sum_{i=1}^{N}X_i(t) -  H(t)\right\vert = O\left(\Delta^{\frac{1}{2}} \right),
    \end{align}
   as desired.

\rightline{\QED}

\noindent{\sc Proof of Theorem \ref{thm:hat:H}.}
Propositions \ref{prop:tild:G}   and \ref{prop:volatility} imply  the limiting result \eqref{e:cons:Hhat-H}.  We now investigate  \eqref{e:cons:Ghat-G}. Observe that the first order Taylor series expansion  of the exponential function $\exp(\cdot)$ gives

\begin{align*}
\nonumber   \left  \vert \widehat{G}(t) - G (t) \right \vert
  = & \left  \vert \exp \widehat{H}(t) - \exp  H (t) \right \vert\\
  = & \left  \vert \widehat{H}(t) -   H (t) \right \vert \exp (\xi(t)), \quad t \in [\alpha , 1],
 \end{align*}
 where $\xi(t)$ lies between $H (t)$ and $\widehat{H}(t)$. Consequently,

 \begin{align}
\nonumber
\underset{t \in [\alpha,1]}{\sup} \left  \vert \widehat{G}(t) - G (t) \right \vert
\leq  &  \underset{t \in [\alpha,1]}{\sup}  \left  \vert \widehat{H}(t) -   H (t) \right \vert \underset{t \in [\alpha,1]}{\sup}  \exp (\xi(t))\\ \label{e:exp:Hhat-H}
  \leq  & \underset{t \in [\alpha,1]}{\sup}  \left  \vert \widehat{H}(t) -   H (t) \right \vert \underset{t \in [\alpha,1]}{\sup}  \exp \left \vert \xi(t) - H(t)\right \vert \underset{t \in [\alpha,1]}{\sup}  \exp \left \vert  H(t)\right \vert
 \end{align}
 The result \eqref{e:cons:Hhat-H} entails
 $\underset{t \in [\alpha,1]}{\sup}  \left  \vert \widehat{H}(t) -   H (t) \right \vert  = O_P\left( N^{-\frac{1}{2}} + \Delta^{\frac{1}{2}} \right)$.
 Combining this with the continuous mapping theorem gives $\underset{t \in [\alpha,1]}{\sup}  \exp \left \vert\widehat{H}(t) - H(t))\right \vert $ is convergent to $1$, in probability. In particular, $\underset{t \in [\alpha,1]}{\sup}  \exp \left \vert\widehat{H}(t) - H(t))\right \vert $ is bounded in probability.
 Inequalities \eqref{e:alpha} entail boundedness of the deterministic term $\exp \left \vert  H(t)\right \vert$. In more detail:
 \begin{align*}
0 < \int_0^{\alpha} \sigma^2(u) du  \leq   G(t) \leq \Vert\sigma \Vert^2_{\infty} < \infty, \qquad t \in [\alpha,1 ].
 \end{align*}
 Consequently,
 \begin{align*}
- \infty <  \log \int_0^{\alpha} \sigma^2(u) du  \leq  H(t) \leq \log\Vert\sigma \Vert^2_{\infty}< \infty, \qquad t \in [\alpha,1 ].
 \end{align*}
Hence,
 \begin{align*}
 \left \vert H(t) \right \vert \leq \max \left \{\left \vert \log\Vert\sigma \Vert^2_{\infty}\right \vert, \left \vert \log \int_0^{\alpha} \sigma^2(u) du \right \vert \right\} < \infty, \qquad t \in [\alpha,1 ].
 \end{align*}
We thus  conclude \eqref{e:exp:Hhat-H} is bounded above as follows:
\begin{align*}
    O_P\left( N^{-\frac{1}{2}} + \Delta^{\frac{1}{2}} \right) O_P(1) O_P(1)
     = & O_P\left( N^{-\frac{1}{2}} + \Delta^{\frac{1}{2}} \right),
\end{align*}
as desired.

\rightline{\QED}
\subsection{Proof of Theorem  \ref{thm:chk:phi:sig} }
The proof of Theorem   \ref{thm:chk:phi:sig}  relies on the   successive Propositions \ref{prop:gam:kapp}  and \ref{prop:Gam:gam} which investigate the differences $ \left (     \gamma_{0,N} - \kappa_0 ,     \gamma_{1,N} - \kappa_1\right)^{\top}$ and $ \left ( \Gamma_{0,N}(t)  -     \gamma_{0,N} , \Gamma_{1,N}(t)  -     \gamma_{1,N} \right)^{\top}$, respectively.


\begin{proposition}\label{prop:gam:kapp}
Assume the stochastic volatility model defined by \eqref{e:R1}--\eqref{e:g:AR1} satisfying conditions \ref{itm:sig}--\ref{itm:err4} and recall  \eqref{e:gam_0} and \eqref{e:gam_1}. Then
    \begin{align}\label{e:gam:kapp}
   N^{\frac{1}{2}} \left(
        \begin{array}{c}
            \gamma_{0,N} -\kappa_0 \\
                \gamma_{1,N} - \kappa_1
        \end{array}
        \right) \overset{\mathcal{L}aw}{\longrightarrow} \mathcal{N} \left(
        \left(
        \begin{array}{c}
            0  \\
             0
        \end{array}
        \right), V
        \right),
    \end{align}
where $V$ is a $2$ by $2$ matrix with entries
\begin{align}\label{e:V}
    V_{k,l} =  (\eta - 3) \kappa_k \kappa_l + \sum_{h= - \infty}^{\infty} \left( \kappa_h \kappa_{h-k+l}+\kappa_{h-k}\kappa_{h+l}\right), \quad k , l = 0,\;1.
\end{align}
\end{proposition}

\noindent{\sc Proof of Proposition \ref{prop:gam:kapp}}
Refer to Proposition 7.3.4 in \citetext{brockwell:davis:1991}.

\rightline{\QED}

\begin{remark}\label{rmk:between}
    Notice that the above  result concerns the dynamics between the random curves $\{R_i(t)\}_t$, $i=1,2,\ldots$ assuming full information of quadratic variation of each random curve.  So, we obtain the rates in terms of $N$ only, not $\Delta$, cf. Remark \ref{rmk:within}. Lemma \ref{lem:Gamm^-1} below relates $N$ and $\Delta$ appropriately.
\end{remark}

\begin{proposition}\label{prop:Gam:gam}
Assume the stochastic volatility model defined by \eqref{e:R1}--\eqref{e:g:AR1} satisfying conditions \ref{itm:sig}--\ref{itm:int:sig} and recall \eqref{e:c0hQ} and \eqref{e:c1hQ}. Then, for any fixed   $0 < \alpha < 1$,
\begin{align*}
    \underset{t \in [\alpha,1]}{\sup} \mathbb{E}   \left \vert     \gamma_{0,N}- \Gamma_{0,N}(t)  \right \vert^2  +  \underset{t \in [\alpha,1]}{\sup} \mathbb{E}   \left \vert     \gamma_{1,N}- \Gamma_{1,N}(t)  \right \vert^2 =  O \left( \frac{\Delta}{N}\right).
\end{align*}
\end{proposition}
\noindent{\sc Proof of Proposition \ref{prop:Gam:gam}}
We first  investigate the  difference $    \gamma_{1,N}- \Gamma_{1,N}(t) $ that reads
 \begin{align}
 \nonumber
         \gamma_{1,N}- \Gamma_{1,N}(t)  = & \frac{1}{4N}\sum_{i=1}^{N-1} \left(\log Q_i(t) - \frac{1}{N}\sum_{i=1}^N \log Q_i(t) \right)\left(\log Q_{i+1}(t) - \frac{1}{N}\sum_{i=1}^N \log Q_i(t) \right)\\  \nonumber
     &- \frac{1}{4N}\sum_{i=1}^{N-1} \left(\log \widehat{Q}_i(t)-\frac{1}{N}\sum_{i=1}^{N} \log \widehat{Q}_i(t)\right)\left(\log \widehat{Q}_{i+1}(t)-\frac{1}{N}\sum_{i=1}^{N} \log \widehat{Q}_i(t)\right)\\ \label{e:QQ-hatQQ}
     = & \frac{1}{4N}\sum_{i=1}^{N-1}  \left[\log Q_i(t) \log Q_{i+1}(t) - \log \widehat{Q}_i(t) \log \widehat{Q}_{i+1}(t)\right]\\  \label{e:QQbar-hatQQ:1}
     &-\frac{1}{4N^2} \sum_{i=1}^{N-1}  \log Q_i(t) \ \sum_{i=1}^{N} \log Q_i(t)  - \frac{1}{4N^2} \sum_{i=1}^{N-1}  \log \widehat{Q}_i(t)    \sum_{i=1}^{N}\log \widehat{Q}_i(t)\\  \label{e:QQbar-hatQQ:2}
     &-\frac{1}{4N^2}\sum_{i=1}^{N-1} \log Q_{i+1}(t) \sum_{i=1}^{N} \log Q_i(t)  - \frac{1}{4N^2}\sum_{i=1}^{N-1} \log \widehat{Q}_{i+1}(t) \sum_{i=1}^{N}\log \widehat{Q}_i(t)\\  \label{e:QQbar-barhatQQ}
     &+ \frac{1}{4} \left (  \frac{1}{N} \sum_{i=1}^{N}\log Q_i(t)
     \right)^2 - \frac{1}{4}\left(  \frac{1}{N}  \sum_{i=1}^{N}\log \widehat{Q}_i(t)\right)^2.
 \end{align}
For the sake of simplicity, we drop coefficient $\frac{1}{4}$ from now on. Recalling \eqref{e:logQ} and \eqref{e:Qhat} 
the summation appearing in \eqref{e:QQ-hatQQ} can be rewritten as
\begin{align}
\nonumber
\frac{1}{N}\sum_{i=1}^{N-1}  
\left[2 H(t)\left(\log g_i + \log g_{i+1}\right) \right. 
&\left.- 2X_{i+1}(t) \log g_i 
-   2X_{i}(t) \log g_{i+1}   +   H^2(t) -  X_i(t)X_{i+1}(t)\right]\\ 
\label{e:(G-X)g}
    =&\frac{2}{N}\sum_{i=1}^{N-1}  \left (  H(t) - X_{i+1}(t) \right) \log g_i \\ \label{e:(X-G)g}
    &+ \frac{2}{N}\sum_{i=1}^{N-1}  \left (  H(t) - X_i(t) \right) \log g_{i+1}\\  \label{e:G2-X2}
    &+ \frac{1}{N}\sum_{i=1}^{N-1}\left[H^2(t)-  X_i(t)X_{i+1}(t)\right].
\end{align}
We now explore decay rate of second moment of \eqref{e:(G-X)g}.
Since for each $t$, the summands $\left (  H(t) - X_{i+1}(t) \right) \log g_i$ 
form a centered stationary process over index $i$,
\begin{align}
\nonumber
  &  \mathbb{E} \left( \frac{1}{N}\sum_{i=1}^{N-1}   \left (  H(t) - X_{i+1}(t) \right) \log g_i \right)^2 \\ \nonumber
 =& \frac{1}{N^2} \sum_{|h|\leq N-2} \mathrm{Cov}\left(\left (  H(t) - X_2(t) \right) \log g_1  ,  \left (  H(t) - X_{2+h}(t) \right) \log g_{1+h}  \right) (N-1-h)\\ \label{e:h=0}
 =& \frac{N-1}{N^2}\mathbb{E}\left[\log g_1\right]^2
 \mathbb{E}\left[ H(t) - X_{2}(t)\right]^2
 \\ \label{e:h:not:0}
 &+ \frac{2(N-1)}{N^2} \sum_{1 \leq |h|\leq N-2} \left(1- \frac{h}{N-1}\right) \mathbb{E}\left( \log g_{1}\log g_{1+h}\right) \mathbb{E}^2\left[  H(t) - X_2(t)  \right],
\end{align}
where we utilize the fact that the processes $\{g_i\}$ and $X_k(\cdot)$, $k=1,2,\ldots$ are independent. Applying   Lemma \ref{lem:2nd:moment} 
to \eqref{e:h=0} and \eqref{e:h:not:0} leads to
\begin{align}  \label{e:XX:1}
    \underset{t \in [\alpha,1]}{\sup} \mathbb{E} \left( \frac{1}{N}\sum_{i=1}^{N-1}   \left (  H(t) - X_{i+1}(t) \right) \log g_i \right)^2  = O\left(\frac{\Delta}{N}\right). 
\end{align}
Likewise, for \eqref{e:(X-G)g} we obtain
\begin{align}  \label{e:XX:2}
    \underset{t \in [\alpha,1]}{\sup} \mathbb{E} \left( \frac{1}{N}\sum_{i=1}^{N-1}   \left (  H(t) - X_{i}(t) \right) \log g_{i+1} \right)^2  = O\left(\frac{\Delta}{N}\right). 
\end{align}
We now turn to \eqref{e:G2-X2}. Observe that
\begin{align*}
    &\frac{1}{N}\sum_{i=1}^{N-1}\left[ H^2(t) -  X_i(t)X_{i+1}(t)\right] \\
    =&  \frac{1}{N}\sum_{i=1}^{N-1}  H(t) \left( H(t) - X_i(t) \right) +  \frac{1}{N}\sum_{i=1}^{N-1}  X_i(t) \left( H(t) - X_{i+1}(t) \right)
\end{align*}
The first summation $\frac{1}{N}\sum_{i=1}^{N-1}  H(t) \left( H(t) - X_i(t) \right)$ can be treated similarly to \eqref{e:(X-G)g}. The  summands appearing in the second term  $\frac{1}{N}\sum_{i=1}^{N-1}  X_i(t) \left( H(t) - X_{i+1}(t) \right)$ are identical and 2-independent. This fact gives
\begin{align*}
    &\mathbb{E}\left(\frac{1}{N}\sum_{i=1}^{N-1}  X_i(t) \left( H(t) - X_{i+1}(t) \right) \right)^2\\
    =& \frac{N-1}{N^2} \mathrm{Var}\left[ X_1(t) \left( H(t) - X_{2}(t) \right)\right]\\
    & + \frac{N-2}{N^2} \mathrm{Cov}\left( X_1(t) \left( H(t) - X_{2}(t) \right) ,
     X_2(t) \left( H(t) - X_{3}(t) \right)
    \right) \\
    = & \frac{N-1}{N^2} \mathbb{E}\left[ X_1(t)\right]^2\mathbb{E}\left[ H(t) - X_{2}(t)\right]^2-\frac{N-1}{N^2}\mathbb{E}^2\left[X_1(t)\right]\mathbb{E}^2\left[ H(t) - X_{2}(t)\right]\\
    &+\frac{N-2}{N^2}
    \mathbb{E}\left[X_1(t)\right]
    \mathbb{E}\left[ H(t) - X_{3}(t)\right]
    \mathbb{E}\left[X_2(t) \left( H(t) - X_{2}(t) \right)\right]
    \\
    &-
    \frac{N-2}{N^2}
    \mathbb{E}\left[X_1(t)\right]
    \mathbb{E}\left[ H(t) - X_{3}(t)\right]
    \mathbb{E}\left[  H(t) - X_{2}(t) \right]
    \mathbb{E}\left[X_2(t) \right].
\end{align*}
Applying Lemma \ref{lem:2nd:moment}, 
we obtain
\begin{align} \label{e:XX:3}
    \underset{t \in [\alpha,1]}{\sup} &\mathbb{E}\left(\frac{1}{N}\sum_{i=1}^{N-1}  X_i(t) \left( H(t) - X_{i+1}(t) \right) \right)^2
    = O \left( \frac{\Delta}{N}\right).
\end{align}
Combining \eqref{e:XX:1}, \eqref{e:XX:2} and \eqref{e:XX:3}, we conclude that \eqref{e:QQ-hatQQ} satisfies
\begin{align}
\underset{t \in [\alpha,1]}{\sup} &\mathbb{E}\left(\frac{1}{N}\sum_{i=1}^{N-1}  \left[\log Q_i(t) \log Q_{i+1}(t) - \log \widehat{Q}_i(t) \log \widehat{Q}_{i+1}(t)\right]
\right)^2  = O \left( \frac{\Delta}{N}\right).
\end{align}
The other expressions \eqref{e:QQbar-hatQQ:1}, \eqref{e:QQbar-hatQQ:2} and \eqref {e:QQbar-barhatQQ}  appearing in the decomposition of $    \gamma_{1,N}- \Gamma_{1,N}(t) $ can be treated similarly. This gives
\begin{align}\label{e:Gamm1}
    \underset{t \in [\alpha,1]}{\sup} &\mathbb{E} \left \vert     \gamma_{1,N}- \Gamma_{1,N}(t)  \right \vert^2 =  O \left( \frac{\Delta}{N}\right).
\end{align}
Following the lines above we obtain
\begin{align}\label{e:Gamm0}
    \underset{t \in [\alpha,1]}{\sup} &\mathbb{E} \left \vert     \gamma_{0,N}- \Gamma_{0,N}(t)  \right \vert^2 =  O \left( \frac{\Delta}{N}\right).
\end{align}
The results \eqref{e:Gamm1}   and \eqref{e:Gamm1} complete the proof.

\rightline{\QED}

The following Corollary is a result of Proposition \ref{prop:gam:kapp}  and Proposition \ref{prop:Gam:gam}, by setting $t=1$.
\begin{corollary}\label{cor:Gam:kapp}
Assume the stochastic volatility model defined by \eqref{e:R1}--\eqref{e:g:AR1}  satisfying conditions \ref{itm:sig}--\ref{itm:err4} and set $t=1$ in \eqref{e:c0hQ} and \eqref{e:c1hQ}. Then, regardless of interplay between $N$ and $\Delta$,
    \begin{align}\label{e:Gam:kapp}
   N^{\frac{1}{2}} \left(
        \begin{array}{c}
            \Gamma_{0,N}(1) -\kappa_0 \\
            \Gamma_{1,N}(1) - \kappa_1
        \end{array}
        \right) \overset{\mathcal{L}aw}{\longrightarrow} \mathcal{N} \left(
        \left(
        \begin{array}{c}
            0  \\
             0
        \end{array}
        \right), V
        \right), \quad \text{as }
 N \rightarrow \infty \text{ and } \Delta \rightarrow 0,    \end{align}
where $V$ is defined in \eqref{e:V}.
\end{corollary}

\noindent{\sc Proof of Theorem \ref{thm:chk:phi:sig}}
We apply  the  bivariate delta method to the vector $\left( \Gamma_{0,N}(1) , \Gamma_{1,N}(1)\right)^{\top}$.  To do so, define the functions
\begin{align*}
    f , g :   (0, \infty) \times \mathbb{R} \longrightarrow \mathbb{R}, \qquad
    f (x,y) = x^{-1}y, \qquad
  \quad
    g (x,y) = x - x^{-1}y^2.
\end{align*}
We now apply  delta method to the limiting result \eqref{e:Gam:kapp}. This gives
\begin{align*}
    N^{1/2}\left(f\left( \Gamma_{0,N}(1) , \Gamma_{1,N}(1)\right) - f\left( \kappa_0 , \kappa_1\right) \right) \overset{\mathcal{L}aw}{\longrightarrow} \mathcal{N} \left(0, \nu \right),
\end{align*}
that is
\begin{align*}
    N^{1/2}\left(\check{\varphi} - \varphi \right) \overset{\mathcal{L}aw}{\longrightarrow} \mathcal{N} \left(0, \nu \right),
\end{align*}
where
\begin{align*}
    \nu  =  \left(\nabla f\left( \kappa_0 , \kappa_1\right)\right)^{\top} V \nabla f\left( \kappa_0 , \kappa_1\right)
         =  \left( -\kappa_0^{-2}\kappa_1 , \kappa_0^{-1}\right) V \left( -\kappa_0^{-2}\kappa_1 , \kappa_0^{-1}\right)^{\top}.
\end{align*}
Similarly,
\begin{align*}
    N^{1/2}\left(g\left( \Gamma_{0,N}(1) , \Gamma_{1,N}(1)\right) - g\left( \kappa_0 , \kappa_1\right) \right) \overset{\mathcal{L}aw}{\longrightarrow} \mathcal{N} \left(0, \tau \right),
\end{align*}
that is
\begin{align*}
    N^{1/2}\left(\check{\sigma}^2_{\varepsilon}- \sigma^2_{\varepsilon}\right) \overset{\mathcal{L}aw}{\longrightarrow} \mathcal{N} \left(0, \tau \right),
\end{align*}
where
\begin{align*}
    \tau  = \left(\nabla g \left( \kappa_0 , \kappa_1\right)\right)^{\top} V \nabla g\left( \kappa_0 , \kappa_1\right)
         =  \left( 1+\kappa_0^{-2}\kappa_1^2 , -\kappa_0^{-1}\right) V \left( 1+\kappa_0^{-2}\kappa_1^2 , -\kappa_0^{-1}\right)^{\top}.
\end{align*}
This completes the proof.

\rightline{\QED}

\subsection{Proof of Theorem \ref{thm:bar:phi:sig}}
We begin with the following Corollary that is  a counterpart of Corollary \ref{cor:Gam:kapp}.
\begin{corollary}\label{cor:int:Gam:kapp}
Assume the stochastic volatility model defined by \eqref{e:R1}--\eqref{e:g:AR1}  satisfying conditions \ref{itm:sig}--\ref{itm:err4}. Then, for any fixed   $0 < \alpha < 1$, regardless of interplay between $N$ and $\Delta$,
    \begin{align}\label{e:int:Gam:kapp}
 N^{\frac{1}{2}}   \frac{1}{1-\alpha} \int_{\alpha}^1  \left(
        \begin{array}{c}
          \Gamma_{0,N}(t)  -\kappa_0 \\
             \Gamma_{1,N}(t)  - \kappa_1
        \end{array}
        \right) dt \overset{\mathcal{L}aw}{\longrightarrow} \mathcal{N} \left(
        \left(
        \begin{array}{c}
            0  \\
             0
        \end{array}
        \right), V
        \right),\quad \text{as }
 N \rightarrow \infty \text{ and } \Delta \rightarrow 0,
    \end{align}
where $V$ is defined in \eqref{e:V}.
\end{corollary}

\noindent{\sc Proof of Corollary \ref{cor:int:Gam:kapp}}
By Proposition \ref{prop:gam:kapp}, it is   enough to prove
\begin{align*}
    N^{\frac{1}{2}}  \int_{\alpha}^1  \left(
        \begin{array}{c}
          \Gamma_{0,N}(t)  -    \gamma_{0,N} \\
             \Gamma_{1,N}(t)  -     \gamma_{1,N}
        \end{array}
        \right) dt  = o_P(1).
\end{align*}
To do so, observe that
\begin{align}
\label{e:E:int:G-g^2}
 \mathbb{E} \left \vert \int_{\alpha}^1  \left( \Gamma_{1,N}(t)  -     \gamma_{1,N}  \right) dt \right \vert^2  \leq  &  \mathbb{E}  \int_{\alpha}^1  \left( \Gamma_{1,N}(t)  -     \gamma_{1,N}  \right)^2 dt \\ \nonumber
 \leq & \underset{t \in [\alpha,1]}{\sup} \mathbb{E}   \left \vert     \gamma_{1,N}- \Gamma_{1,N}(t)  \right \vert^2 \\ \label{e:Delt/N}
 = & O \left( \frac{\Delta}{N}\right),
\end{align}
where \eqref{e:Delt/N} is a result of Proposition \ref{prop:Gam:gam}.   Consequently,
\begin{align*}
  N^{\frac{1}{2}}     \int_{\alpha}^1  \left( \Gamma_{1,N}(t)  -     \gamma_{1,N}  \right) dt  =  O_P(\Delta).
\end{align*}
Similarly, we have
\begin{align*}
  N^{\frac{1}{2}}     \int_{\alpha}^1  \left( \Gamma_{0,N}(t)  -     \gamma_{0,N}  \right) dt  =  O_P(\Delta).
\end{align*}
This completes the proof.

\rightline{\QED}

\noindent{\sc Proof of Theorem \ref{thm:bar:phi:sig}}
 The proof follows the lines of Proof of Theorem \ref{thm:chk:phi:sig};  cf. \eqref{e:int:Gam:kapp} and \eqref{e:Gam:kapp}.
\rightline{\QED}

\subsection{Proof of Theorem   \ref{thm:hat:phi:sig}}

We commence the proof of Theorem   \ref{thm:hat:phi:sig} with Proposition \ref{prop:unif:Gam:gam} and Lemma \ref{lem:Gamm^-1} below.

\begin{proposition}\label{prop:unif:Gam:gam}
Assume the stochastic volatility model defined by \eqref{e:R1}--\eqref{e:g:AR1} satisfying conditions \ref{itm:sig}--\ref{itm:int:sig} and recall \eqref{e:c0hQ} and \eqref{e:c1hQ}. Then,  for any fixed   $0 < \alpha < 1$,
\begin{align*}
     \mathbb{E} \left( \underset{t \in [\alpha,1]}{\sup} \left \vert     \gamma_{0,N}- \Gamma_{0,N}(t)  \right \vert^2 \right) +  \mathbb{E} \left( \underset{t \in [\alpha,1]}{\sup}  \left \vert     \gamma_{1,N}- \Gamma_{1,N}(t)  \right \vert^2 \right)=  O \left( \Delta \right).
\end{align*}
\end{proposition}
\noindent{\sc Proof of Proposition \ref{prop:unif:Gam:gam}}
We  prove that
\begin{align*}
      \mathbb{E} \left( \underset{t \in [\alpha,1]}{\sup}  \left \vert     \gamma_{1,N}- \Gamma_{1,N}(t)  \right \vert^2 \right)= O \left( \Delta \right).
\end{align*}
The argument for lag--0 autocovariance is  similar. Following the lines \eqref{e:QQ-hatQQ}--\eqref{e:QQbar-barhatQQ}, the distance $  \gamma_{1,N} - \Gamma_{1,N}(t) $ admits the decomposition
\begin{align}
 \nonumber
         \gamma_{1,N}- \Gamma_{1,N}(t)  =
     &- \frac{1}{4N}\sum_{i=1}^{N-1} \left(\log \widehat{Q}_i(t)-\frac{1}{N}\sum_{i=1}^{N} \log \widehat{Q}_i(t)\right)\left(\log \widehat{Q}_{i+1}(t)-\frac{1}{N}\sum_{i=1}^{N} \log \widehat{Q}_i(t)\right)\\ \label{e:unif:QQ-hatQQ}
     = & \frac{1}{4N}\sum_{i=1}^{N-1}  \left[\log Q_i(t) \log Q_{i+1}(t) - \log \widehat{Q}_i(t) \log \widehat{Q}_{i+1}(t)\right]\\  \label{e:unif:QQbar-hatQQ:1}
     &-\frac{1}{4N^2}  \sum_{i=1}^{N-1}  \log Q_i(t) \ \sum_{i=1}^{N} \log Q_i(t)  - \frac{1}{4N^2} \sum_{i=1}^{N-1}  \log \widehat{Q}_i(t)    \sum_{i=1}^{N}\log \widehat{Q}_i(t)\\  \label{e:unif:QQbar-hatQQ:2}
     &-\frac{1}{4N^2}\sum_{i=1}^{N-1} \log Q_{i+1}(t) \sum_{i=1}^{N} \log Q_i(t)  - \frac{1}{4N^2}\sum_{i=1}^{N-1} \log \widehat{Q}_{i+1}(t) \sum_{i=1}^{N}\log \widehat{Q}_i(t)\\  \label{e:unif:QQbar-barhatQQ}
     &+ \frac{1}{4} \left (  \frac{1}{N} \sum_{i=1}^{N}\log Q_i(t)
     \right)^2- \frac{1}{4}\left( \frac{1}{N}  \sum_{i=1}^{N}\log \widehat{Q}_i(t)\right)^2.
 \end{align}
For the sake of simplicity, we drop coefficient $\frac{1}{4}$ from now on.
The summation appearing in \eqref{e:unif:QQ-hatQQ} can be rewritten as
\begin{align}
\nonumber
    &\frac{1}{N}\sum_{i=1}^{N-1}  \left[2 H(t)\left(\log g_i + \log g_{i+1}\right) - 2X_{i+1}(t) \log g_i -   2X_{i}(t) \log g_{i+1}   +   H^2(t) -  X_i(t)X_{i+1}(t)\right]\\ \label{e:unif:(G-X)g}
    =&\frac{2}{N}\sum_{i=1}^{N-1}  \left (  H(t) - X_{i+1}(t) \right) \log g_i \\ \label{e:unif:(X-G)g}
    &+ \frac{2}{N}\sum_{i=1}^{N-1}  \left (  H(t) - X_i(t) \right) \log g_{i+1}\\  \label{e:unif:G2-X2}
    &+ \frac{1}{N}\sum_{i=1}^{N-1}\left[\left( H(t)\right)^2 -  X_i(t)X_{i+1}(t)\right].
\end{align}
We now explore the uniform decay rate of the second moment of \eqref{e:unif:(G-X)g}:
\begin{align*}
  \mathbb{E} \underset{t \in [\alpha,1]}{\sup}  \left \vert \frac{1}{N}\sum_{i=1}^{N-1}   \left (  H(t) - X_{i+1}(t) \right ) \log g_i \right\vert^2
  \leq &  \mathbb{E}  \left ( \frac{1}{N}\sum_{i=1}^{N-1} \underset{t \in [\alpha,1]}{\sup}   \left \vert  H(t) - X_{i+1}(t)\right \vert \left \vert \log g_i\right\vert  \right)^2 \\
  =: & \mathbb{E}  \left ( \frac{1}{N}\sum_{i=1}^{N-1} S_{i+1} \left \vert \log g_i\right\vert  \right)^2.
\end{align*}
Since the sequence $\left\{S_{i+1} \left \vert \log g_i\right\vert\right\}_i$ is stationary,
\begin{align*}
    \mathbb{E}  \left ( \frac{1}{N}\sum_{i=1}^{N-1} S_{i+1} \left \vert \log g_i\right\vert  \right)^2 = &  \frac{1}{N^2} \sum_{|h|\leq N-2} \mathbb{E}  \left (S_{2} \left \vert \log g_1\right\vert   S_{2+h} \left \vert \log g_{1+h}\right\vert\right) (N-1-h)\\
    = & \frac{N-1}{N^2}  \mathbb{E}  \left (  S_{2} \left \vert \log g_1\right\vert  \right)^2\\
    & + \frac{1}{N^2} \sum_{1 \leq |h|\leq N-2} \mathbb{E}  \left (S_{2} \left \vert \log g_1\right\vert   S_{2+h} \left \vert \log g_{1+h}\right\vert\right) (N-1-h).
\end{align*}
The independence of the $S_{i}$, $i=1,2,\ldots, N$, and $\{\log g_i \}$ gives
\begin{align*}
    \mathbb{E}  \left ( \frac{1}{N}\sum_{i=1}^{N-1} S_{i+1} \left \vert \log g_i\right\vert  \right)^2
    = & \frac{N-1}{N^2} \mathbb{E}  \left (  S_{2}  \right)^2\mathbb{E}  \left (   \left \vert \log g_1\right\vert  \right)^2\\
    & + \frac{N-1}{N^2} \sum_{1 \leq |h|\leq N-2} \left(1- \frac{h}{N-1}\right) \mathbb{E}  \left (S_{2}     \right) \mathbb{E}  \left (   S_{2+h} \right)
    \mathbb{E}  \left ( \left \vert \log g_1\right\vert   \left \vert \log g_{1+h}\right\vert\right)\\
    = & O \left ( \frac{\Delta}{N}\right ) + O \left ( \frac{\Delta}{N}\right )\sum_{1 \leq |h|\leq N-2} \left(1- \frac{h}{N-1}\right)
    \mathbb{E}  \left ( \left \vert \log g_1\right\vert   \left \vert \log g_{1+h}\right\vert\right)\\= &  O \left ( \Delta\right ).
\end{align*}
Consequently, \eqref{e:unif:(G-X)g} satisfies
\begin{align}\label{e:unif:XX:1}
    \mathbb{E}   \underset{t \in [\alpha,1]}{\sup}\left( \frac{1}{N}\sum_{i=1}^{N-1}   \left (  H(t) - X_{i+1}(t) \right) \log g_{i} \right)^2  = O\left(\Delta\right).
\end{align}
Likewise, for \eqref{e:unif:(X-G)g} we obtain
\begin{align}  \label{e:unif:XX:2}
   \mathbb{E}   \underset{t \in [\alpha,1]}{\sup}\left( \frac{1}{N}\sum_{i=1}^{N-1}   \left (  H(t) - X_{i}(t) \right) \log g_{i+1} \right)^2  = O\left(\Delta\right). 
\end{align}
We now turn to \eqref{e:unif:G2-X2}. Observe that
\begin{align}
\nonumber
    &\frac{1}{N}\sum_{i=1}^{N-1}\left[ H^2(t) -  X_i(t)X_{i+1}(t)\right] \\ \label{e:unif:GG-XX}
    =&  \frac{1}{N}\sum_{i=1}^{N-1}  H(t) \left( H(t) - X_i(t) \right) +  \frac{1}{N}\sum_{i=1}^{N-1}  X_i(t) \left( H(t) - X_{i+1}(t) \right)
\end{align}
The first summand $\frac{1}{N}\sum_{i=1}^{N-1}  H(t) \left( H(t) - X_i(t) \right)$ satisfies
\begin{align*}
   & \mathbb{E}  \underset{t \in [\alpha,1]}{\sup}\left(\frac{1}{N}\sum_{i=1}^{N-1}  H(t) \left( H(t) - X_i(t) \right) \right)^2 \\ \leq &     \mathbb{E}  \left(\frac{1}{N}\sum_{i=1}^{N-1} \underset{t \in [\alpha,1]}{\sup} \left \vert  H(t) \right \vert  \underset{t \in [\alpha,1]}{\sup}  \left\vert H(t) - X_i(t) \right\vert\right)^2\\
   \leq & \frac{1}{N^2}
   \mathbb{E}  \left(\sum_{i=1}^{N-1} \underset{t \in [\alpha,1]}{\sup} \left \vert  H(t) \right \vert^2
   \sum_{i=1}^{N-1} \underset{t \in [\alpha,1]}{\sup}  \left\vert H(t) - X_i(t) \right\vert^2\right)\\
   = & \frac{1}{N^2} \sum_{i=1}^{N-1} \underset{t \in [\alpha,1]}{\sup} \left \vert  H(t) \right \vert^2
   \mathbb{E}  \left(
   \sum_{i=1}^{N-1} \underset{t \in [\alpha,1]}{\sup}  \left\vert H(t) - X_i(t) \right\vert^2\right)
\end{align*}
The boundedness of $\left \vert  H(t) \right \vert$ together with Corollary \ref{cor:unif:log:2nd} imply that
\begin{align}\label{e:unif:G(G-X)}
    \mathbb{E}  \underset{t \in [\alpha,1]}{\sup}\left(\frac{1}{N}\sum_{i=1}^{N-1}  H(t) \left( H(t) - X_i(t) \right) \right)^2 = O\left ( \Delta \right ).
\end{align}
The second summand appearing in \eqref{e:unif:GG-XX} satisfies
\begin{align*}
 &  \mathbb{E}  \underset{t \in [\alpha,1]}{\sup}\left( \frac{1}{N}\sum_{i=1}^{N-1}  X_i(t) \left( H(t) - X_{i+1}(t) \right)  \right)^2\\
   \leq & \frac{1}{N^2} \mathbb{E} \left( \sum_{i=1}^{N-1}  \underset{t \in [\alpha,1]}{\sup}  \left \vert X_i(t)\right \vert \underset{t \in [\alpha,1]}{\sup} \left\vert H(t) - X_{i+1}(t) \right\vert  \right)^2\\
   \leq & \frac{1}{N^2} \mathbb{E} \left( \sum_{i=1}^{N-1}  \underset{t \in [\alpha,1]}{\sup}  \left \vert X_i(t)\right \vert^2 \sum_{i=1}^{N-1} \underset{t \in [\alpha,1]}{\sup} \left\vert H(t) - X_{i+1}(t) \right\vert^2  \right)\\
   = & \frac{1}{N^2} \mathbb{E}  \sum_{i=1}^{N-1}  \underset{t \in [\alpha,1]}{\sup}  \left \vert X_i(t)\right \vert^2 \mathbb{E} \sum_{i=1}^{N-1} \underset{t \in [\alpha,1]}{\sup} \left\vert H(t) - X_{i+1}(t) \right\vert^2.
\end{align*}
Applying Corollary \eqref{cor:unif:log:2nd} completes the proof of
\begin{align}\label{e:unif:XX:3}
    \mathbb{E}  \underset{t \in [\alpha,1]}{\sup}\left( \frac{1}{N}\sum_{i=1}^{N-1}  X_i(t) \left( H(t) - X_{i+1}(t) \right)  \right)^2 = O \left ( \Delta\right ).
\end{align}
Combining \eqref{e:unif:XX:1} , \eqref{e:unif:XX:2} , \eqref{e:unif:G(G-X)} and \eqref{e:unif:XX:3}, we conclude that \eqref{e:unif:QQ-hatQQ} satisfies
\begin{align}
 &\mathbb{E}\underset{t \in [\alpha,1]}{\sup} \left(\frac{1}{N}\sum_{i=1}^{N-1}  \left[\log Q_i(t) \log Q_{i+1}(t) - \log \widehat{Q}_i(t) \log \widehat{Q}_{i+1}(t)\right]
\right)^2  = O \left( \Delta\right).
\end{align}

Applying a similar arguments on the  expressions \eqref{e:unif:QQbar-hatQQ:1}, \eqref{e:unif:QQbar-hatQQ:2} and \eqref {e:unif:QQbar-barhatQQ}  appearing in the decomposition of $    \gamma_{1,N}- \Gamma_{1,N}(t) $ we obtain
\begin{align}\label{e:unif:Gamm1}
    \mathbb{E} \underset{t \in [\alpha,1]}{\sup}  \left \vert     \gamma_{1,N}- \Gamma_{1,N}(t)  \right \vert^2 =  O \left( \Delta \right).
\end{align}
A similar argument gives
\begin{align}\label{e:unif:Gamm0}
    \mathbb{E} \underset{t \in [\alpha,1]}{\sup}  \left \vert     \gamma_{0,N}- \Gamma_{0,N}(t)  \right \vert^2 =  O \left( \Delta \right).
\end{align}
This completes the proof.

\rightline{\QED}

\begin{lemma}\label{lem:Gamm^-1}  Assume the stochastic volatility model defined by \eqref{e:R1}--\eqref{e:g:AR1} and  conditions \ref{itm:sig}--\ref{itm:err4}.  Then,
    \begin{align}\label{e:Gamm^-1}
        \int_{\alpha}^1 \left \vert  \Gamma_{0,N}^{-1}(t) - \kappa_0^{-1}\right \vert^2   dt = o_P(1)
    \end{align}
\end{lemma}

\noindent{\sc Proof of Lemma \ref{lem:Gamm^-1}}
We  use an argument implicit in the proof of
Theorem 1.1 in \citetext{berkes_2006}.
First observe that according to Propositions     \ref{prop:unif:Gam:gam} and \ref{prop:gam:kapp} we obtain

\begin{align}\label{e:Gamm0:kapp0}
    \underset{t \in [\alpha,1]}{\sup}  \left \vert  \Gamma_{0,N}(t) - \kappa_0\right \vert \leq  \underset{t \in [\alpha,1]}{\sup}  \left \vert  \Gamma_{0,N}(t) -     \gamma_{0,N}\right \vert +  \left \vert    \gamma_{0,N} - \kappa_0 \right \vert = o_P(1), \; \mathrm{as} \;\Delta \rightarrow 0 \;\mathrm{and} \; N \rightarrow \infty.
\end{align}
Now define the functions $T(\cdot)$ and $T_L(\cdot)$, for $L>0$, of the following form
\begin{align*}
    T(x) = x^{-1}, \qquad x \in (0,\infty),
\end{align*}
and
\begin{align*}
 T_L(x) = \left\{
\begin{array}{cc}
       x^{-1},& \qquad x \in \left[ \frac{1}{L},L\right],\\
        0,& \qquad x \in (0,\infty) \setminus \left[ \frac{1}{L},L\right]
\end{array}
\right. , \qquad L>0.
\end{align*}
Observe that, for each $L>0$, the integral appearing in \eqref{e:Gamm^-1} admits the following decomposition
\begin{align}
\nonumber
            \int_{\alpha}^1 \left \vert  \Gamma_{0,N}^{-1}(t) - \kappa_0^{-1}\right \vert^2   dt = & \int_{\alpha}^1 \left \vert  T \left(\Gamma_{0,N}\right)(t) - T(\kappa_0)\right \vert^2   dt
            \\ \label{e:TL:T:kap} \leq & 8\int_{\alpha}^1 \left \vert  T_L(\kappa_0) - T(\kappa_0)\right \vert^2 dt\\ \label{e:TL:T:Gam}
            &+ 8\int_{\alpha}^1 \left \vert  T \left(\Gamma_{0,N}\right)(t) - T_L \left(\Gamma_{0,N}\right)(t)\right \vert^2 dt \\ \label{e:TL:Gamkap}
            & + 8\int_{\alpha}^1 \left \vert  T_L\left(\Gamma_{0,N}\right)(t) - T_L(\kappa_0)\right \vert^2 dt.
\end{align}
In order to conclude \eqref{e:Gamm^-1}, we study the  terms  appearing in the decomposition \eqref{e:TL:T:kap}--\eqref{e:TL:Gamkap} separately.
Since $\kappa_0$ is a positive fixed parameter, there always exists $L $ sufficiently large such that
\begin{align*}
  0 <  \frac{1}{L} < \frac{2}{L}< \frac{\kappa_0}{2} < \kappa_0 < \frac{3}{2}\kappa_0< 2 \kappa_0< L.
\end{align*}
In other words, there exists $L>0$ sufficiently large such that, for any $\epsilon >0$,
 \begin{align}\label{e:P:kapp0}
     \mathbb{P}\left( \kappa_0 \geq \frac{L}{2} \;\;\; \mathrm{or} \;\;\; \kappa_0 \leq \frac{2}{L}\right) = 0 \leq \epsilon.
 \end{align}
Relation \eqref{e:P:kapp0} obviously implies that there exists $L>0$ sufficiently large such that $T_L(\kappa_0) = T(\kappa_0)$. Equivalently, there is  $L$ sufficiently large such that, for any $\epsilon >0$,
\begin{align}
\nonumber
     \mathbb{P}\left(  T (  \kappa_0 ) \neq T_L (  \kappa_0 ) \right)    = 0  \leq  \epsilon.
 \end{align}
This gives there exists $L>0$ such that for any $\epsilon >0$,  any $\zeta >0$, any step size $\Delta$ and any sample size $N$
\begin{align}
\label{e:L:kapp0}
    \mathbb{P}\left (\int_{\alpha}^1 \left \vert  T_L(\kappa_0) - T(\kappa_0)\right \vert^2 dt > \zeta  \right) \leq  \mathbb{P}\left(  T (  \kappa_0 ) \neq T_L (  \kappa_0 ) \right)    = 0  \leq  \epsilon.
\end{align}
This entails \eqref{e:TL:T:kap} is convergent  to zero in probability.  We now turn to the term \eqref{e:TL:T:Gam}. Relations \eqref{e:Gamm0:kapp0} and \eqref{e:P:kapp0} together imply there exist $L>0$, obtained above, such that for any $\epsilon >0$, there exist  step size $\Delta_{\varepsilon}$ and sample size $N_{\epsilon}$ such that
 \begin{align*}
     \nonumber
     &\mathbb{P}\left( \underset{t \in [\alpha,1]}{\sup}  \left \vert  \Gamma_{0,N}(t)  \right \vert  \geq L \;\;\; \mathrm{or} \;\;\;  \underset{t \in [\alpha,1]}{\inf} \left \vert  \Gamma_{0,N}(t)  \right \vert  \leq \frac{1}{L}\right)  \\
      \nonumber
     \leq &    \mathbb{P}\left(\underset{t \in [\alpha,1]}{\sup}  \left \vert  \Gamma_{0,N}(t) - \kappa_0\right \vert \geq \frac{\kappa_0}{2} \right) \\
     \leq & \epsilon, \quad \forall N \geq N_{\epsilon}, \; \Delta \leq \Delta_{\varepsilon}.
 \end{align*}
Consequently, there exists $L>0$ such that for any $\epsilon >0$, there exist  step size $\Delta_{\varepsilon}$ and sample size $N_{\epsilon}$ such that
 \begin{align}
 \nonumber
    & \mathbb{P}\left( \exists t \in [\alpha, 1];\;\;\; T (  \Gamma_{0,N}(t)  ) \neq T_L (  \Gamma_{0,N}(t)  ) \right)
    \\  \nonumber =&  \mathbb{P}\left( \exists t \in [\alpha, 1];\;\;\; \Gamma_{0,N}(t) \notin  \left[ \frac{1}{L},L\right]\right)
    \\  \nonumber \leq & \mathbb{P}\left( \underset{t \in [\alpha,1]}{\sup}  \left \vert  \Gamma_{0,N}(t)  \right \vert  > L \;\;\; \mathrm{or} \;\;\; \underset{t \in [\alpha,1]}{\inf}  \left \vert  \Gamma_{0,N}(t)  \right \vert  < \frac{1}{L} \right)   \\ \nonumber \leq &  \epsilon, \quad \forall N \geq N_{\epsilon}, \; \Delta \leq \Delta_{\varepsilon}.
 \end{align}
This implies there exist $L>0$ such that for any $\epsilon >0$, any $\zeta >0$, there exist  step size $\Delta_{\varepsilon}$ and sample size $N_{\epsilon}$ such that
\begin{align}
\nonumber
       & \mathbb{P}\left ( \int_{\alpha}^1 \left \vert  T \left(\Gamma_{0,N}\right)(t) - T_L \left(\Gamma_{0,N}\right)(t)\right \vert^2 dt> \zeta \right) \\ \nonumber
        \leq &  \mathbb{P}\left(  \exists t \in [\alpha, 1];\;\;\; T (  \Gamma_{0,N}(t)  ) \neq T_L (  \Gamma_{0,N}(t)  ) \right)     \\ \label{e:L:Gamm0}
        \leq &   \epsilon, \quad \forall N \geq N_{\epsilon}, \; \Delta \leq \Delta_{\varepsilon}.
\end{align}
This implies  \eqref{e:TL:T:Gam} tends to zero in probability.
 By  \eqref{e:L:kapp0} and \eqref{e:L:Gamm0}, in order to conclude \eqref{e:Gamm^-1}, it is  enough to prove for the positive $L$ obtained above
    \begin{align*}
        \int_{\alpha}^1 \left \vert  T_L(\Gamma_{0,N}(t) )- T_L(\kappa_0)\right \vert^2   dt = o_P(1).
    \end{align*}
To do so, define
\begin{align*}
A_N= \left\{     \underset{t \in [\alpha,1]}{\sup}  \left \vert  \Gamma_{0,N}(t)  - \kappa_{0} \right \vert   > \frac{\kappa_0}{2} \right\}.
\end{align*}
Then we have
    \begin{align} \label{e:int:T_L:T_L}
    \nonumber
        \int_{\alpha}^1 \left \vert  T_L(\Gamma_{0,N}(t) )- T_L(\kappa_0)\right \vert^2   dt  = &
           \int_{\alpha}^1 \left \vert  T_L(\Gamma_{0,N}(t) )- T_L(\kappa_0)\right \vert^2 \mathbb{I} \{A^c_N\}  dt
           \\ \nonumber
           &+     \int_{\alpha}^1 \left \vert  T_L(\Gamma_{0,N}(t) )- T_L(\kappa_0)\right \vert^2 \mathbb{I} \{A_N\}  dt\\ \nonumber
           = & \int_{\alpha}^1 \left \vert  \Gamma_{0,N}^{-1}(t) - \kappa_0^{-1}\right \vert^2 \mathbb{I} \{A^c_N\}  dt
           \\ \nonumber
           &+     \int_{\alpha}^1 \left \vert  T_L(\Gamma_{0,N}(t) )- \kappa_0^{-1}\right \vert^2 \mathbb{I} \{A_N\}  dt
           \\ \nonumber
           \leq & 2^4\int_{\alpha}^1 \frac{\left \vert  \Gamma_{0,N}(t) - \kappa_0\right \vert^2}{\kappa_0^4}  \mathbb{I} \{A^c_N\}  dt
           \\ \nonumber
           &+     \int_{\alpha}^1 \left \vert  T_L(\Gamma_{0,N}(t) )- \kappa_0^{-1}\right \vert^2 \mathbb{I} \{A_N\}  dt\\ \nonumber
           \leq & \frac{2^4}{\kappa_0^4}
           \underset{t \in [\alpha,1]}{\sup} \left \vert  \Gamma_{0,N}(t) - \kappa_0\right \vert^2+4L^2 \mathbb{I} \{A_N\} \\
           = & o_P(1)+  o_P(1), \; \mathrm{as} \;\Delta \rightarrow 0 \;\mathrm{and} \; N \rightarrow \infty,
    \end{align}
where \eqref{e:int:T_L:T_L} is a conclusion of \eqref{e:Gamm0:kapp0}. This completes the proof.

\rightline{\QED}

\noindent{\sc Proof of Theorem \ref{thm:hat:phi:sig}}
We utilize   Propositions  \ref{prop:gam:kapp} and \ref{prop:Gam:gam} and Lemma \ref{lem:Gamm^-1}. Observe that
\begin{align}
     \nonumber
     \hat{\varphi} - \varphi  =&   \frac{1}{1-\alpha}\int_{\alpha}^1\hat{\varphi}(t)dt - \varphi  \\
     \nonumber
     = & \frac{1}{1-\alpha} \int_{\alpha}^1 \Gamma_{0,N}^{-1}(t)  \Gamma_{1,N}(t) dt  -\kappa_0^{-1}\kappa_1  \\
     \nonumber
     = & \frac{1}{1-\alpha}  \int_{\alpha}^1 \kappa_0^{-1}\left( \Gamma_{1,N}(t) - \kappa_1\right) dt
     + \frac{1}{1-\alpha} \int_{\alpha}^1 \left( \Gamma_{0,N}^{-1}(t) - \kappa_0^{-1}\right) \Gamma_{1,N}(t) dt
     \\ \nonumber 
     =: & A+ D
\end{align}
According to Corollary \ref{cor:int:Gam:kapp},  the first term $A$ converges in distribution with rate $N^{\frac{1}{2}}$. More precisely, $N^{\frac{1}{2}} A $
is convergent in law, as long as $N$ tends to infinity and $\Delta$ tends to zero, regardless of interplay between $N$ and $\Delta$. The second term $D$ is dominated by
\begin{align}
\nonumber
    \vert D \vert \leq &   \frac{1}{1-\alpha} \left( \int_{\alpha}^1  \left \vert  \Gamma_{1,N}(t)  \right \vert^2  dt \right)^{\frac{1}{2}}
      \left ( \int_{\alpha}^1 \left \vert  \Gamma_{0,N}^{-1}(t) - \kappa_0^{-1}\right \vert^2   dt \right)^{\frac{1}{2}}   \\  \label{e:BC}
     =: & BC
\end{align}

In order to obtain consistency of  the term $B$ observe that $B^2$ reads
\begin{align*}
    B^2 = \int_{\alpha}^1  \left \vert  \Gamma_{1,N}(t)  \  \right \vert^2  dt  \leq  &
    16  \int_{\alpha}^1  \left \vert  \Gamma_{1,N}(t)  -     \gamma_{1,N} \right \vert^2  dt + 16  \int_{\alpha}^1  \left \vert       \gamma_{1,N} - \kappa_1 \right \vert^2  dt + 16  \int_{\alpha}^1  \left \vert    \kappa_1 \right \vert^2  dt \\
   \leq &  16 \int_{\alpha}^1  \left \vert  \Gamma_{1,N}(t)  -     \gamma_{1,N} \right \vert^2  dt
   + 16 \left \vert       \gamma_{1,N} - \kappa_1 \right \vert^2
   + \left \vert   \kappa_1 \right \vert^2
    \\
    =: & B_1 + B_2 + B_3
\end{align*}
According to \eqref{e:E:int:G-g^2} and \eqref{e:Delt/N},  we have   $B_1 = O_P(\frac{\Delta}{N})$. According to  Proposition \ref{prop:gam:kapp} the second term $B_2$ converges in distribution with rate $N$. That is $NB_2$
is convergent in law. The last term $B_3$ is constant. These altogether imply the term $B^2$ is convergent in distribution with rate $N$. That is  $ N^{\frac{1}{2}} B $
is convergent in law, as long as $N$ tends to infinity and $\Delta$ tends to zero, regardless of interplay between $N$ and $\Delta$ .
We now turn to the term $C$ appearing in \eqref{e:BC}.
Lemma \ref{lem:Gamm^-1} implies $C = o_P(1)$. Consequently, $ N^{\frac{1}{2}}(BC) = o_P(1) $. Hence, $ N^{\frac{1}{2}}D = o_P(1) $. Altogether, we have $ N^{\frac{1}{2}}(A+D) $ is convergent in law. Moreover, the limiting distribution of $ N^{\frac{1}{2}}(A+D) $ is the same as the limiting distribution of $ N^{\frac{1}{2}}A $ obtained in Corollary \ref{cor:int:Gam:kapp}. This gives \eqref{e:lim:hatphi}.
 Consistency of $\left( \hat{\sigma}^2_{\varepsilon}- \sigma^2_{\varepsilon} \right)$ is established by a similar argument. Observe that
\begin{align}
\nonumber
\hat{\sigma}^2_\eg - \sigma^2_\eg =& \frac{1}{1- \alpha} \int_{\alpha}^1 \hat{\sigma}^2_\eg (t) dt - \sigma^2_\eg \\ \nonumber
= & \frac{1}{1- \alpha} \int_{\alpha}^1 \left( \Gamma_{0,N}(t)  - \hat{\varphi} \Gamma_{1,N}(t)\right) dt - \left(\kappa_0 - \varphi \kappa_1 \right) \\ \nonumber
=& - \frac{\left(\hat{\varphi} - \varphi \right)}{1- \alpha} \int_{\alpha}^1 \Gamma_{1,N}(t) dt
\\ \nonumber
 & +\frac{1}{1- \alpha} \int_{\alpha}^1 \left( \Gamma_{0,N}(t)  - \kappa_0 \right)dt - \frac{ \varphi }{1- \alpha} \int_{\alpha}^1 \left( \Gamma_{1,N}(t) -\kappa_1 \right)  dt \\ \nonumber
 =:& E \\ \nonumber
 &+ F
\end{align}
The limiting result \eqref{e:lim:hatphi} and Corollary \ref{cor:int:Gam:kapp} imply that $N^{\frac{1}{2}}E$ is  $o_P(1)$. Corollary \ref{cor:int:Gam:kapp} implies  $N^{\frac{1}{2}}F$ is convergent in law, as long as $N$ tends to infinity and $\Delta$ tends to zero, regardless of interplay between $N$ and $\Delta$. Moreover, $ N^{\frac{1}{2}}\left( \hat{\sigma}^2_{\varepsilon}- \sigma^2_{\varepsilon} \right)$ admits the same limiting law as  of $N^{\frac{1}{2}}F$. This completes the proof.

\rightline{\QED}

\section{Proofs of the results of Section \ref{s:ext}}\label{s:proofs:ARp}

\subsection{Proof of Theorem \ref{t:p}}

\noindent{\sc Proof of Theorem \ref{t:p}}
The left hand side of \eqref{e:ARpcons:Hhat-H}   is upper bounded by
\begin{align*}
    \mathbb{E} \underset{t \in [\alpha,1]}{\sup} \left\vert  \widehat{H}(t)-  \widetilde{H}(t)\right\vert  +  \mathbb{E} \underset{t \in [\alpha,1]}{\sup} \left\vert   \widetilde{H}(t) - H \right\vert
\end{align*}
Recall the result of Proposition \ref{prop:volatility}:
  \begin{align*}
   \mathbb{E}  \underset{t \in [\alpha,1]}{\sup} \left\vert  \widehat{H}(t) -   \widetilde{H}(t) \right\vert  = O\left( \Delta^{\frac{1}{2}} \right),
   \end{align*}
where   $\widetilde{H}(t)$ and $\widehat{H}(t)$  are defined in \eqref{e:tild:H}  and \eqref{e:hatH}, respectively. So, in order to verify \eqref{e:ARpcons:Hhat-H} its enough to prove
    \begin{align*}
         \mathbb{E} \underset{t \in [0,1]}{\sup }\left \vert \widetilde{H}(t) -  H(t)\right \vert^2  = O(N^{-1}).
    \end{align*}
To do so, we present an argument similar to Proposition \ref{prop:tild:G}. In details

\begin{align}
 \nonumber
N\mathbb{E} \underset{t \in (0,1]}{\sup }\left \vert \widetilde{H}(t) -  H(t)\right \vert^2 =&
   N  \mathbb{E} \underset{t \in (0,1]}{\sup } \left \vert \frac{1}{N}\sum_{i=1}^{N} \log Q_i(t) - H(t)\right \vert^2\\  \label{e:ARp:logQ/G}
   =& N  \mathbb{E} \underset{t \in (0,1]}{\sup } \left \vert  \frac{1}{N}\sum_{i=1}^{N} \log \left( g^2_i \int_0^t\sigma^2(u)du \right) - \log  \int_0^t\sigma^2(u)du\right \vert^2
   \\ \nonumber 
   =& N  \mathbb{E} \left \vert  \frac{2}{N}\sum_{i=1}^{N} \log g_i \right \vert^2
   \\ \nonumber
   \leq & 4 \sum_{\vert h \vert \leq N-1} \left \vert \mathrm{Cov}\left(\log g_{\vert h \vert +1}, \log g_1 \right) \right \vert \left( 1- \frac{\vert h \vert}{N}\right)
    \\   \label{eq:ARp:covg}
   < & \infty,
   \end{align}
   where the  equality \eqref{e:ARp:logQ/G} follows  from  \eqref{e:QV} and \eqref{eq:ARp:covg} is a consequence of  the condition \eqref{e:causal} which implies that the autocovariances decay exponentially fast with  lag $h$.
   This completes the proof of \eqref{e:ARpcons:Hhat-H}. The limiting result \eqref{e:ARp:cons:Ghat-G} is a consequence of \eqref{e:ARpcons:Hhat-H}. See proof of Theorem \ref{thm:hat:H}

\rightline{\QED}

\subsection{Proof of Theorems \ref{t:pA} and \ref{t:pB}}

\begin{proposition}\label{prop:ARp:gam:kapp}
Assume the stochastic volatility model defined by \eqref{e:AR(p)}--\eqref{e:g:ARp} satisfying conditions \ref{itm:sig}--\ref{itm:err4}, except that we replace condition \ref{itm:phi} by \eqref{e:causal}, and recall  \eqref{e:ARp:gam_h}. Then
    \begin{align*}
   N^{\frac{1}{2}} \left(
        \begin{array}{c}
            \gamma_{0,N} -\kappa_0 \\
               \vdots  \\
                 \gamma_{p,N} - \kappa_p
        \end{array}
        \right) \overset{\mathcal{L}aw}{\longrightarrow} \mathcal{N} \left(
        \left(
        \begin{array}{c}
            0  \\
            \vdots \\
             0
        \end{array}
        \right), V
        \right),
    \end{align*}
where $V$ is a $(p+1) \times (p+1)$ matrix with entries
\begin{align}\label{e:ARp:V}
    V_{k,l} =  (\eta - 3) \kappa_k \kappa_l + \sum_{h= - \infty}^{\infty} \left( \kappa_h \kappa_{h-k+l}+\kappa_{h-k}\kappa_{h+l}\right), \quad k , l = 0,\ldots, p.
\end{align}
\end{proposition}

\noindent{\sc Proof of Proposition \ref{prop:ARp:gam:kapp}}
Refer to Proposition 7.3.4 in \citetext{brockwell:davis:1991}.

\rightline{\QED}

\begin{proposition}\label{prop:ARp:Gam:gam}
Assume the stochastic volatility model defined by \eqref{e:AR(p)}--\eqref{e:g:ARp} satisfying conditions \ref{itm:sig}--\ref{itm:int:sig}, except that we replace condition \ref{itm:phi} by \eqref{e:causal}, and recall \eqref{e:ARp:Gamm}. Then, for any fixed   $0 < \alpha < 1$,
\begin{align*}
    \underset{t \in [\alpha,1]}{\sup} \mathbb{E}   \left \vert     \gamma_{0,N}- \Gamma_{0,N}(t)  \right \vert^2  + \ldots + \underset{t \in [\alpha,1]}{\sup} \mathbb{E}   \left \vert     \gamma_{p,N}- \Gamma_{p,N}(t)  \right \vert^2 =  O \left( \frac{\Delta}{N}\right),
\end{align*}
and
\begin{align*}
     \mathbb{E} \left( \underset{t \in [\alpha,1]}{\sup} \left \vert     \gamma_{0,N}- \Gamma_{0,N}(t)  \right \vert^2 \right) + \dots + \mathbb{E} \left( \underset{t \in [\alpha,1]}{\sup}  \left \vert     \gamma_{p,N}- \Gamma_{p,N}(t)  \right \vert^2 \right)=  O \left( \Delta \right).
\end{align*}
\end{proposition}

\noindent{\sc Proof of Proposition \ref{prop:ARp:Gam:gam}}
Follow the lines of proofs of Propositions \ref{prop:Gam:gam} and \ref{prop:unif:Gam:gam}.

\rightline{\QED}

\begin{corollary}\label{cor:ARp:int:Gam:kapp}
Assume the stochastic volatility model defined by \eqref{e:AR(p)}--\eqref{e:g:ARp} satisfying conditions \ref{itm:sig}--\ref{itm:err4}, except that we replace condition \ref{itm:phi} by \eqref{e:causal}. Then, for any fixed   $0 < \alpha < 1$,  regardless of interplay between $N$ and $\Delta$,
 if we set $t=1$ in \eqref{e:ARp:Gamm}, then
        \begin{align}\label{e:ARp:Gam:kapp}
   N^{\frac{1}{2}} \left(
        \begin{array}{c}
            \Gamma_{0,N}(1) -\kappa_0 \\
            \vdots \\
            \Gamma_{p,N}(1) - \kappa_p
        \end{array}
        \right) \overset{\mathcal{L}aw}{\longrightarrow} \mathcal{N} \left(
        \left(
        \begin{array}{c}
            0  \\
            \vdots \\
             0
        \end{array}
        \right), V
        \right), \quad \text{as }
 N \rightarrow \infty \text{ and } \Delta \rightarrow 0,    \end{align}
if we take integral over $t$, then
    \begin{align}\label{e:ARp:int:Gam:kapp}
 N^{\frac{1}{2}}   \frac{1}{1-\alpha} \int_{\alpha}^1  \left(
        \begin{array}{c}
          \Gamma_{0,N}(t)  -\kappa_0 \\
          \vdots\\
             \Gamma_{p,N}(t)  - \kappa_p
        \end{array}
        \right) dt \overset{\mathcal{L}aw}{\longrightarrow} \mathcal{N} \left(
        \left(
        \begin{array}{c}
            0  \\
            \vdots \\
             0
        \end{array}
        \right), V
        \right),\quad \text{as }
 N \rightarrow \infty \text{ and } \Delta \rightarrow 0,
    \end{align}
where $V$ is defined in \eqref{e:ARp:V}.
\end{corollary}

\noindent{\sc Proof of Corollary \ref{cor:ARp:int:Gam:kapp}}
Follow the lines of proof of Corollary \ref{cor:int:Gam:kapp}.

\rightline{\QED}

\noindent{\sc Proof of Theorems \ref{t:pA} and \ref{t:pB}}
Applying\textit{ multivariate} delta method to the limiting results obtained in Corollary \ref{cor:ARp:int:Gam:kapp} implies the claims of these theorems.
Doing so, first recall the domain $\mathcal{D} \subset \mathbb{R}^p$:
\begin{align*}
 \mathcal{D} = \left \{ \left(x_0,\ldots,x_{p-1} \right) \;|\; X = \left(x_{i-j}\right)_{i,j = 1}^p \text{ is positive definite}\right\},
\end{align*}
and the \textit{multivariate} functions
\begin{align*}
    f  :    \mathcal{D} \times \mathbb{R} \longrightarrow \mathbb{R}^p, \;
    f (x_0,\ldots,x_p) = X^{-1}a,
  \quad
    g : \mathcal{D} \times \mathbb{R} \longrightarrow \mathbb{R}, \;
    g (x_0,\ldots,x_p) = x_0 - \left( X^{-1}a\right)^{T}a , \qquad
\end{align*}
where
\begin{align*}
    X = \left(x_{i-j}\right)_{i,j = 1}^p, \qquad a = \left(x_1,\ldots,x_p \right)^{\top}.
\end{align*}
The above functions $f(\cdot)$ and $g(\cdot)$ are continuously differentiable on their  respective domains. Applying these functions on the limiting result  \eqref{e:ARp:Gam:kapp} implies the claims of Theorem  \ref{t:pA} :
\begin{align*}
    N^{\frac{1}{2}} \left( f\left( \Gamma_{0,N}(1), \boldsymbol{\Gamma}_{p,N}(1)\right) - f\left( \kappa_0,\boldsymbol{\kappa}_p\right) \right)\overset{\mathcal{L}aw}{\longrightarrow} \mathcal{N} \left(
        \left(
        \begin{array}{c}
            0  \\
            \vdots \\
             0
        \end{array}
        \right),  \left(\nabla f\left( \kappa_0 ,\boldsymbol{\kappa}_p\right)\right)^{\top} V \nabla f\left( \kappa_0 ,\boldsymbol{\kappa}_p\right)
        \right).
\end{align*}
Likewise,
\begin{align*}
    N^{\frac{1}{2}} \left( g\left( \Gamma_{0,N}(1), \boldsymbol{\Gamma}_{p,N}(1)\right) - g\left( \kappa_0,\boldsymbol{\kappa}_p\right) \right)\overset{\mathcal{L}aw}{\longrightarrow} \mathcal{N} \left(
        \left(
        \begin{array}{c}
            0  \\
            \vdots \\
             0
        \end{array}
        \right),  \left(\nabla g\left( \kappa_0 ,\boldsymbol{\kappa}_p\right)\right)^{\top} V \nabla g\left( \kappa_0 ,\boldsymbol{\kappa}_p\right)
        \right).
\end{align*}

Applying these functions on the limiting result  \eqref{e:ARp:int:Gam:kapp} implies the claims of Theorem  \ref{t:pB}:
\begin{align*}
    N^{\frac{1}{2}} \left( f\left( \bar\Gamma_{0,N}, \bar{\boldsymbol{\Gamma}}_{p,N}\right) - f\left( \kappa_0, \boldsymbol{\kappa}_p\right) \right)\overset{\mathcal{L}aw}{\longrightarrow} \mathcal{N} \left(
        \left(
        \begin{array}{c}
            0  \\
            \vdots \\
             0
        \end{array}
        \right),  \left(\nabla f\left( \kappa_0, \boldsymbol{\kappa}_p\right)\right)^{\top} V \nabla f\left( \kappa_0, \boldsymbol{\kappa}_p\right)
        \right).
\end{align*}
Likewise,
\begin{align*}
    N^{\frac{1}{2}} \left( g\left( \bar\Gamma_{0,N}, \bar{\boldsymbol{\Gamma}}_{p,N}\right) - g\left( \kappa_0,\boldsymbol{\kappa}_p\right) \right)\overset{\mathcal{L}aw}{\longrightarrow} \mathcal{N} \left(
        \left(
        \begin{array}{c}
            0  \\
            \vdots \\
             0
        \end{array}
        \right),  \left(\nabla g\left( \kappa_0 ,\boldsymbol{\kappa}_p\right)\right)^{\top} V \nabla g\left( \kappa_0 ,\boldsymbol{\kappa}_p\right)
        \right).
\end{align*}

\rightline{\QED}

\subsection{Proof of Theorem \ref{t:pC}}
\noindent{\sc Proof of Theorem \ref{t:pC}}
Proof of this theorem is similar to the proof of Theorem \ref{thm:hat:phi:sig}. So, we  just present the sketch of the proof. Consider the decomposition:
\begin{align}
\nonumber
     \hat{\boldsymbol{\varphi}}- \boldsymbol{\varphi}  =& \frac{1}{1- \alpha}\int_{\alpha}^1\mathbf{\Sigma}_{p,N}^{-1}(t)  \boldsymbol{\Gamma}_{p,N}(t)dt -  \boldsymbol{\varphi}\\ \nonumber
     = & \frac{1}{1- \alpha}\int_{\alpha}^1\mathbf{\Sigma}_{p,N}^{-1}(t)  \boldsymbol{\Gamma}_{p,N}(t)dt -  \mathbf{\Xi}_p^{-1}\boldsymbol{\kappa}_p\\ \nonumber
= & \frac{1}{1- \alpha}\int_{\alpha}^1
     \mathbf{\Xi}_p^{-1} \left( \boldsymbol{\Gamma}_{p,N}(t) -\boldsymbol{\kappa}_p\right)dt
     + \frac{1}{1- \alpha}\int_{\alpha}^1
     \left( \mathbf{\Sigma}_{p,N}^{-1}(t)  - \mathbf{\Xi}_p^{-1}\right) \boldsymbol{\Gamma}_{p,N}(t)dt\\
    =& A+D. \label{e:ARp:decom:hatPhi}
\end{align}
According to \eqref{e:ARp:int:Gam:kapp}, for the term $A$ we have
\begin{align}\label{e:ARp:limA}
  N^{\frac{1}{2}}A  \overset{\mathcal{L}aw}{\longrightarrow} \mathcal{N} \left(0, \mathbf{\Xi}_p^{-1}  W V   W^{\top}\mathbf{\Xi}_p^{-1}\right),
\end{align}
where $W$ is the $p\times (p+1)$ defined in \eqref{e:mtrx:W}.
The second term $D$ appearing in \eqref{e:ARp:decom:hatPhi} is bounded by
\begin{align*}
    \frac{1}{1- \alpha} \left(\int_{\alpha}^1
    \left \Vert  \boldsymbol{\Gamma}_{p,N}(t) \right \Vert^2_E dt\right)^{\frac{1}{2}}
    \left(\int_{\alpha}^1
    \left \Vert \left( \mathbf{\Sigma}_{p,N}^{-1}(t)  - \mathbf{\Xi}_p^{-1}\right)  \right \Vert^2_E dt\right)^{\frac{1}{2}}  
    =:BC,
\end{align*}
where $\Vert \cdot\Vert_E$ stands for Euclidean norm. Proposition \ref{prop:ARp:gam:kapp} and the limiting result \eqref{e:ARp:int:Gam:kapp} imply $N^{\frac{1}{2}}B$ is convergent in law. Extending Lemma \ref{lem:Gamm^-1}  to the elements appearing in the  integrand of $C$ implies $C =  o_P(1)$ is . This gives $N^{\frac{1}{2}}BC =  o_P(1)$. So, $ N^{\frac{1}{2}} \left( \hat{\boldsymbol{\varphi}}- \boldsymbol{\varphi}\right) $ has the same limiting law as of $N^{\frac{1}{2}} A$. See \eqref{e:ARp:limA}.
We now obtain the limiting distribution of $ N^{1/2} \left(  \bar{\sigma}^2_{\varepsilon} -  \sigma^2_{\varepsilon} \right)$.

\begin{align}
\nonumber
\hat{\sigma}^2_\eg - \sigma^2_\eg =& \frac{1}{1- \alpha} \int_{\alpha}^1 \hat{\sigma}^2_\eg (t) dt - \sigma^2_\eg \\ \nonumber
= & \frac{1}{1- \alpha} \int_{\alpha}^1 \left( \Gamma_{0,N}(t)  - \hat{\boldsymbol{\varphi}}^{\top}\boldsymbol{\Gamma}_{p,N}(t)\right) dt - \left(\kappa_0 - \boldsymbol{\varphi}^{\top} \kappa_1 \right) \\ \nonumber
=& - \frac{\left(\hat{\boldsymbol{\varphi}}^{\top} - \boldsymbol{\varphi}^{\top} \right)}{1- \alpha} \int_{\alpha}^1 \boldsymbol{\Gamma}_{p,N}(t) dt
\\ \nonumber
 & +\frac{1}{1- \alpha} \int_{\alpha}^1 \left( \Gamma_{0,N}(t)  - \kappa_0 \right)dt - \frac{\boldsymbol{\varphi}^{\top} }{1- \alpha} \int_{\alpha}^1 \left( \boldsymbol{\Gamma}_{p,N}(t) -\kappa_1 \right)  dt \\ \nonumber
 =:& E \\ \nonumber
 &+ F
\end{align}\
The results \eqref{e:ARp:lim:hatphi} and \eqref{e:ARp:int:Gam:kapp} imply $N^{\frac{1}{2}}E = o_P(1)$. The result \eqref{e:ARp:int:Gam:kapp} also implies
\begin{align*}
    N^{\frac{1}{2}}F \overset{\mathcal{L}aw}{\longrightarrow} \mathcal{N} \left(0, \rho_p \right),\quad \text{where }
    \rho_p = \left( 1, -\boldsymbol{\varphi}^{\top}\right) V \left(\begin{array}{c}
         1  \\
          -\boldsymbol{\varphi}
    \end{array} \right).
\end{align*}
This completes the proof.

\rightline{\QED}

\section{Data cleaning}\label{s:dataclean}
To ensure the data quality of the U.S. stock prices, we clean the data based on the following rules:
\begin{enumerate}\setlength\itemsep{0em}
	\item Prices are adjusted by cumulative price adjust factor downloaded from the CRSP database.
	\item Stocks with over 5\% missing data are eliminated. Otherwise, the missing price is replaced by the latest available price.
	\item Prices that are outside the interval $[0.98\times \mbox{Low}, 1.02 \times \mbox{High}]$ are treated as missing value and replaced by the latest available price, where Low and High denotes the daily lowest and highest prices downloaded from CRSP. We eliminate stocks that have such error more than 5\% of total dates.
	\item Holiday trading days (with shorter opening hours) are eliminated.
	\item Stocks listed less than 1 year are eliminated.
\end{enumerate}

\section{Volatility functions used in simulations}\label{s:HG}
\begin{figure}
	\centering
	\includegraphics[width=\linewidth]{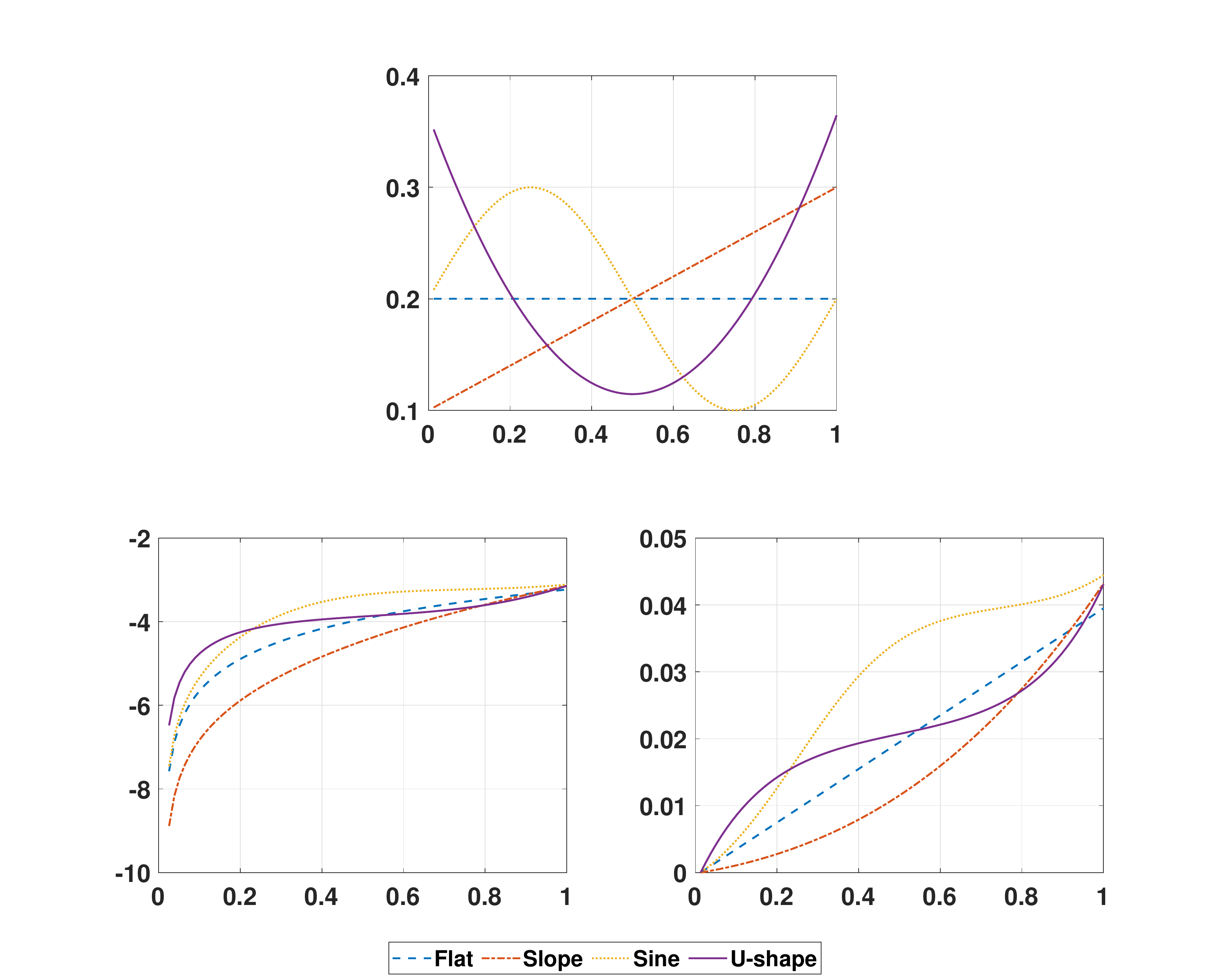}
	\caption{Top: Four choices of simulated $\sigma$ function;
Bottom Left: Corresponding $H$; Bottom Right: Corresponding $G$.}
	\label{fig:simsigmahg}
\end{figure}

Figure \ref{fig:simsigmahg} shows the four volatility functions $\sg$
used in the simulations presented in Section \ref{ss:sim} together
with the corresponding functions $H$ and $G$.

\section{Bias Corrected Estimators}\label{s:bs_est}
Inspired by the bias correction in the scalar AR(1) process, we propose the bias corrected estimators for $\varphi$ and $\sigma_{\varepsilon}$ with detailed steps below:
\begin{itemize}
	\item Compute $\hat{\varphi}(t)$ and  $\hat{\sigma}_\varepsilon(t)$
	$$
	\hat{\varphi}(t) = \Gamma_{0,N}^{-1} (t) \Gamma_{1,N} (t),  \quad \quad  \hat{\sigma}_\varepsilon^2(t) = \Gamma_{0,N} (t) -  \hat{\varphi} (t) \Gamma_{1,N} (t),
	$$
	and their bias corrected version:
	$$
	\hat{\varphi}_{BC}(t) = \hat{\varphi}(t) + \frac{3 \hat{\varphi}(t)}{N},  \quad \quad  \hat{\sigma}_{\varepsilon,BC}^2(t) = \Gamma_{0,N} (t) -  \hat{\varphi}_{BC} (t) \Gamma_{1,N} (t).
	$$
	\item \underline{PROCEDURE A}:
	$$
	\check{\boldsymbol{\theta}}_{BC} = \left[ \widehat{G}(t), \ \check{\varphi}_{BC}, \ \check{\sigma}_{\varepsilon,BC}^2\right],
	$$
	where
	$$
	\check{\varphi}_{BC} = \hat{\varphi}_{BC}(1), \qquad \qquad \check{\sigma}_\varepsilon^2 = \hat{\sigma}_{\varepsilon,BC}^2(1).
	$$
	
	\item \underline{PROCEDURE B}: First calculate
	$$
	\bar{\Gamma}_{0,N} =  \frac{1}{1-\alpha}\int_{\alpha}^{1} \Gamma_{0,N} (t) dt,  \qquad \qquad \bar{\Gamma}_{1,N} =  \frac{1}{1-\alpha}\int_{\alpha}^{1} \Gamma_{1,N} (t) dt.
	$$
	Then compute
	$$
	\bar{\boldsymbol{\theta}}_{BC} = \left[ \widehat{G}(t), \ \bar{\varphi}_{BC}, \ \bar{\sigma}_{\varepsilon,BC}^2\right],
	$$
	where
	$$
	\bar{\varphi} = \bar{\Gamma}_{0,N}^{-1} \bar{\Gamma}_{1,N}, \qquad \qquad \bar{\sigma}_\varepsilon^2 = \bar{\Gamma}_{0,N} - \bar{\varphi} \bar{\Gamma}_{1,N},
	$$
	and their bias corrected version:
	$$
	\bar{\varphi}_{BC} = \bar{\varphi} + \frac{3 \bar{\varphi}}{N}, \qquad \qquad \bar{\sigma}_{\varepsilon,BC}^2 = \bar{\Gamma}_{0,N} - \bar{\varphi}_{BC} \bar{\Gamma}_{1,N},
	$$
	
	\item \underline{PROCEDURE C}:
	$$
	\hat{\boldsymbol{\theta}} = \left[ \widehat{G}(t), \ \hat{\varphi}_{BC}, \ \hat{\sigma}_{\varepsilon, BC}^2\right],
	$$
	where
	$$
	\hat{\varphi}_{BC} = \frac{1}{1-\alpha}\int_{\alpha}^{1} \hat{\varphi}_{BC}(t) dt, \qquad \qquad\hat{\sigma}_\varepsilon^2 = \frac{1}{1-\alpha}\int_{\alpha}^{1} \left[ \Gamma_{0,N} (t) -  {\color{black} \hat{\varphi}_{BC}} \Gamma_{1,N} (t) \right] dt.
	$$
\end{itemize}
\bigskip

Table \ref{tab:est_phi_sigma_bc} presents the results of 
our bias corrected estimators. As expected, the bias corrected estimators show a smaller bias compared to those in Table \ref{tab:est_phi_sigma}. This is particularly prominent in small sample sizes, although some bias can still be observed. In larger samples, the difference in bias between the bias corrected estimators and the original estimators is not obvious. This is because the bias correction is based on the reciprocal of the sample size. In summary, the bias corrected estimators can help to reduce the bias to some extent, particularly in small samples.

\begin{table}[htbp]
\footnotesize 
	\centering
	\caption{Estimation error of the bias corrected estimators 
for $\varphi$ and $\sigma_\varepsilon^2$ with $\alpha =\Delta$}
	\resizebox{\columnwidth}{!}{
		\begin{tabular}{llrccccccc}
			\toprule
			\toprule
			&       &       & \multicolumn{3}{c}{$\varphi$} &       & \multicolumn{3}{c}{$\sigma^2$} \\
			\midrule
			&       &       & Proc. A & Proc. B & Proc. C &       & Proc. A & Proc. B & Proc. C \\
			\midrule
			\textbf{Flat} &       &       &       &       &       &       &       &       &  \\
			\midrule
			\multirow{4}[2]{*}{EB} & $N=100$ &       & -0.027 & -0.072 & -0.057 &       & 0.003 & 0.042 & 0.040 \\
			& $N=500$ &       & -0.014 & -0.058 & -0.043 &       & 0.007 & 0.047 & 0.045 \\
			& $N=1000$ &       & -0.012 & -0.055 & -0.041 &       & 0.008 & 0.048 & 0.046 \\
			& $N=2000$ &       & -0.011 & -0.054 & -0.040 &       & 0.008 & 0.049 & 0.046 \\
			\midrule
			\multirow{4}[2]{*}{ERMSE} & $N=100$ &       & 0.093 & 0.114 & 0.105 &       & 0.036 & 0.057 & 0.055 \\
			& $N=500$ &       & 0.041 & 0.069 & 0.058 &       & 0.018 & 0.050 & 0.048 \\
			& $N=1000$ &       & 0.030 & 0.062 & 0.049 &       & 0.014 & 0.050 & 0.047 \\
			& $N=2000$ &       & 0.022 & 0.057 & 0.044 &       & 0.012 & 0.049 & 0.047 \\
			\midrule
			\midrule
			\textbf{Slope} &       &       &       &       &       &       &       &       &  \\
			\midrule
			\multirow{4}[2]{*}{EB} & $N=100$ &       & -0.029 & -0.072 & -0.058 &       & 0.005 & 0.044 & 0.042 \\
			& $N=500$ &       & -0.016 & -0.058 & -0.044 &       & 0.010 & 0.049 & 0.047 \\
			& $N=1000$ &       & -0.014 & -0.057 & -0.043 &       & 0.010 & 0.050 & 0.047 \\
			& $N=2000$ &       & -0.014 & -0.056 & -0.042 &       & 0.010 & 0.050 & 0.048 \\
			\midrule
			\multirow{4}[2]{*}{ERMSE} & $N=100$ &       & 0.093 & 0.113 & 0.105 &       & 0.036 & 0.058 & 0.056 \\
			& $N=500$ &       & 0.042 & 0.070 & 0.059 &       & 0.019 & 0.052 & 0.050 \\
			& $N=1000$ &       & 0.031 & 0.063 & 0.051 &       & 0.016 & 0.052 & 0.049 \\
			& $N=2000$ &       & 0.024 & 0.059 & 0.046 &       & 0.013 & 0.051 & 0.048 \\
			\midrule
			\midrule
			\textbf{Sine} &       &       &       &       &       &       &       &       &  \\
			\midrule
			\multirow{4}[2]{*}{EB} & $N=100$ &       & -0.029 & -0.071 & -0.057 &       & 0.005 & 0.044 & 0.041 \\
			& $N=500$ &       & -0.017 & -0.059 & -0.045 &       & 0.010 & 0.049 & 0.047 \\
			& $N=1000$ &       & -0.015 & -0.057 & -0.043 &       & 0.011 & 0.050 & 0.047 \\
			& $N=2000$ &       & -0.014 & -0.056 & -0.042 &       & 0.011 & 0.050 & 0.047 \\
			\midrule
			\multirow{4}[2]{*}{ERMSE} & $N=100$ &       & 0.095 & 0.114 & 0.106 &       & 0.036 & 0.058 & 0.056 \\
			& $N=500$ &       & 0.041 & 0.070 & 0.058 &       & 0.020 & 0.052 & 0.050 \\
			& $N=1000$ &       & 0.031 & 0.063 & 0.050 &       & 0.016 & 0.052 & 0.049 \\
			& $N=2000$ &       & 0.024 & 0.059 & 0.046 &       & 0.014 & 0.051 & 0.048 \\
			\midrule
			\midrule
			\textbf{U-Shape} &       &       &       &       &       &       &       &       &  \\
			\midrule
			\multirow{4}[2]{*}{EB} & $N=100$ &       & -0.033 & -0.077 & -0.063 &       & 0.006 & 0.047 & 0.044 \\
			& $N=500$ &       & -0.017 & -0.061 & -0.048 &       & 0.011 & 0.053 & 0.050 \\
			& $N=1000$ &       & -0.017 & -0.061 & -0.048 &       & 0.012 & 0.054 & 0.051 \\
			& $N=2000$ &       & -0.015 & -0.060 & -0.046 &       & 0.013 & 0.054 & 0.052 \\
			\midrule
			\multirow{4}[2]{*}{ERMSE} & $N=100$ &       & 0.096 & 0.119 & 0.110 &       & 0.037 & 0.061 & 0.059 \\
			& $N=500$ &       & 0.042 & 0.073 & 0.062 &       & 0.020 & 0.056 & 0.053 \\
			& $N=1000$ &       & 0.032 & 0.067 & 0.055 &       & 0.017 & 0.055 & 0.053 \\
			& $N=2000$ &       & 0.025 & 0.063 & 0.050 &       & 0.015 & 0.055 & 0.052 \\
			\bottomrule
			\bottomrule
		\end{tabular}}
	\label{tab:est_phi_sigma_bc}
\end{table}

\end{document}